\documentclass[a4paper,fleqn]{cas-sc}

\usepackage[numbers, compress]{natbib}
\usepackage{bm}
\usepackage[ruled,linesnumbered]{algorithm2e}
\usepackage{makecell}
\usepackage{threeparttable}

\def\tsc#1{\csdef{#1}{\textsc{\lowercase{#1}}\xspace}}
\tsc{WGM}
\tsc{QE}

\begin{document}
\let\WriteBookmarks\relax
\def\floatpagepagefraction{1}
\def\textpagefraction{.001}

\let\printorcid\relax

\shorttitle{pyHB: an open-source automatic-differentiation-enhanced semi-analytical solver for nonlinear dynamics}    

\shortauthors{Yuhong Jin et~al.}  

\title [mode = title]{pyHB: an open-source automatic-differentiation-enhanced semi-analytical solver for nonlinear dynamics}  

\tnotemark[1] 

\tnotetext[1]{Source code of PyHB is available at \url{https://github.com/shuizidesu/pyhb}}

%

\author[1]{Yuhong Jin}[type=editor,
      style=chinese
]

\author[1]{Qi Liu}[type=editor,
      style=chinese
]

\author[2]{Lei Hou}[type=editor,
      style=chinese
]







\author[3]{Yi Chen}[type=editor,
      style=chinese
]

\cormark[1]


\ead{chenyi@hit.edu.cn}

\author[2]{Qingye Meng}[type=editor,
      style=chinese
]

\author[1]{Jun Xu}[type=editor,
      style=chinese
]

\author[4]{Hongyuan Fang}[type=editor,
      style=chinese
]





\affiliation[1]{organization={Department of Civil Engineering and Architecture},
            addressline={Nanyang Normal University}, 
            city={Nanyang},
            postcode={473061}, 
            country={P. R. China}}

\affiliation[2]{organization={School of Astronautics},
            addressline={Harbin Institute of Technology}, 
            city={Harbin},
            postcode={150001}, 
            country={P. R. China}}

\affiliation[3]{organization={School of Civil Engineering},
            addressline={Harbin Institute of Technology}, 
            city={Harbin},
            postcode={150090}, 
            country={P. R. China}}

\affiliation[4]{organization={School of Water Conservancy and Transportation},
            addressline={Zhengzhou University}, 
            city={Zhengzhou},
            postcode={450001}, 
            country={P. R. China}}

\cortext[1]{Corresponding author}



\begin{abstract}
      The Harmonic Balance (HB) method is widely used to compute and analyze the periodic responses of nonlinear systems. However, its application to high-dimensional complex systems is limited by the burden of handling the partial derivatives of the nonlinearities. To address this issue, this work presents pyHB, an open-source, automatic-differentiation-enhanced semi-analytical framework that integrates the complete HB workflow for general user-defined nonlinear systems. The proposed formulation exploits localized nonlinearities and applies PyTorch-based automatic differentiation (AD) only to the reduced nonlinear force, thereby avoiding the need for user-supplied derivatives of the nonlinear force and maintaining controllable GPU memory usage. Weighted arc-length continuation, sparse matrix assembly, a blocked solution strategy for the augmented continuation equations, and Floquet-based stability analysis are incorporated within a modular architecture that separates model definition from reusable numerical procedures. Hence, pyHB is capable of providing a complete landscape of the nonlinear system's periodic response based solely on the user-defined dynamical equations. Four examples, including a quasi-zero-stiffness isolator, a nonlinear piezoelectric energy harvester, a 284 degrees of freedom (DOFs) aeroengine model, and a 2000 DOFs Bernoulli beam, demonstrate the ability of pyHB to trace stable and unstable solution branches and capture subharmonic resonance, combination resonance, and mixed-order electromechanical responses. Notably, in the Bernoulli beam example with 202000 HB unknowns, the AD-enhanced solver requires approximately 0.44s per continuation point, achieving several-hundred-fold speedup compared to the Newmark-$\beta$ method and remaining 637.8MB of additional RAM and 243.5MB of GPU memory. The proposed pyHB provides a general, one-stop benchmark platform for HB-based nonlinear dynamics analysis.
\end{abstract}

\begin{graphicalabstract}
      \includegraphics[width=1.0\textwidth]{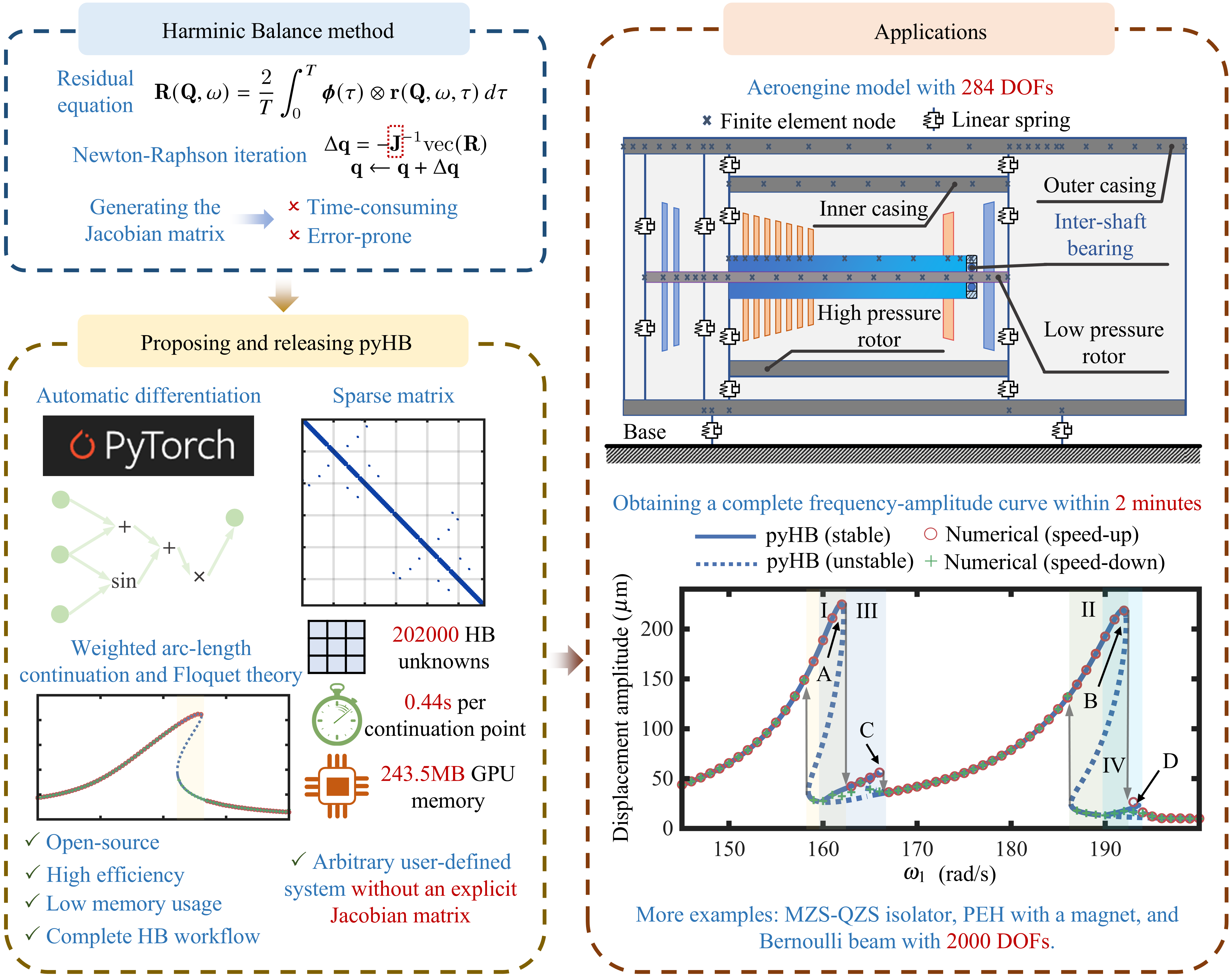}
\end{graphicalabstract}

\begin{highlights}
\item AD automates nonlinear-force Jacobian evaluation without explicit derivatives.
\item Sparse matrix techniques ensure high efficiency and controllable memory usage.
\item pyHB integrates the complete HB workflow, from solving to stability analysis.
\item Open-source code includes extensive examples and reusable model templates.
\end{highlights}


\maketitle

\section{Introduction}
The motion and vibration of dynamical systems are typically described by ordinary differential equations after discretization \cite{MEI2026111521, ZHANG2026115115}. In real-world engineering applications, the periodic response of a system during steady-state operation is of particular interest \cite{THEODOSIOU20093565, SU20261}. To accurately model and characterize the system's behavior, complex nonlinearities are generally incorporated, making it highly challenging to determine the periodic response \cite{dongComprehensiveStudyCoupled2022}. Therefore, developing highly versatile and efficient solvers for obtaining the periodic response of nonlinear systems holds significant theoretical importance.

Numerical time-integration methods are widely used in analyzing the periodic responses of nonlinear systems due to their great generality \cite{karamLowcostRungeKuttaIntegrators2021, Wang2026115162, SOKOLOV2026117017}. Based on differences in the stepping format, these approaches are categorized as explicit, implicit, and explicit-implicit hybrid methods. Explicit approaches, including the forward Euler method \cite{ZHANG2022108048, HANNA19881083}, the explicit Runge-Kutta method \cite{FALEICHIK2026117061}, and the explicit linear multistep method \cite{MEI20129547}, depend solely on known results from previous steps to compute the solution at the next time step. They do not require solving high-dimensional systems of linear or nonlinear equations and are highly computationally efficient, hence have widespread applications in rotor dynamics \cite{changMonolithicApproachesTransient2025}, fluid mechanics \cite{QIN2024106089}, and nonlinear circuit systems \cite{elabbassiControlNonlinearDynamic2025}.s However, the time step of explicit methods is constrained by stability conditions; when dealing with high-dimensional stiff problems, extremely small time steps may be required to ensure stability, resulting in a significant decrease in computational efficiency \cite{liIdenticalSecondorderSingle2021}. In contrast, implicit methods, including the backward Euler method \cite{LIU201220}, the implicit Runge-Kutta method \cite{GUI2026110088}, and the generalized-$\alpha$ method and Newmark-$\beta$ method which are widely used in structural dynamics \cite{liRigidflexibleCouplingDynamic2025, liuNonlinearDynamicsThreedimensional2024, chowdhuryNonlinearStiffenedInertial2025}, require the solution of a system of algebraic equations (typically nonlinear) at each time step. Although their per-step computational efficiency is lower than that of explicit methods, they offer ideal numerical stability and are therefore indispensable for handling high-dimensional, strongly nonlinear systems \cite{CHEN2026113858, WANG2026110403, YIN2025104022}. Explicit-implicit hybrid methods effectively combine the strengths of the two paradigms mentioned above, achieving an excellent trade-off between computational efficiency, stability, and accuracy, providing a more advanced route for designing numerical integrators \cite{PITERSKAYA2026100683, YIN2026126751, ZHANG2026106977}. However, the above numerical integration methods share a common inherent drawback when solving the system's steady-state periodic response: waiting for the free oscillation to gradually decay, resulting in a lengthy solution process and significant waste of computation resources \cite{cardonaMultiharmonicMethodNonlinear1994}.

Analytical methods derive closed-form solutions for the periodic response of nonlinear systems, providing a rigorous foundation for understanding nonlinear behavior. Represented by the perturbation method \cite{keQuantitativeAnalysisLimit2024}, the multiple time scale method \cite{SU2027117138}, and the Harmonic Balance (HB) method \cite{kangInvestigationTunedClutch2026a}, these techniques have found important applications in the analysis and interpretation of nonlinear phenomena, including internal resonance \cite{changInvestigation122022}, superharmonic resonance \cite{liStudySuperharmonicparametricCombined2026}, subharmonic resonance \cite{WU2020112056}, multistability \cite{linNonlinearVibrationStability2024a}, and bifurcation mechanisms \cite{liCoupledmodeSubharmonicResonance2025}. However, the perturbation method and the multiple time scale method rely on the assumption of small-parameter expansion, and their applicability is limited to cases of weak nonlinearity \cite{xuDynamicCharacteristicsReliability2020}. Furthermore, the computational processes for both methods entail complex, extensive symbolic derivations, which are feasible only for low-dimensional systems \cite{yangNonlinearDynamicsAxially2014}. In contrast, the HB method, which transforms the ODE into an algebraic equation in terms of harmonic coefficients by expanding the system's periodic solution into a truncated Fourier series \cite{wangThreemagnetringQuasizeroStiffness2024b}, has attracted researchers' widespread attention due to its relatively simple derivation and its ability to handle high-dimensional and strongly nonlinear systems \cite{al-shudeifatGeneralHarmonicBalance2010, mohamedAutoencoderbasedReductionHarmonic2025}. However, the classical HB method adopts algebraic relationships among trigonometric functions to perform Fourier series expansions and is unable to handle non-polynomial nonlinear factors, which severely limits its applicability in complex engineering systems. 

To address the shortcomings of the classical HB method, Lau and Cheung proposed the Incremental Harmonic Balance (IHB) method \cite{lauAmplitudeIncrementalVariational1981}, while Kim and Noah developed the HB-Alternating Frequency/Time domain (HB-AFT) method \cite{kimStabilityBifurcationAnalysis1991}. The former introduces linearized incremental equations and Galerkin projections, whereas the latter uses the discrete Fourier transform (often implemented via the fast Fourier transform (FFT)) to handle the harmonic components of nonlinear forces. Both of the aforementioned foundational works fall within the framework of the HB-Newton iteration \cite{chopraNonlinearDynamicResponse1979}; hence, researchers commonly refer to this approach as the semi-analytical method \cite{yanHarmonicBalanceMethods2023}. The semi-analytical method is well-suited and robust for complex nonlinearities of non-polynomial types \cite{zhangDynamicModelingParameter2024} and can track complete solution branches, including unstable periodic orbits, by combining with numerical continuation techniques \cite{litewkaNonlinearHarmonicVibrations2026}, thereby supporting a comprehensive understanding of a system's nonlinear characteristics and has been widely applied in various fields, including aerospace \cite{changModelingNonlinearAnalysis2024, linSynchronousImpactPhenomenon2023}, vibration isolation \cite{mengQuasizeroStiffnessVibration2025, leiEmbeddingIntelligentExcitation2025}, metamaterials \cite{wuRotationalinertiaenhancedQuasizerostiffnessChiral2026, WEI2020105433}, non-smooth network systems \cite{CHEN2023104256}, and nonlinear system identification \cite{jinPhysicsinformedHarmonicBalance2026}. In recent years, researchers have further developed improved methods, such as the reconstruction HB method \cite{daiTimeDomainCollocation2014, daiCollocationbasedHarmonicBalance2023a} and indirect HB method based on the frequency response function \cite{liangIntegratedApproachResponse2024, chenIndirectHarmonicBalance2025a}, and also conducted a series of fruitful experiments aimed at mitigating the Gibbs phenomenon in non-smooth systems \cite{colaitisHarmonicBalanceMethod2021}, adaptively selecting initial guesses \cite{liGeneralizedIncrementalHarmonic2025c}, enhancing the convergence of iterative processes \cite{liEnhancedIncrementalHarmonic2025, wuImprovingSolutionProcedure2025}, and extending these methods to the solution of differential-algebraic equations \cite{juEfficientGalerkinAveragingincremental2021a, zhouConstraintEliminationbasedHarmonic2026b} and quasi-periodic responses \cite{HUANG2021103767, liTwotimescaleConvergenceenhancedEfficient2026a}, resulting in numerous outstanding contributions. 

Despite these advances, two important limitations remain. First, although FFT-based harmonic projection and chain-rule formulations have substantially reduced the number of symbolic operations in the IHB and HB-AFT methods, the partial derivatives of the nonlinear force are still generally required to be provided explicitly. Their manual derivation is time-consuming and error-prone for systems with complex nonlinearities. Second, software support and reproducibility have not kept pace with methodological development. Most studies focus on theoretical formulations and case-specific results. Consequently, their code implementations are either unavailable or tightly coupled to a particular model, making reuse for arbitrary user-defined systems difficult. Although ANSYS provides APDL support for the HB-AFT method in its new version \cite{ansys}, and Malte and Johann have developed the NLvib toolbox \cite{nlvib} for the HB method, both are limited to specific nonlinear types and lack versatility. Our previous work \cite{chenHarmonicBalanceautomaticDifferentiation2026} innovatively introduced automatic differentiation (AD) into the IHB solving process, eliminating the need to derive the derivatives of the nonlinear force. We also released a complete code implementation. However, because automatic differentiation is coupled with discrete response reconstruction, FFT projection, and other operations, the computational graph size and GPU memory usage increase rapidly with system scale and the number of harmonic orders, leading to a GPU memory bottleneck in high-dimensional systems. Besides, this work does not provide a corresponding implementation for stability determination and fails to cover the complete HB workflow.

This work substantially extends our previous work \cite{chenHarmonicBalanceautomaticDifferentiation2026} and introduces pyHB, an open-source, automatic-differentiation-enhanced, semi-analytical solver for nonlinear dynamics. AD is applied to the reduced nonlinear force associated with the local nonlinear degrees of freedom (DOFs), avoiding the drawbacks of symbolic differentiation (expression explosion) and numerical differentiation (the trade-off between truncation and round-off errors) while maintaining controllable GPU memory usage. By integrating model definition, HB precomputation, arc-length continuation, and Floquet multiplier calculation, pyHB requires users to provide only the system's dynamical equations via its modular interface to complete the full HB workflow, from periodic response calculation to branch tracking and stability analysis, without reimplementing any core algorithms. Sparse matrix techniques and a carefully designed blocked solving strategy are introduced to ensure pyHB's computational efficiency and memory scalability. Overall, pyHB provides a modified implementation and a benchmark platform for HB-based nonlinear analysis, enabling researchers to focus on developing new methods and applications without being hindered by implementation complexities.

\section{Preliminary theories}
This section explains the notations and basic theories needed for the subsequent content. Note that no new methodological or mathematical contributions are presented here. 

\subsection{Problem definition}
Generally, a multi-degree-of-freedom (MDOF) nonlinear system can be expressed as
\begin{equation}
      \label{eq:MODF_nonlinear_system}
      \mathbf{M}\ddot{\mathbf{x}} + (\mathbf{C} + \omega \mathbf{G}) \dot{\mathbf{x}} + \mathbf{K}\mathbf{x} + \mathbf{f}_{nl}(\ddot{\mathbf{x}}, \dot{\mathbf{x}}, \mathbf{x}, t, \omega) = \mathbf{f}_{ex}(t, \omega),
\end{equation}
where $t$ is the time variable, and $\mathbf{x}, \dot{\mathbf{x}}, \ddot{\mathbf{x}} \in \mathbb{R}^n$ represent the system's displacement, velocity, and acceleration vectors, respectively. $n$ is the number of DOF. $\dot{()}, \ddot{()}$ denote the first and second derivatives with respect to time, and $\mathbf{f}_{ex}(t, \omega) \in \mathbb{R}^n$ is the external force. Specifically, the steady-state response of the system under harmonic excitation at base frequency $\omega$ is considered; hence, $\mathbf{f}_{ex}$ is periodic. $\mathbf{M}, \mathbf{C}, \mathbf{K}$ are the mass, damping, and stiffness matrices adopted to describe the system's linear properties. Besides, Eq.\eqref{eq:MODF_nonlinear_system} also considers the gyroscopic matrix $\mathbf{G}$ utilized to describe the gyroscopic effect, which is common in rotating machinery \cite{chenCombinationResonancesDualrotorbearingcasing2024, linNonlinearVibrationStability2024a, wangDynamicCharacteristicsHorizontal2026}. $\mathbf{f}_{nl} \in \mathbb{R}^{n}$ encapsulates general nonlinear forces, which may depend on $\mathbf{x}, \dot{\mathbf{x}}, \ddot{\mathbf{x}}, t, \omega$, and their coupling. Introducing the dimensionless time variable $\tau = \omega t$, Eq.\eqref{eq:MODF_nonlinear_system} can be rewritten as
\begin{equation}
      \label{eq:MODF_nonlinear_system_dimensionless}
      \omega^2 \mathbf{M} \mathbf{x}'' + (\mathbf{C} + \omega \mathbf{G}) \omega \mathbf{x}' + \mathbf{K}\mathbf{x} + \mathbf{f}_{nl}(\mathbf{x}'', \mathbf{x}', \mathbf{x}, \tau, \omega) = \mathbf{f}_{ex}(\tau, \omega),
\end{equation}
where $()' = d() / d \tau$ and $()'' = d^{2}() / d \tau^{2}$ represent the first and second derivatives with respect to $\tau$. We focus on the periodic response analysis problem of Eq.\eqref{eq:MODF_nonlinear_system_dimensionless}, particularly the potential bifurcation and unstable solutions controlled by $\mathbf{f}_{nl}$.

\subsection{Tensor computation conventions}
The subsequent theoretical derivations will involve contraction, rearrangement, and differentiation of matrices and higher-order tensors. This subsection clarifies conventions for the relevant operations to ensure the manuscript flows smoothly and the notation is clear. Consider a matrix $\mathbf{G} \in \mathbb{R}^{a \times b}$, whose column-priority vectorization $\text{vec}()$ is defined as
\begin{equation}
      \label{eq:vectorization}
      \mathbf{q} = \text{vec}(\mathbf{G}) = [\mathbf{G}_{:,1}^{T}, \mathbf{G}_{:,2}^{T}, \cdots, \mathbf{G}_{:,b}^{T}]^{T} \in \mathbb{R}^{ab},
\end{equation}
where $\mathbf{G}_{:,i}$ is the $i$-th column of $\mathbf{G}$. Correspondingly, the inverse operation $\text{mat}()$ is defined as
\begin{equation}
      \label{eq:inverse_vectorization}
      \mathbf{G} = \text{mat}(\mathbf{q}; a, b) = [\mathbf{q}_{1:a}, \mathbf{q}_{a+1:2a}, \cdots, \mathbf{q}_{(b-1)a+1:ba}] \in \mathbb{R}^{a \times b}.
\end{equation}
Furthermore, for a matrix function $\mathbf{H}(\mathbf{G}) \in \mathbb{R}^{c \times d}$, its partial derivative with respect to $\mathbf{G}$ can be represented as a fourth-order tensor $\partial \mathbf{H} / \partial \mathbf{G} \in \mathbb{R}^{c \times d \times a \times b}$, and the following relation holds
\begin{equation}
      \label{eq:partial_derivative_tensor}
      \mathbf{J} = \text{Mat}(\frac{\partial \mathbf{H}}{\partial \mathbf{G}} ; cd, ab) = \frac{\partial \text{vec}(\mathbf{H})}{\partial \text{vec}(\mathbf{G})} \in \mathbb{R}^{cd \times ab},
\end{equation}
where $\text{Mat}()$ denotes the matrixization operation, and $\mathbf{J}$ is known as the Jacobian matrix. 

Moreover, let $\otimes$ denote the Kronecker product of two matrices and the outer product of two higher-order tensors or vectors, and let $\bullet^{i}_{j}$ represent the contraction between the $i$-th mode of the first tensor and the $j$-th mode of the second tensor. For example, for two third-order tensors $\mathbf{A} \in \mathbb{R}^{a \times b \times c}$ and $\mathbf{B} \in \mathbb{R}^{c \times d \times e}$, the contraction $\mathbf{C} = \mathbf{A} \bullet^{3}_{1} \mathbf{B}$ yields a fourth-order tensor $\mathbf{C} \in \mathbb{R}^{a \times b \times d \times e}$, with its elements defined as
\begin{equation}
      \label{eq:tensor_contraction}
      \mathbf{C}_{i,j,k,l} = \sum_{m=1}^{c} \mathbf{A}_{i,j,m} \mathbf{B}_{m,k,l}, \quad i=1,\cdots,a, j=1,\cdots,b, k=1,\cdots,d, l=1,\cdots,e.
\end{equation}
One of the most common forms of Eq.\eqref{eq:tensor_contraction} is the contraction over the last mode of $\mathbf{A}$ and the first mode of $\mathbf{B}$. In this case, for the sake of brevity, the sub- and super-scripts of $\bullet^{i}_{j}$ can be omitted, that is
\begin{equation}
      \label{eq:tensor_contraction_simplified}
      \mathbf{C} = \mathbf{A} \bullet \mathbf{B}.
\end{equation}
The permutation operation $\text{perm}()$ is defined to rearrange the tensor's modes. For instance, for a fourth-order tensor $\mathbf{C} \in \mathbb{R}^{a \times b \times c \times d}$, the permutation
\begin{equation}
      \label{eq:tensor_permutation}
      \mathbf{D} = \text{perm}(\mathbf{C}; 3, 1, 4, 2)
\end{equation}
yields a fourth-order tensor $\mathbf{D} \in \mathbb{R}^{c \times a \times d \times b}$. The above conventions will be adopted repeatedly in the subsequent theoretical derivations.

\subsection{Harmonic balance method for the nonlinear system}
The Harmonic Balance (HB) method is a widely used semi-analytical method for analyzing the periodic response of nonlinear systems \cite{arangomontoyaHarmonicBalanceMethod2026a, leeComputationalFrameworkBased2026a, zhangTheoreticalAnalysisCoupled2025a}. Compared to the numerical technique that performs time integration step-by-step, the HB method directly processes the system's steady-state response in the frequency domain, offers high computational efficiency, and has gained popularity among researchers \cite{saadatmandNonlinearVibrationAnalysis2025a, zhouMultiPassageHarmonicBalance2024a, mohamedMultifidelityHarmonicBalance2025a}. Formally, the HB method represents the periodic steady-state response of Eq.\eqref{eq:MODF_nonlinear_system_dimensionless} as a truncated Fourier series, yielding
\begin{subequations}
      \label{eq:fourier_series}
      \begin{align}
            \bm{\phi} (\tau) &= [1, \cos(h_{1} \tau), \cdots, \cos(h_{m} \tau), \sin(h_{1} \tau), \cdots, \sin(h_{m} \tau)] \in \mathbb{R}^{2m+1}, \label{eq:fourier_series_1} \\
            \mathbf{x} &= \bm{\phi}(\tau) \mathbf{Q}, \label{eq:fourier_series_2} \\
            \mathbf{x}' &= \bm{\phi}'(\tau) \mathbf{Q}, \label{eq:fourier_series_3} \\
            \mathbf{x}'' &= \bm{\phi}''(\tau) \mathbf{Q}, \label{eq:fourier_series_4}
      \end{align}
\end{subequations}
where $\bm{\phi}(\tau)$ is the Fourier basis vector, $h_{i}, i = 1, \cdots, m$ denote the harmonic frequencies, and $m$ is the number of harmonics. Typically, $h_{i}$ takes integer values, i.e., $h_{i} = i$; other cases will be discussed later. For convenience, denote $o = 2m + 1$ and $\mathbf{Q} \in \mathbb{R}^{o \times n}$ as the harmonic coefficient matrix, which is the unknown to be solved. Substituting Eq.\eqref{eq:fourier_series} into Eq.\eqref{eq:MODF_nonlinear_system_dimensionless} and applying the Galerkin projection method, we can obtain the HB residual equation as follows
\begin{subequations}
      \label{eq:HB_residual_equation}
      \begin{align}
            \mathbf{R}(\mathbf{Q}, \omega) &= \frac{2}{T} \int_{0}^{T} \bm{\phi}(\tau) \otimes \mathbf{r}(\mathbf{Q}, \omega, \tau)\, d\tau \in \mathbb{R}^{o \times n}, \label{eq:HB_residual_equation_1} \\
            \mathbf{r}(\mathbf{Q}, \omega, \tau) &= \mathbf{f}_{ex}(\tau, \omega)-(\omega^2 \bm{\phi}(\tau) \mathbf{Q} \mathbf{M}^{T} + \omega \bm{\phi}'(\tau) \mathbf{Q} \mathbf{C}^{T} + \omega \bm{\phi}'(\tau) \mathbf{Q} \mathbf{G}^{T} + \bm{\phi}(\tau) \mathbf{Q} \mathbf{K}^{T}) \notag \\
            &\quad - \mathbf{f}_{nl}(\bm{\phi}''(\tau) \mathbf{Q}, \bm{\phi}'(\tau) \mathbf{Q}, \bm{\phi}(\tau) \mathbf{Q}, \tau, \omega) \in \mathbb{R}^{n}, \label{eq:HB_residual_equation_2}
      \end{align}
\end{subequations}
where $T = 2\pi$ represents the maximum period of $\bm{\phi}(\tau)$ with integer $h_{i}$. 

The HB method aims to find the harmonic coefficient matrix $\mathbf{Q}$ that satisfies $\mathbf{R}(\mathbf{Q}, \omega) = 0$, and the Newton-Raphson iteration method is a common approach for solving this problem, which can be expressed as
\begin{subequations}
      \label{eq:NR_iteration}
      \begin{align}
            \mathbf{q} &= \text{vec}(\mathbf{Q}), \label{eq:NR_iteration_1} \\
            \mathbf{J} &= \frac{\partial \text{vec}(\mathbf{R})}{\partial \text{vec}(\mathbf{Q})} = \frac{\partial \text{vec}(\mathbf{R})}{\partial \mathbf{q}}, \label{eq:NR_iteration_2} \\
            \Delta \mathbf{q} &= -\mathbf{J}^{-1} \text{vec}(\mathbf{R}), \\
            \mathbf{q} &\leftarrow \mathbf{q} + \Delta \mathbf{q}, \label{eq:NR_iteration_3} \\
            \mathbf{Q} &= \text{mat}(\mathbf{q}; o, n), \label{eq:NR_iteration_4}
      \end{align}
\end{subequations}
where $\mathbf{q}$ is the vectorized form of $\mathbf{Q}$, initiallizing by a random guess or the given condition. Threshold of the residual and increment norm, $\|\text{vec} (\mathbf{R})\|$ and $\|\Delta \mathbf{q}\|$, are set to determine convergence. 

\textbf{Remark 1: } As shown in Eqs.\eqref{eq:fourier_series} to \eqref{eq:NR_iteration}, the core of the HB method is the efficient computation of the projected residual equation and its corresponding Jacobian matrix. Depending on the implementation, researchers have further classified the HB method into two categories: the Incremental Harmonic Balance (IHB) method \cite{liTwotimescaleConvergenceenhancedEfficient2026a, wangModifiedIncrementalHarmonic2015a, huangQuasiperiodicMotionsHighdimensional2019} and the HB-Alternating Frequency/Time domain (HB-AFT) method \cite{yuanHarmonicBalanceApproach2019, sunComputationallyEfficientMethod2024, chenNonlinearDynamicsAnalysis2023a}. However, Ju et al. \cite{juComparisonIncrementalHarmonic2021} rigorously proved that the IHB and HB-AFT methods are mathematically equivalent, with the former replacing the latter's explicit Galerkin projection calculations with the Fast Fourier Transform (FFT) algorithm. Therefore, this manuscript does not distinguish between the two methods and collectively refers to them as the HB method. 

\subsection{Arc-length continuation method}
When solving the problem initially, $\omega$ in Eq. \eqref{eq:HB_residual_equation} is typically specified. The system's amplitude-frequency response is a key dynamic characteristic; however, due to phenomena such as multistability, subharmonic resonance, and superharmonic resonance caused by nonlinearity, using a fixed $\omega$ increment is generally hard to track the system's complete solution branch and may even cause the Newton-Raphson iteration to fail to converge near the bifurcation point \cite{linNovelAdaptiveHarmonic2023, speksnijderApplicationHarmonicBalance2025a, liModelingDynamicAnalysis2026a}. To address this issue, the arc-length continuation method is employed to track the continuous solution branch of Eq. \eqref{eq:HB_residual_equation} by treating $\omega$ as an unknown variable \cite{zhangNonlinearVibrationsPoroushyperelastic2024, woiwodeComparisonTwoAlgorithms2020, leeProperGeneralizedDecompositionbased2023}.

Let $\mathbf{q}^{(0)}$ and $\mathbf{Q}^{(0)}$ be the convergent solution obtained by Eq.\eqref{eq:NR_iteration} at a given initial frequency $\omega^{(0)}$, and the augmented varible is defined as $\mathbf{z} = [\mathbf{q}, \omega] \in \mathbb{R}^{on+1}$. The arc-length continuation method introduces an additional constraint equation to solve the next solution, which can be expressed as
\begin{equation}
      \label{eq:arc_length_constraint}
      h(\mathbf{z}) = \langle  \mathbf{z} - \mathbf{z}^{(0)}, \mathbf{v}^{(0)} \rangle  - s,
\end{equation}
where $\langle \rangle$ denotes the inner product of the vectors. $\mathbf{v}^{(0)} \in \mathbb{R}^{on+1}$ is the unit tangent vector at $\mathbf{z}^{(0)} = [\mathbf{q}^{(0)}, \omega^{(0)}]$, which lies in the null space of the augmented Jacobian matrix $[\mathbf{J}, \partial \text{vec}(\mathbf{R}) / \partial \omega] \in \mathbb{R}^{on \times (on+1)}$ at $\mathbf{z}^{(0)}$, and $s$ is the arc-length step size. Thus, the iteration process of the augmented variable $\mathbf{z}$ can be given as
\begin{subequations}
      \label{eq:arc_length_NR_iteration}
      \begin{align}
            \mathbf{q} &= \mathbf{z}_{1: on}, \\
            \mathbf{Q} &= \text{mat}(\mathbf{q}; o, n), \\
            \omega &= \mathbf{z}_{on+1}, \\
            \Delta \mathbf{z}
            &= -
            \begin{bmatrix}
            \mathbf{J}
            &
            \partial \text{vec}(\mathbf{R}) / \partial \omega \\
            \multicolumn{2}{c}{\mathbf{v}^{(0)}}
            \end{bmatrix}^{-1}
            \begin{bmatrix}
            \text{vec}(\mathbf{R}) \\
            h(\mathbf{z})
            \end{bmatrix}, \\
            \mathbf{z} &\leftarrow \mathbf{z} + \Delta \mathbf{z},
      \end{align}
\end{subequations}
where $\mathbf{z}$ is initialized as $\mathbf{z}^{(0)} + s \mathbf{v}^{(0)}$, know as the predictor-corrector scheme. After Eq.\eqref{eq:arc_length_NR_iteration} converges, the tangent vector $\mathbf{v}^{(0)}$ is updated by
\begin{subequations}
      \label{eq:tangent_vector_update}
      \begin{align}
            \mathbf{v}^{(0)} &\leftarrow \begin{bmatrix}
            \mathbf{J}
            &
            \partial \text{vec}(\mathbf{R}) / \partial \omega \\
            \multicolumn{2}{c}{\mathbf{v}^{(0)}}
            \end{bmatrix}^{-1} \begin{bmatrix}
            \mathbf{0} \\ 1
            \end{bmatrix}, \label{eq:tangent_vector_update_1} \\
            \mathbf{v}^{(0)} &\leftarrow \frac{\mathbf{v}^{(0)}}{\|\mathbf{v}^{(0)}\|}. \label{eq:tangent_vector_update_2}
      \end{align}
\end{subequations}
Eq.\eqref{eq:tangent_vector_update_2} ensures that the new $\mathbf{v}^{(0)}$ is normalized. Moreover, the arc-length step size $s$ can be adaptively adjusted based on the convergence status of Eq.\eqref{eq:arc_length_NR_iteration} to improve the efficiency of the continuation process. 

\subsection{Free-frequency problem}
Eq.\eqref{eq:MODF_nonlinear_system_dimensionless} addresses the forced vibration problem, in which $\omega$ is explicitly specified by $\mathbf{f}_{ex}$. By contrast, in self-excited or free-vibration systems, the oscillation frequency is an intrinsic property of the system and remains unknown even at the start of the solution process (as shown in Eq.\eqref{eq:NR_iteration}) \cite{wuNonlinearDynamicAnalysis2025, wuImprovingSolutionProcedure2025, guoSemianalyticalHopfBifurcation2026}. This manuscript refers to this type of problem as the free-frequency problem, and the augmented variable $\mathbf{z} = [\mathbf{q}, \omega]$ to be solved initially is the same as in the arc-length continuation process. 

A typical characteristic of free-frequency problems is that the system's solution depends on the specific initial conditions, and its representation in the frequency domain is not unique. To restore uniqueness, a phase condition, denoted as $g(\mathbf{z}) = 0$, is introduced, and the most common form is to specify a particular harmonic coeffcient in $\mathbf{Q}$ to be a constant value \cite{zhouVariablecoefficientHarmonicBalance2015, huangIncrementalHarmonicBalance2021, klingerVanPolBehavior1995}, given by
\begin{equation}
      \label{eq:phase_condition}
      g_{fix}(\mathbf{z}) = \mathbf{Q}_{i,j} - c,
\end{equation}
where $i$ and $j$ correspond to the indices of the DOF and the harmonic coefficient, respectively. $c$ is set to zero generally. Eq.\eqref{eq:phase_condition} essentially constrains the initial displacement or velocity of the periodic response.

The arc-length continuation method can also be applied to the free-frequency problem. However, the continuation variable is not $\omega$; it may be the damping ratio, the excitation amplitude, or other system parameters (denoted as $\lambda$). Hence, the further-augmented variable is defined as $\bar{\mathbf{z}} = [\mathbf{q}, \omega, \lambda]$, and the continuation process can be realized using the phase condition and Eqs.\eqref{eq:arc_length_constraint} to \eqref{eq:tangent_vector_update}. Note that Eq.\eqref{eq:phase_condition} may result in phase drift during the continuation process, which may introduce numerical instability. Hence, a more general and robust phase condition used in AUTO software \cite{seydelPracticalBifurcationStability2010, Doedel1997AUTO2} is adopted, described as
\begin{equation}
      \label{eq:phase_condition_general}
      g_{gen}(\bar{\mathbf{z}}) = \langle \mathbf{q}^{(0)}, (\mathbf{I}_{n}  \otimes \mathbf{D}_{1}) \mathbf{q} \rangle,
\end{equation}
where $\mathbf{D}_{1} \in \mathbb{R}^{o \times o}$ is the first-order differential matrix of the Fourier basis vector $\bm{\phi}(\tau)$, satisfying $\bm{\phi}'(\tau) = \bm{\phi}(\tau) \mathbf{D}_{1}$, and $\mathbf{I}_{n}$ is the $n \times n$ identity matrix. Similarly, the second-order differential matrix $\mathbf{D}_{2} \in \mathbb{R}^{o \times o}$ is defined by $\bm{\phi}''(\tau) = \bm{\phi}(\tau) \mathbf{D}_{2}$. $\mathbf{D}_{1}$ and $\mathbf{D}_{2}$ will be utilized in the subsequent sections. 

\subsection{Stability of the periodic response}
Floquet theory is a classical method for analyzing the stability of periodic responses in nonlinear systems \cite{TAEGE2025680, WANG2025136279, RIAHI2025109507}. Consider a small perturbation $\delta \mathbf{x}(\tau)$ around the periodic solution $\mathbf{x}(\tau)$, the linearized perturbation equation of Eq.\eqref{eq:MODF_nonlinear_system_dimensionless} is given by
\begin{equation}
      \label{eq:linearized_perturbation_equation}
      \underbrace{(\omega^2 \mathbf{M} + \frac{\partial \mathbf{f}_{nl}}{\partial \mathbf{x}''})}_{\bar{\mathbf{M}}(\tau)} \delta \mathbf{x}'' + \underbrace{(\omega \mathbf{C} + \omega^{2} \mathbf{G} + \frac{\partial \mathbf{f}_{nl}}{\partial \mathbf{x}''})}_{\bar{\mathbf{C}}(\tau)} \delta \mathbf{x}' + \underbrace{(\mathbf{K} + \frac{\partial \mathbf{f}_{nl}}{\partial \mathbf{x}})}_{\bar{\mathbf{K}}(\tau)} \delta \mathbf{x} = 0,
\end{equation}
where $\partial \mathbf{f}_{nl} / \partial \mathbf{x}''$, $\partial \mathbf{f}_{nl} / \partial \mathbf{x}'$, and $\partial \mathbf{f}_{nl} / \partial \mathbf{x} \in \mathbb{R}^{n \times n}$ are the Jacobian matrices of the nonlinear force with respect to $\mathbf{x}''$, $\mathbf{x}'$, and $\mathbf{x}$, respectively, known as the tangent mass, damping, and stiffness matrices. $\bar{\mathbf{M}}(\tau)$, $\bar{\mathbf{C}}(\tau)$, and $\bar{\mathbf{K}}(\tau)$ are periodic matrix functions. Introducing the first-order state variable $\mathbf{y} = [\delta \mathbf{x}, \delta \mathbf{x}']^{T} \in \mathbb{R}^{2n}$, Eq.\eqref{eq:linearized_perturbation_equation} can be reformulated as
\begin{subequations}
      \label{eq:linearized_perturbation_equation_state}
      \begin{align}
            \mathbf{y}' &= \mathbf{A}(\tau) \mathbf{y}, \label{eq:linearized_perturbation_equation_state_1} \\
            \mathbf{A}(\tau) &= 
            \begin{bmatrix}
            \mathbf{0} & \mathbf{I}_{n} \\
            -\bar{\mathbf{M}}^{-1}(\tau) \bar{\mathbf{K}}(\tau) & -\bar{\mathbf{M}}^{-1}(\tau) \bar{\mathbf{C}}(\tau)
            \end{bmatrix} \in \mathbb{R}^{2n \times 2n}. \label{eq:linearized_perturbation_equation_state_2}
      \end{align}
\end{subequations}
Floquet theory states that for an ordinary differential equation of the form given in Eq.\eqref{eq:linearized_perturbation_equation_state}, there exists a constant matrix $\bm{\Lambda}$ such that
\begin{equation}
      \label{eq:floquet_theory}
      \mathbf{y}(\tau + T) = \bm{\Lambda} \mathbf{y}(\tau),
\end{equation}
where $\bm{\Lambda}$ represents the monodromy matrix, and its eigenvalues $\zeta_{i}, i = 1, \cdots, 2n$ are known as the Floquet multipliers. The stability of the periodic response is determined by the spectral radius of $\bm{\Lambda}$, denoted as
\begin{equation}
      \label{eq:spectral_radius}
      \rho(\bm{\Lambda}) = \max_{i} |\zeta_{i}|.
\end{equation}
If $\rho(\bm{\Lambda}) < 1$, the resulting periodic response is asymptotically stable; if $\rho(\bm{\Lambda}) > 1$, it is unstable. Furthermore, the path of $\zeta_{i}$ intersecting the unit circle in the complex plane corresponds to bifurcation modes, such as saddle-node, period-doubling, and Neimark-Sacker bifurcations. More detailed discussions are available in reference \cite{changModelingNonlinearAnalysis2024}.

Because computing $\bm{\Lambda}$ directly is challenging \cite{ricciDiscussionDynamicStability2012}, the iterative algorithm developed by Hsu et al. \cite{Hsu1973ApplicationsOT} is more commonly used in practice. Specifically, Eq.\eqref{eq:linearized_perturbation_equation_state} is discretized into $N_{2}$ segments, and the monodromy matrix for each segment is computed via matrix exponentiation. Finally, $\bm{\Lambda}$ is obtained by multiplying the monodromy matrices across all segments. Further algorithmic details can be found in references \cite{Hsu1973ApplicationsOT, cheungApplicationIncrementalHarmonic1990}. 

\section{Algorithm optimization and programming implementation}
As established in the previous section, the HB method combined with arc-length continuation consists of three core components: (1) System and parameter definition (preprocessing), which includes specifying the form of Eq.\eqref{eq:MODF_nonlinear_system_dimensionless} and defining the method parameters; (2) Solving, which includes first solving at a given $\omega$ or $\lambda$ and then tracking the solution branch using arc-length continuation; (3) Postprocessing, which includes computing the Floquet multipliers and analyzing stability. Following the motivation outlined in the \textbf{Section 1.}, this manuscript develops and releases pyHB (\url{https://github.com/shuizidesu/pyhb}), an open-source Python repository that integrates the entire HB analysis workflow and supports arbitrary user-defined nonlinear systems without requiring explicit Jacobian matrix computations. Notably, pyHB achieves highly efficient computation through a series of code-level optimizations. This section introduces the key optimization strategies and implementation details of pyHB. For clarity, this section uses the forced vibration problem as an example; the solution process for the free-frequency problems is essentially the same, except for the phase condition.

\subsection{FFT-based residual projection}
The HB residual equation, as defined in Eq.\eqref{eq:HB_residual_equation}, involves the Galerkin projection of the residual vector $\mathbf{r}(\mathbf{Q}, \omega, \tau)$ onto the Fourier basis vector $\bm{\phi}(\tau)$. Direct computation of this projection requires numerical integration over one period, which is computationally expensive. Reexamining Eq.\eqref{eq:HB_residual_equation_2}, we observe that $\mathbf{r}$ is periodic in $\tau$ and can therefore be expressed as a truncated Fourier series, given by
\begin{equation}
      \label{eq:fourier_series_residual}
      \mathbf{r} = \bm{\phi} \mathbf{H},
\end{equation}
where $\mathbf{H} \in \mathbb{R}^{o \times n}$ is the coefficient matrix. Substituting Eq.\eqref{eq:fourier_series_residual} into Eq.\eqref{eq:HB_residual_equation_1} yields
\begin{equation}
      \label{eq:HB_residual_equation_FFT}
      \mathbf{R} = \frac{2}{T} \int_{0}^{T} \bm{\phi} \otimes (\bm{\phi} \mathbf{H})\, d\tau = (\frac{2}{T} \int_{0}^{T} \bm{\phi} \otimes \bm{\phi}\, d\tau) \mathbf{H} = \mathbf{D}_{0} \mathbf{H},
\end{equation}
where the integral term is a constant matrix, which can be easily computed based on the orthogonality of the Fourier basis vector $\bm{\phi}$, given by
\begin{equation}
      \label{eq:fourier_basis_orthogonality}
      \mathbf{D}_{0} = \frac{2}{T} \int_{0}^{T} \bm{\phi} \otimes \bm{\phi}\, d\tau = 
      \begin{bmatrix}
      2 & 0 & \cdots & 0 \\
      0 & 1 & \cdots & 0 \\
      \vdots & \vdots & \ddots & \vdots \\
      0 & 0 & \cdots & 1
      \end{bmatrix} \in \mathbb{R}^{o \times o}.
\end{equation}
Hence, the computation of $\mathbf{R}$ reduces to determining the Fourier coefficient matrix $\mathbf{H}$ for the current $\mathbf{Q}$ and $\omega$.

FFT is a highly efficient algorithm to address the above issue. Specifically, considering the discretation of $T$ into $N_{1}$ equally spaced points, denoted as $\hat{\bm{\tau}} = \{ \tau_i \}_{i=1}^{N_1}$, then the corresponding discrete version of Eq.\eqref{eq:fourier_series} is expressed as
\begin{subequations}
      \label{eq:fourier_series_discrete}
      \begin{align}
            \hat{\bm{\phi}} &= [\mathbf{1}, \cos(h_{1} \hat{\bm{\tau}}), \cdots, \cos(h_{m} \hat{\bm{\tau}}), \sin(h_{1} \hat{\bm{\tau}}), \cdots, \sin(h_{m} \hat{\bm{\tau}})] \in \mathbb{R}^{N_{1} \times o}, \label{eq:fourier_series_discrete_1} \\
            \hat{\mathbf{x}} &= \hat{\bm{\phi}} \mathbf{Q} \in \mathbb{R}^{N_{1} \times n}, \label{eq:fourier_series_discrete_2} \\
            \hat{\mathbf{x}}' &= \hat{\bm{\phi}}' \mathbf{Q} \in \mathbb{R}^{N_{1} \times n}, \label{eq:fourier_series_discrete_3} \\
            \hat{\mathbf{x}}'' &= \hat{\bm{\phi}}'' \mathbf{Q} \in \mathbb{R}^{N_{1} \times n}, \label{eq:fourier_series_discrete_4}
      \end{align}
\end{subequations}
where $\hat{\bm{\phi}}' = \hat{\bm{\phi}} \mathbf{D}_{1}$ and $\hat{\bm{\phi}}'' = \hat{\bm{\phi}} \mathbf{D}_{2}$. $\hat{\mathbf{\tau}}$, $\hat{\bm{\phi}}$, $\hat{\bm{\phi}'}$, and $\hat{\bm{\phi}''}$ need to be computed only once, so they can be included in the preprocessing. Based on Eq.\eqref{eq:fourier_series_discrete}, the discrete version of $\mathbf{r}$ is discribed as
\begin{equation}
      \label{eq:fourier_series_residual_discrete}
      \hat{\mathbf{r}} = \hat{\mathbf{f}}_{ex}(\hat{\bm{\tau}}, \omega) - (\omega^2 \hat{\mathbf{x}} \mathbf{M}^{T} + \omega \hat{\mathbf{x}}' \mathbf{C}^{T} + \omega^2 \hat{\mathbf{x}}' \mathbf{G}^{T} + \hat{\mathbf{x}} \mathbf{K}^{T}) - \hat{\mathbf{f}}_{nl}(\hat{\mathbf{x}}'', \hat{\mathbf{x}}', \hat{\mathbf{x}}, \hat{\bm{\tau}}, \omega) \in \mathbb{R}^{N_{1} \times n},
\end{equation}
then the Fourier coefficient matrix $\mathbf{H}$ can be efficiently computed using the FFT algorithm, given by
\begin{equation}
      \label{eq:fourier_coefficient_FFT}
      \mathbf{H} = \begin{bmatrix}
      \text{Re}(\text{FFT}(\hat{\mathbf{r}}))_{1:m+1, :} \\
      \text{Im}(\text{FFT}(\hat{\mathbf{r}}))_{2:m+1, :}
      \end{bmatrix},
\end{equation}
where $\text{Re}()$ and $\text{Im}()$ denote the real and imaginary parts of the result, respectively. The computational complexity of the FFT is $\mathcal{O}(N_{1} \log N_{1})$, and the modern FFT algorithm supports parallel batch processing; hence, Eq.\eqref{eq:fourier_coefficient_FFT} is highly efficient. 

\subsection{Precomputation of the Jacobian matrix's linear part}
The required augmented Jacobian matrix in Eq.\eqref{eq:arc_length_NR_iteration} consists of two parts, $\mathbf{J} = \partial \text{vec}(\mathbf{R}) / \partial \mathbf{q}$ and $\partial \text{vec}(\mathbf{R}) / \partial \omega$; we will discuss them separately. It can be observed that $\mathbf{J}$ can be decomposed into a linear part and a nonlinear part, given by
\begin{equation}
      \label{eq:jacobian_decomposition}
      \mathbf{J} = -(\mathbf{J}_{L} + \mathbf{J}_{N}),
\end{equation}
where $\mathbf{J}_{L}$ is associated with the system's linear properties, and $\mathbf{J}_{N}$ is related to $\mathbf{f}_{nl}$. Based on Eq.\eqref{eq:HB_residual_equation}, the residual equation with respect to $\mathbf{J}_{L}$ can be given by
\begin{equation}
      \label{eq:jacobian_linear_part}
      \mathbf{R}_{L} = \omega^{2} \mathbf{B}_{3} \mathbf{Q} \mathbf{M}^{T} + \omega \mathbf{B}_{2} \mathbf{Q} \mathbf{C}^{T} + \omega^{2} \mathbf{B}_{2} \mathbf{Q} \mathbf{G}^{T} + \mathbf{B}_{1} \mathbf{Q} \mathbf{K}^{T},
\end{equation}
where $\mathbf{R} = \mathbf{F} - \mathbf{R}_{L} - \mathbf{R}_{N}$, and
\begin{subequations}
      \label{eq:jacobian_linear_part_B}
      \begin{align}
            \mathbf{B}_{1} &= \frac{2}{T} \int_{0}^{T} \bm{\phi} \otimes \bm{\phi}\, d\tau = \mathbf{D}_{0}, \label{eq:jacobian_linear_part_B_1} \\
            \mathbf{B}_{2} &= \frac{2}{T} \int_{0}^{T} \bm{\phi} \otimes \bm{\phi}'\, d\tau = \frac{2}{T} \int_{0}^{T} \bm{\phi} \otimes (\bm{\phi} \mathbf{D}_{1})\, d\tau = \mathbf{D}_{0} \mathbf{D}_{1}, \label{eq:jacobian_linear_part_B_2} \\
            \mathbf{B}_{3} &= \frac{2}{T} \int_{0}^{T} \bm{\phi} \otimes \bm{\phi}''\, d\tau = \frac{2}{T} \int_{0}^{T} \bm{\phi} \otimes (\bm{\phi} \mathbf{D}_{2})\, d\tau = \mathbf{D}_{0} \mathbf{D}_{2}.  \label{eq:jacobian_linear_part_B_3}
      \end{align}
\end{subequations}
$\mathbf{B}_{1}$, $\mathbf{B}_{2}$, and $\mathbf{B}_{3}$ are constant matrices that can be precomputed in the preprocessing stage. For column-priority vectorization, the following equality holds:
\begin{equation}
      \label{eq:jacobian_linear_part_vectorization}
      \text{vec}(\mathbf{R}_{L}) = (\omega^{2} \mathbf{M} \otimes \mathbf{B}_{3} + \omega \mathbf{C} \otimes \mathbf{B}_{2} + \omega^{2} \mathbf{G} \otimes \mathbf{B}_{2} + \mathbf{K} \otimes \mathbf{B}_{1}) \text{vec}(\mathbf{Q}).
\end{equation}
Thus, $\mathbf{J}_{L}$ is obtained as
\begin{equation}
      \label{eq:jacobian_linear_part_final}
      \mathbf{J}_{L} = \omega^{2} \mathbf{M} \otimes \mathbf{B}_{3} + \omega \mathbf{C} \otimes \mathbf{B}_{2} + \omega^{2} \mathbf{G} \otimes \mathbf{B}_{2} + \mathbf{K} \otimes \mathbf{B}_{1}.
\end{equation}
$\mathbf{M} \otimes \mathbf{B}_{3}$, $\mathbf{C} \otimes \mathbf{B}_{2}$, $\mathbf{G} \otimes \mathbf{B}_{2}$, and $\mathbf{K} \otimes \mathbf{B}_{1}$ need to be computed only once; thereafter, the full $\mathbf{J}_{L}$ can be generated quickly by combining polynomial operations in $\omega$. 

Simlarly, the computation of $\partial \text{vec}(\mathbf{R}) / \partial \omega$ can be divided into three parts, given by
\begin{equation}
      \label{eq:jacobian_omega_decomposition}
      \frac{\partial \text{vec}(\mathbf{R})}{\partial \omega} = \frac{\partial \text{vec}(\mathbf{F})}{\partial \omega} - (\frac{\partial \text{vec}(\mathbf{R}_{L})}{\partial \omega} + \frac{\partial \text{vec}(\mathbf{R}_{N})}{\partial \omega}),
\end{equation}
where $\partial \text{vec}(\mathbf{F}) / \partial \omega$ is associated with $\mathbf{f}_{ex}$. Note that $\mathbf{f}_{ex}$ is periodic and can be expressed in the standard form
\begin{equation}
      \label{eq:jacobian_omega_decomposition_F}
      \mathbf{f}_{ex}(\tau, \omega) = \sum_{i = 0}^{k} \omega ^ {i} \mathbf{f}_{ex}^{(i)}(\tau),
\end{equation}
where $k$ is the highest power of $\omega$ and, generally, $k \leq 2$. For example, the unbalance excitation in rotating machinery can be expressed as $\mathbf{f}_{ex} = \omega^{2} \mathbf{f}_{ex}^{(2)}(\tau)$ \cite{YANG2025116240, JIANG2025110482, YAO2026104180}. $\mathbf{f}_{ex}^{(i)}$ is periodic and independent of $\omega$, which can be expressed as a truncated Fourier series, given by
\begin{equation}
      \label{eq:jacobian_omega_decomposition_F_fourier}
      \mathbf{f}_{ex}(\tau, \omega) = \sum_{i = 0}^{k} \omega ^ {i} \bm{\phi}(\tau) \mathbf{H}_{ex}^{(i)},
\end{equation}
where $\mathbf{H}_{ex}^{(i)} \in \mathbb{R}^{o \times n}$ is the Fourier coefficient matrix, and can be computed via FFT, formulated as
\begin{equation}
      \label{eq:jacobian_omega_decomposition_F_fourier_coefficient}
      \mathbf{H}_{ex}^{(i)} = \begin{bmatrix}
      \text{Re}(\text{FFT}(\hat{\mathbf{f}}_{ex}^{(i)}))_{1:m+1, :} \\
      \text{Im}(\text{FFT}(\hat{\mathbf{f}}_{ex}^{(i)}))_{2:m+1, :}
      \end{bmatrix},
\end{equation}
where $\hat{\mathbf{f}}_{ex}^{(i)}$ is the discrete version of $\mathbf{f}_{ex}^{(i)}$. Eq.\eqref{eq:jacobian_omega_decomposition_F_fourier_coefficient} can be precomputed in the preprocessing stage and adopted for anbitrary $\omega$ during the iteration process. Substituting Eq.\eqref{eq:jacobian_omega_decomposition_F_fourier} into Eq.\eqref{eq:HB_residual_equation_1} yields
\begin{subequations}
      \label{eq:jacobian_omega_decomposition_F_final}
      \begin{align}
            \partial \text{vec}(\mathbf{F}) &= \sum_{i = 0}^{k} \omega ^ {i} \text{vec}(\int_{0}^{T} \bm{\phi} \otimes (\bm{\phi} \mathbf{H}_{ex}^{(i)})\, d\tau) = \sum_{i = 0}^{k} \omega ^ {i} \text{vec}(\mathbf{D}_{0} \mathbf{H}_{ex}^{(i)}), \\
            \frac{\partial \text{vec}(\mathbf{F})}{\partial \omega} &= \sum_{i = 0}^{k} i \omega ^ {i-1} \text{vec}(\int_{0}^{T} \bm{\phi} \otimes (\bm{\phi} \mathbf{H}_{ex}^{(i)})\, d\tau) = \sum_{i = 0}^{k} i \omega ^ {i-1} \text{vec}(\mathbf{D}_{0} \mathbf{H}_{ex}^{(i)}).
      \end{align}
\end{subequations}
Thus, $\partial \text{vec}(\mathbf{F}) / \partial \omega$ can be obtained efficiently. Moreover, reconsidering Eq.\eqref{eq:jacobian_linear_part_vectorization}, we can obtain
\begin{equation}
      \label{eq:jacobian_omega_decomposition_RL_final}
      \frac{\partial \text{vec}(\mathbf{R}_{L})}{\partial \omega} = (2 \omega \mathbf{M} \otimes \mathbf{B}_{3} + \mathbf{C} \otimes \mathbf{B}_{2} + 2 \omega \mathbf{G} \otimes \mathbf{B}_{2}) \text{vec}(\mathbf{Q}),
\end{equation}
where the Kronecker product terms have been computed. Overall, most computations of $\mathbf{J}_{L}$ and $\partial \text{vec}(\mathbf{R}_{L}) / \partial \omega$ can be precomputed during preprocessing, and only FFT and a few polynomial operations are required during iteration.

\subsection{Tensor-contraction-based computation of the Jacobian matrix's nonlinear part}
Next, we need to compute the remaining nonlinear components, $\mathbf{J}_{N}$ and $\partial \text{vec}(\mathbf{R}_{N}) / \partial \omega$. $\mathbf{J}_{N}$ is written as
\begin{subequations}
      \label{eq:jacobian_nonlinear_part}
      \begin{align}
            \mathbf{J}_{N} &= \frac{\partial \text{vec}(\mathbf{R}_{N})}{\partial \mathbf{q}} = \text{Mat}(\frac{\partial \mathbf{R}_{N}}{\partial \mathbf{Q}}; on, on), \label{eq:jacobian_nonlinear_part_1} \\
            \frac{\partial \mathbf{R}_{N}}{\partial \mathbf{Q}} &= \underbrace{\frac{2}{T} \int_{0}^{T} \bm{\phi} \otimes \frac{\partial \mathbf{f}_{nl}} {\partial \mathbf{x}''} \bullet \frac{\partial \mathbf{x}''}{\partial \mathbf{Q}} \, d\tau}_{\mathbf{N}_{3}} + \underbrace{\frac{2}{T} \int_{0}^{T} \bm{\phi} \otimes \frac{\partial \mathbf{f}_{nl}}{\partial \mathbf{x}'} \bullet \frac{\partial \mathbf{x}'}{\partial \mathbf{Q}} \, d\tau}_{\mathbf{N}_{2}} + \underbrace{\frac{2}{T} \int_{0}^{T} \bm{\phi} \otimes \frac{\partial \mathbf{f}_{nl}}{\partial \mathbf{x}} \bullet \frac{\partial \mathbf{x}}{\partial \mathbf{Q}} \, d\tau}_{\mathbf{N}_{1}}, \label{eq:jacobian_nonlinear_part_2}
      \end{align}  
\end{subequations}
where $\mathbf{N}_{1}$, $\mathbf{N}_{2}$, and $\mathbf{N}_{3} \in \mathbb{R}^{o \times n \times o \times n}$ are fourth-order tensors that follow similar computation procedures. For brevity, only $\mathbf{N}_{1}$ is discussed here. Based on Eq.\eqref{eq:fourier_series_2}, we have
\begin{equation}
      \label{eq:jacobian_nonlinear_part_N1}
      \mathbf{N}_{1} = \frac{2}{T} \int_{0}^{T} \bm{\phi} \otimes \frac{\partial \mathbf{f}_{nl}}{\partial \mathbf{x}} \bullet \text{perm} (\mathbf{I}_{n}  \otimes \bm{\phi}; 1, 3, 2)\, d\tau = \frac{2}{T} \int_{0}^{T} \bm{\phi} \otimes \text{perm}(\frac{\partial \mathbf{f}_{nl}}{\partial \mathbf{x}} \otimes \bm{\phi}; 1, 3, 2) d\tau.
\end{equation}
With an appropriate permutation, Eq.\eqref{eq:jacobian_nonlinear_part_N1} can be reformulated as
\begin{equation}
      \label{eq:jacobian_nonlinear_part_N1_permutation}
      \mathbf{N}_{1} = \text{perm} ( \frac{2}{T} \int_{0}^{T} \bm{\phi} \otimes \bm{\phi} \otimes \frac{\partial \mathbf{f}_{nl}}{\partial \mathbf{x}}\, d\tau ; 1, 3, 2, 4 ) = \text{perm} (\hat{\mathbf{N}}_{1}; 1, 3, 2, 4),
\end{equation}
where $\hat{\mathbf{N}}_{1} \in \mathbb{R}^{o \times o \times n \times n}$ is a fourth-order tensor, and its sorting makes subsequent computations more convenient. Since $\partial \mathbf{f}_{nl} / \partial \mathbf{x}$ is periodic, it can also be expanded into a truncated Fourier series. However, it is important to note that, to accurately represent the nonlinear property, a higher-order Fourier basis vector should be adopted, given by
\begin{subequations}
      \label{eq:fourier_series_nonlinear}
      \begin{align}
            \bm{\psi} &= [\mathbf{1}, \cos(h_{1} \tau), \cdots, \cos(h_{p} \tau), \sin(h_{1} \tau), \cdots, \sin(h_{p} \tau)] \in \mathbb{R}^{2p+1}, \label{eq:fourier_series_nonlinear_1} \\
            \frac{\partial \mathbf{f}_{nl}}{\partial \mathbf{x}} &= \bm{\psi} \bullet \mathbf{H}_{nl}^{(1)} \in \mathbb{R}^{n \times n}, \label{eq:fourier_series_nonlinear_2}
      \end{align}
\end{subequations}
where $p >> m$, and $\mathbf{H}_{nl}^{(1)} \in \mathbb{R}^{(2p+1) \times n \times n}$ is the Fourier coefficient tensor. Substituting Eq.\eqref{eq:fourier_series_nonlinear_2} into Eq.\eqref{eq:jacobian_nonlinear_part_N1_permutation} yields
\begin{equation}
      \label{eq:jacobian_nonlinear_part_N1_permutation_final}
      \hat{\mathbf{N}}_{1} = \frac{2}{T} \int_{0}^{T} \bm{\phi} \otimes \bm{\phi} \otimes (\bm{\psi} \bullet \mathbf{H}_{nl}^{(1)})\, d\tau = \underbrace{\frac{2}{T} \int_{0}^{T} (\bm{\phi} \otimes \bm{\phi} \otimes \bm{\psi}) \, d\tau}_{\mathbf{S}_{1}} \bullet \mathbf{H}_{nl}^{(1)},
\end{equation}
where $\mathbf{S}_{1} \in \mathbb{R}^{o \times o \times (2p+1)}$ is a third-order tensor independent of $\mathbf{Q}$ and $\omega$ and can therefore be precomputed in the preprocessing stage. Noting that the maximum frequency in $\bm{\phi} \otimes \bm{\phi}$ is $2 h_{m}$, $h_{p}$ only needs to be set to $2 h_{m}$, given the orthogonality of trigonometric functions. Then, utilizing the discretization stratehy similar to Eq.\eqref{eq:fourier_series_discrete} and Eq.\eqref{eq:fourier_series_residual_discrete}, $\mathbf{H}_{nl}^{(1)}$ can be computed by
\begin{equation}
      \label{eq:fourier_coefficient_nonlinear_FFT}
      \mathbf{H}_{nl}^{(1)} = \begin{bmatrix}
      \text{Re}(\text{FFT}(\hat{\partial \mathbf{f}_{nl}} / \partial \hat{\mathbf{x}}))_{1:p+1, :, :} \\
      \text{Im}(\text{FFT}(\hat{\partial \mathbf{f}_{nl}} / \partial \hat{\mathbf{x}}))_{2:p+1, :, :}
      \end{bmatrix},
\end{equation}
where $\hat{\partial \mathbf{f}_{nl}} / \partial \hat{\mathbf{x}}$ is the discrete version of $\partial \mathbf{f}_{nl} / \partial \mathbf{x}$, which essentially computes the tangent stiffness matrix pointwise at each $\tau_{i}$. The $n \times n$-times FFT and the evaluation of $N_{1}$-times tangent stiffness matrices can both be performed in parallel, making Eq.\eqref{eq:fourier_coefficient_nonlinear_FFT} highly efficient. Similarly, $\mathbf{S}_{2}$, $\mathbf{S}_{3}$, $\mathbf{H}_{nl}^{(2)}$, and $\mathbf{H}_{nl}^{(3)}$ can be computed, and the final $\mathbf{J}_{N}$ is obtained by
\begin{equation}
      \label{eq:jacobian_nonlinear_part_final}
      \mathbf{J}_{N} = \text{Mat}(\text{perm}(\mathbf{S}_{1} \bullet \mathbf{H}_{nl}^{(1)} + \mathbf{S}_{2} \bullet \mathbf{H}_{nl}^{(2)} + \mathbf{S}_{3} \bullet \mathbf{H}_{nl}^{(3)}; 1, 3, 2, 4); on, on),
\end{equation}
where
\begin{subequations}
      \label{eq:jacobian_nonlinear_part_final_S}
      \begin{align}
            \mathbf{S}_{2} &= \frac{2}{T} \int_{0}^{T} (\bm{\phi} \otimes \bm{\phi}' \otimes \bm{\psi}) \, d\tau \in \mathbb{R}^{o \times o \times (2p+1)}, \label{eq:jacobian_nonlinear_part_final_S_2} \\
            \mathbf{S}_{3} &= \frac{2}{T} \int_{0}^{T} (\bm{\phi} \otimes \bm{\phi}'' \otimes \bm{\psi}) \, d\tau \in \mathbb{R}^{o \times o \times (2p+1)}.  \label{eq:jacobian_nonlinear_part_final_S_3}
      \end{align}
\end{subequations}

Based on the above derivation, $\partial \text{vec}(\mathbf{R}_{N}) / \partial \omega$ can be computed similarly. The final result is given by
\begin{equation}
      \label{eq:jacobian_nonlinear_part_final_omega}
      \frac{\partial \text{vec}(\mathbf{R}_{N})}{\partial \omega} = \text{vec}(\frac{2}{T} \int_{0}^{T} \bm{\phi} \otimes \bm{\phi} \, d\tau \bullet \mathbf{H}_{nl}^{4}) = \text{vec}(\mathbf{D}_{0} \bullet \mathbf{H}_{nl}^{(4)}),
\end{equation}
where $\mathbf{H}_{nl}^{(4)} \in \mathbb{R}^{o \times n}$ is the Fourier coefficient matrix of $\partial \mathbf{f}_{nl} / \partial \omega$ (the corresponding discrete version is $\hat{\partial \mathbf{f}_{nl}} / \partial \omega$). Overall, during the iteration process, it is sufficient to compute the tensor contraction using $\mathbf{S}_{1}$, $\mathbf{S}_{2}$, and $\mathbf{S}_{3}$ according to Eq.\eqref{eq:jacobian_nonlinear_part_final}, and then perform the appropriate rearrangement to generate $\mathbf{J}_{N}$. Modern computer CPUs feature multi-level caches \cite{baglioniMultibodyModellingDOF2016}; this type of tensor operation fully leverages hardware capabilities, significantly reduces memory read/write overhead, and markedly improves computation speed. 

\subsection{Exploiting localized nonlinearities}
In the derivation above, the dimension of $\mathbf{f}_{nl}$ is $n$, and the tangent stiffness, damping, and mass matrices are all $n \times n$, indicating that the nonlinearities are distributed throughout the system. However, in the vast majority of practical engineering systems, nonlinearity is typically localized, such as at hinged joints or bearing supports \cite{sunComputationallyEfficientMethod2024, liangIntegratedApproachResponse2024, chenIndirectHarmonicBalance2025a}. Exploiting this property helps further reduce computational overhead. Specifically, given that $\mathbf{f}_{nl}$ has only $r << n$ non-zero terms and is related to only $q << n$ DOFs, denoted as
\begin{equation}
      \label{eq:localized_nonlinearity}
      \mathbf{f}_{nl}(\mathbf{x}'', \mathbf{x}', \mathbf{x}, \tau, \omega) = \bm{\Xi} \mathbf{f}_{nl}^{(r)}(\bm{\Theta} \mathbf{x}'', \bm{\Theta} \mathbf{x}', \bm{\Theta} \mathbf{x}, \tau, \omega) = \bm{\Xi} \mathbf{f}_{nl}^{(r)}(\mathbf{x}''_{r}, \mathbf{x}'_{r}, \mathbf{x}_{r}, \tau, \omega)
\end{equation}
where $\mathbf{f}_{nl}^{(r)} \in \mathbb{R}^{r}$ is the reduced nonlinear force, and $\bm{\Xi} \in \mathbb{R}^{n \times r}$ and $\bm{\Theta} \in \mathbb{R}^{q \times n}$ are selection matrices consisting of 0s and 1s. Then, it can be observed that $\mathbf{R}_{N}$ and $\mathbf{J}_{N}$ only need to be computed baed on the $\mathbf{f}_{nl}^{(r)}$. Taking $\hat{\mathbf{N}}_{1}$ shown in Eq.\eqref{eq:jacobian_nonlinear_part_N1_permutation_final} as an example, its computation process can be reformulated to first calculate a smaller block given by
\begin{equation}
      \label{eq:jacobian_nonlinear_part_N1_permutation_final_localized}
      \hat{\mathbf{N}}_{1}^{(r)} = \frac{2}{T} \int_{0}^{T} \bm{\phi} \otimes \bm{\phi} \otimes (\bm{\psi} \bullet \mathbf{H}_{nl}^{(1, r)})\, d\tau = \mathbf{S}_{1} \bullet \mathbf{H}_{nl}^{(1, r)},
\end{equation}
where $\mathbf{H}_{nl}^{(1, r)} \in \mathbb{R}^{(2p+1) \times r \times q}$ is the Fourier coefficient tensor computed from the discrete version of $\partial \mathbf{f}_{nl}^{(r)} / \partial \mathbf{x}_{r}$ (denoted as $\hat{\partial \mathbf{f}}_{nl}^{(r)} / \partial \hat{\mathbf{x}}_{r}$), given by
\begin{equation}
      \label{eq:fourier_coefficient_nonlinear_FFT_localized}
      \mathbf{H}_{nl}^{(1, r)} = \begin{bmatrix}
      \text{Re}(\text{FFT}(\partial \hat{\mathbf{f}}_{nl}^{(r)} / \partial \hat{\mathbf{x}}_{r}))_{1:p+1, :, :} \\
      \text{Im}(\text{FFT}(\partial \hat{\mathbf{f}}_{nl}^{(r)} / \partial \hat{\mathbf{x}}_{r}))_{2:p+1, :, :}
      \end{bmatrix}.
\end{equation}
Then, the full $\hat{\mathbf{N}}_{1}$ can be obtained by scattering $\hat{\mathbf{N}}_{1}^{(r)}$ with $\bm{\Xi}$ and $\bm{\Theta}$. The remaining terms in Eq.\eqref{eq:jacobian_nonlinear_part_final} and Eq.\eqref{eq:jacobian_nonlinear_part_final_omega} can be handled similarly.

This treatment offers three main benefits: (1) Significantly reduces the number of tensor contractions and FFTs required; (2) Calculating $\mathbf{R}_{N}$ does not require full-DOF discrete-time operations, which is expensive for large-scale systems; instead, it only requires constructing $\hat{\mathbf{x}}''_{r}$, $\hat{\mathbf{x}}'_{r}$, and $\hat{\mathbf{x}}_{r}$; (3) Makes it easier for users to define the model, eliminating the need to check the DOFs' order repeatedly. Besides, 

\subsection{Automatic differentiation for the Jacobian matrix}
Although the above derivations have greatly simplified the calculation of $\mathbf{J}$ and $\partial \text{vec}(\mathbf{R}) / \partial \omega$, information on the partial derivative of $\mathbf{f}_{nl}^{(r)}$ is still required. For simple forms, such as polynomials, manual derivation is straightforward; however, for more general cases, particularly when $\mathbf{f}_{nl}^{(r)}$ involves complex operations and the nonlinear coupling of stiffness and damping, accurately providing the expression for the partial derivative of $\mathbf{f}_{nl}^{(r)}$ is often difficult, time-consuming, and error-prone. Furthermore, in many pratical problems, $\mathbf{f}_{nl}^{(r)}$ has a non-smooth form, such as gaps, piecewise linear stiffness, and hysteretic damping \cite{caoPerturbationFunctionIteration2025, deshpandeOptimizationSecondarySuspension2006, teloliNewWayHarmonic2019}; manual derivation also requires careful handling of the non-smooth point \cite{wangVibrationAnalysisNonlinear2023, wangApplicationsIncrementalHarmonic2019}. 

Automatic differentiation (AD) offers a new technique to overcome the above challenges. As a cornerstone of modern deep learning frameworks, AD enables efficient computation of the gradient of a complex function by constructing static or dynamic computational graphs. By representing the computation of a nonlinear function as a combination of basic operations (such as addition, multiplication, and trigonometric functions), AD records the intermediate results of each operation and efficiently computes the required derivatives based on the chain rule \cite{wengertSimpleAutomaticDerivative1964, hansonAnalyticDifferentiationComputer1962, 3122009_3242010}. Compared to traditional symbolic and numerical differentiation, AD avoids issues such as numerical errors and expression bloat, thereby achieving excellent trade-off between accuracy and efficiency. 

One of pyHB's highlights is its support for AD to compute the partial derivative of $\mathbf{f}_{nl}^{(r)}$. Leveraging the high-level AD implementations provided by mature deep learning frameworks, such as PyTorch \cite{paszke2019pytorchimperativestylehighperformance} and JAX, this functionality is easy to implement. This feature of pyHB is built on PyTorch. Specifically, users only need to provide the discrete version of the PyTorch-style $\mathbf{f}_{nl}^{(r)}$ during preprocessing, and the solver will internally call the \texttt{torch.func.jacrev} API to automatically compute the corresponding Jacobian matrix. For example, the psuedocode for computing the tangent stiffness matrix utilized in Eq.\eqref{eq:fourier_coefficient_nonlinear_FFT_localized} is presented in Alg.\ref{alg:tangent_stiffness}. $\hat{\partial \mathbf{f}}_{nl}^{(r)} / \partial \hat{\mathbf{x}}'_{r}$, $\hat{\partial \mathbf{f}}_{nl}^{(r)} / \partial \hat{\mathbf{x}}''_{r}$, and $\hat{\partial \mathbf{f}}_{nl}^{(r)} / \partial \omega$ can be obtained in the same manner. However, it should be emphasized that $q$ and $r$ are generally similar in size; therefore, the sampled tangent matrix is better suited to calculation via reverse-mode AD. For $\partial \mathbf{f}_{nl}^{(r)} / \partial \omega$, however, since its input is a scalar, forward-mode AD (corresponding to the \texttt{torch.func.jacfwd} API) is more efficient.

\begin{algorithm}
    \caption{PyTorch-style automatic differentiation for the tangent stiffness matrix of $\mathbf{f}_{nl}^{(r)}$}
    \label{alg:tangent_stiffness}

    \KwIn{
        Discrete time sequence $\hat{\bm{\tau}}$; sampled local states
        $\hat{\mathbf{x}}''_{r}$, $\hat{\mathbf{x}}'_{r}$, and $\hat{\mathbf{x}}_{r}$;
        frequency $\omega$; PyTorch-style nonlinear force function
        $\hat{\mathbf{f}}_{nl}^{(r)}$
    }

    \KwOut{
        Sampled tangent stiffness matrices
        $\hat{\partial \mathbf{f}}_{nl}^{(r)} / \partial \hat{\mathbf{x}}_{r}$
    }

    Convert $\hat{\bm{\tau}} \in \mathbb{R}^{N_{1}}$,
    $\hat{\mathbf{x}}''_{r} \in \mathbb{R}^{N_{1} \times q}$,
    $\hat{\mathbf{x}}'_{r} \in \mathbb{R}^{N_{1} \times q}$,
    $\hat{\mathbf{x}}_{r} \in \mathbb{R}^{N_{1} \times q}$, and $\omega$
    to PyTorch tensors\;

    Define the summed nonlinear force function
    $\mathbf{g}(\hat{\mathbf{x}}_{r}) =
    \sum_{i = 1}^{N_{1}}
    [\hat{\mathbf{f}}_{nl}^{(r)}
    (\hat{\mathbf{x}}''_{r}, \hat{\mathbf{x}}'_{r},
    \hat{\mathbf{x}}_{r}, \hat{\bm{\tau}}, \omega)]_{i, :}$,
    where
    $\hat{\mathbf{f}}_{nl}^{(r)}
    (\hat{\mathbf{x}}''_{r}, \hat{\mathbf{x}}'_{r},
    \hat{\mathbf{x}}_{r}, \hat{\bm{\tau}}, \omega)
    \in \mathbb{R}^{N_{1} \times r}$\;

    Compute the derivative using reverse-mode AD as
    $\hat{\partial \mathbf{f}}_{nl}^{(r)} / \partial \hat{\mathbf{x}}_{r}
    =
    \texttt{torch.func.jacrev}(\mathbf{g})(\hat{\mathbf{x}}_{r})$,
    where
    $\hat{\partial \mathbf{f}}_{nl}^{(r)} / \partial \hat{\mathbf{x}}_{r}
    \in \mathbb{R}^{r \times N_{1} \times q}$\;

    Permute the tensor axes to obtain the desired tangent stiffness:
    $\hat{\partial \mathbf{f}}_{nl}^{(r)} / \partial \hat{\mathbf{x}}_{r}
    \leftarrow
    \text{perm}
    (\hat{\partial \mathbf{f}}_{nl}^{(r)} / \partial \hat{\mathbf{x}}_{r}; 2, 1, 3)
    \in \mathbb{R}^{N_{1} \times r \times q}$\;

    \Return $\hat{\partial \mathbf{f}}_{nl}^{(r)} / \partial \hat{\mathbf{x}}_{r}$\;
\end{algorithm}

Supported by the PyTorch framework, Alg.\ref{alg:tangent_stiffness} can perform efficient parallel computation on a GPU using CUDA. Furthermore, the AD procedure involves only the nonlinear force, which is completely separate from the other computations required in the HB method, such as tensor contraction, rearrangement, and FFT, and whose problem scale is independent of the number of harmonics. Meanwhile, $\mathbf{f}_{nl}^{(r)}$ is defined and implemented in terms of local states, and its input and output dimensions are smaller; therefore, GPU memory usage during computational graph construction can be kept within a manageable range. This strategy differs fundamentally from our previous work \cite{chenHarmonicBalanceautomaticDifferentiation2026}, in which the AD process is coupled to the entire residual computation, causing GPU memory usage to increase sharply with the number of DOFs and harmonics, thereby significantly limiting its application in high-dimensional systems. Besides, non-differentiable functions are also common in modern neural network models, such as Rectified Linear Units (ReLU) and max pooling. PyTorch's AD mechanism provides a robust solution for computing derivatives using subgradients and predefined rules \cite{pytorch_autograd_mechanics_2026}. Therefore, the differentiation of non-smooth, nonlinear force can be computed directly using AD, without additional manual processing. Overall, the introduction of AD provides an efficient, automated approach for handling the Jacobian matrix, greatly reducing users' preprocessing effort and significantly enhancing pyHB's universality and ease of use.

\subsection{Weighted arc-length continuation strategy}
In some systems, the amplitude of the steady-state response and the continuation parameter can differ widely in scale. For example, the vibration amplitude of the aeroengine's dual-rotor system is on the order of micrometers, while its operating rotational speed exceeds 150 rad/s \cite{chenGeneralEfficientHarmonic2024}. As a result, the harmonic coefficient vector and the continuation parameter may have significantly different physical dimensions and numerical magnitudes. Consequently, the constraint equation shown in Eq.\eqref{eq:arc_length_constraint} is dominated by the variable with the larger numerical scale. In contrast, increments of the other variables are poorly represented, leading to a nominally small arc-length step size that may still correspond to an excessively large change in the large-scale variable, causing the predicted solution to leave the convergence region of the iteration. To maintain convergence, the arc-length step size must be reduced to a very small value, which severely reduces the efficiency of tracing the solution branch. 

To address this issue, a weighted arc-length constraint is introduced in pyHB, which is inspired by the work of Riks \cite{riksIncrementalApproachSolution1979}, Crisfield \cite{crisfieldFastIncrementalIterative1981}, and the AUTO software \cite{doedel2007auto07p}. Specifically, Eq.\eqref{eq:arc_length_constraint} is optimized to
\begin{equation}
      \label{eq:arc_length_constraint_weighted}
      h(\mathbf{z}) = \langle \mathbf{z}-\mathbf{z}_{0}, \mathbf{v}^{(0)} \rangle_{\mathbf{W}} - s,
\end{equation}
where $\langle \mathbf{a}, \mathbf{b} \rangle_{\mathbf{W}} = \mathbf{a}^{T} \mathbf{W} \mathbf{b}$ is the weighted inner product. $\mathbf{W}$ is a given weighting matrix, given by
\begin{equation}
      \label{eq:arc_length_constraint_weighted_W}
      \mathbf{W} = \text{diag} (\underbrace{1/q_{s}^{2}, \cdots, 1 / q_{s}^{2}}_{no}, 1 / \omega_{s}^{2}),
\end{equation}
where $\text{diag}()$ denotes the diagonal matrix. $q_{s}$ and $\omega_{s}$ are the scaling factors for the harmonic coefficient vector and the continuation parameter, respectively. These factors can be determined from the expected numerical ranges of $\mathbf{q}$ and $\omega$. By appropriately choosing $q_{s}$ and $\omega_{s}$, the weighted arc-length constraint ensures that both the harmonic coefficients and the continuation parameter contribute comparably to the arc-length step size, thereby improving convergence and enabling larger $s$ during continuation. 

\subsection{Sparse linear solver for the Newton-Raphson iteration}
\begin{figure}[htbp]
      \centering
      \includegraphics[width=1.0\textwidth]{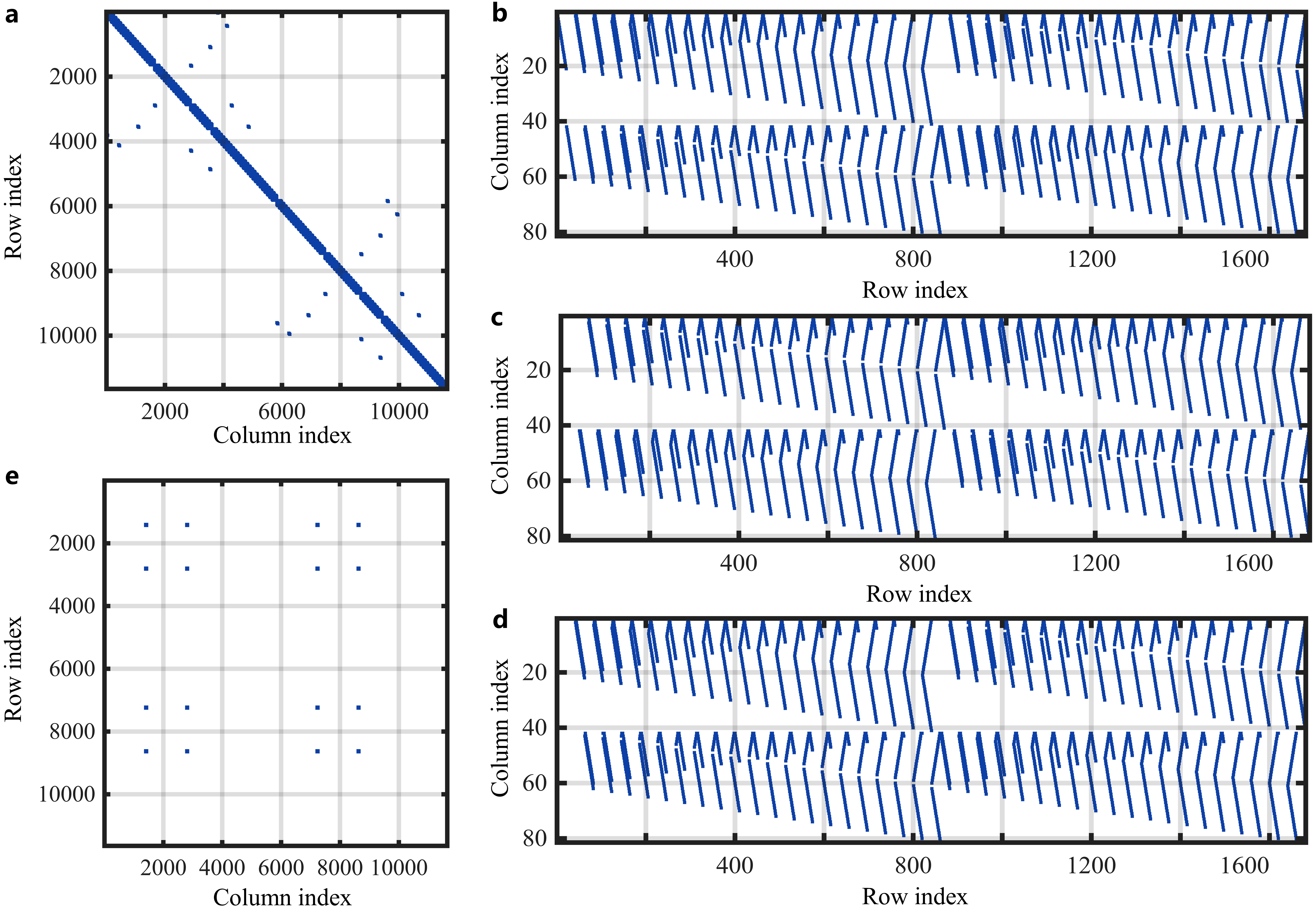}
      \caption{Sparse structure of the matrices utilized in the HB method with $m=20$, where the blue dots represent non-zero elements. \textbf{a} $\mathbf{J}_{L}$, whose fill-in ratio is 0.05\%. \textbf{b} Compressed $\mathbf{S}_{1}$, whose fill-in ratio is 2.38\%. \textbf{c} Compressed $\mathbf{S}_{2}$, whose fill-in ratio is 2.35\%. \textbf{d} Compressed $\mathbf{S}_{3}$, whose fill-in ratio is 2.35\%. \textbf{e} $\mathbf{J}_{N}$, whose fill-in ratio is 0.02\%.}
      \label{fig:matrix_structure}
\end{figure}

The size of the system of linear equations in Eqs.\eqref{eq:NR_iteration} and \eqref{eq:arc_length_NR_iteration} increases with $m$ and the number of the system's DOFs; solving these equations is a major computational bottleneck for the HB method. For large-scale problems, solving them directly with dense matrices is unacceptable in terms of both memory usage and computational efficiency. pyHB uses sparse matrix techniques to build a linear solver, making it suitable for large-scale and even extremely large-scale problems. Specifically, given the orthogonality of the Fourier basis vector, the fill-in ratios of $\mathbf{B}_{1}$, $\mathbf{B}_{2}$, and $\mathbf{B}_{3}$ are $1/o$. Additionally, the linear mass, stiffness, and damping matrices derived using the finite element method also exhibit a sparse banded structure. Hence, the linear part of $\mathbf{J}$ is generally highly sparse. Taking the dual-rotor system model provided in reference \cite{chenHarmonicBalanceautomaticDifferentiation2026} as an example, Fig.\ref{fig:matrix_structure}\textbf{a} presents the structure of $\mathbf{J}_{L}$ with the harmonic order set to 20. As can be seen, it exhibits a typical banded sparse structure, with a fill-in ratio of only 0.05\%. Similarly, Fig.\ref{fig:matrix_structure}\textbf{b}-\textbf{d} show the structure of $\mathbf{S}_{1}$, $\mathbf{S}_{2}$, and $\mathbf{S}_{3}$, respectively. Note that for clarity, the first two dimensions of the above third-order tensors are compressed. It can be observed that they are also highly sparse, with fill-in ratios of 2.38\%, 2.35\%, and 2.35\%. Furthermore, the system's nonlinearity is concentrated at the inter-shaft bearing; therefore, the structure of $\mathbf{J}_{N}$ obtained based on the localized strategy described earlier is shown in Fig.\ref{fig:matrix_structure}\textbf{e}, whose fill-in ratio is only 0.02\%. 

Based on the above insights, pyHB employs a sparse linear solver supported by the scipy package for the Newton-Raphson iteration. Specifically, during the preprocessing stage, both the system's linear matrices and the constant matrices in Eq.\eqref{eq:jacobian_linear_part_B} are processed as Compressed Sparse Column (CSC) format sparse matrices, and the subsequent Kronecker product is also performed in its sparse version. Then, $\mathbf{J}_{L}$'s sparse version is obtained by performing a polynomial combination with respect to $\omega$. Note that the sizes of $\mathbf{S}_{1}$, $\mathbf{S}_{2}$, and $\mathbf{S}_{3}$ depend only on the harmonic settings and do not scale with the number of DOFs. Furthermore, the nonlinear force are confined to a localized version, so the Fourier coefficient tensor (shown in Eq.\eqref{eq:fourier_coefficient_nonlinear_FFT_localized}) is low-dimensional but relatively dense. Considering the above two characteristics, along with the underlying BLAS library's efficient support for dense matrix multiplication, pyHB completes the computations for $\mathbf{S}_{1}$, $\mathbf{S}_{2}$, and $\mathbf{S}_{3}$ during the preprocessing stage and compresses them into dense matrices for later use. The Fourier coefficient tensors computed via AD and FFT are also compressed into dense matrices, and tensor contraction is performed according to Eq.\eqref{eq:jacobian_nonlinear_part_N1_permutation_final_localized} (which essentially involves matrix multiplication). Subsequently, based on the local coordinates of the nonlinear forces, the computation results are scattered into sparse matrices in Coordinate (COO) format, then converted to CSC format to obtain the final sparse version of $\mathbf{J}_{N}$. 

The complete $\mathbf{J}$ is obtained through sparse matrix addition of $\mathbf{J}_{L}$ and $\mathbf{J}_{N}$. The above method avoids constructing a large, dense matrix, thereby ensuring manageable memory usage. Furthermore, when solving systems of linear equations, given the high sparsity of $\mathbf{J}$ and its CSC format, pyHB employs the sparse LU decomposition solver provided by \texttt{scipy.sparse.linalg.splu} API. It should be emphasized that \texttt{splu} can be directly adopted in Eq.\eqref{eq:NR_iteration}, but cannot be directly used in Eq.\eqref{eq:arc_length_NR_iteration} due to the additional row and column introduced by the arc-length constraint. Specifically, the core of Eq.\eqref{eq:arc_length_NR_iteration} is to solve the following augmented system of linear equations
\begin{equation}
      \label{eq:arc_length_NR_iteration_augmented}
      \underbrace{
      \begin{bmatrix}
            \mathbf{J}
            &
            \partial \text{vec}(\mathbf{R}) / \partial \omega \\
            \multicolumn{2}{c}{\mathbf{v}^{(0)}}
            \end{bmatrix}}_{\mathbf{J}_{a}} \Delta \mathbf{z}
            \begin{bmatrix} = 
            \text{vec}(\mathbf{R}) \\
            h(\mathbf{z})
      \end{bmatrix},
\end{equation}
where $h(\mathbf{z})$ is shown in Eq.\eqref{eq:arc_length_constraint_weighted}. The tangent vector $\mathbf{v}^{(0)}$ is dense; hence, if \texttt{splu} is directly applied to $\mathbf{J}_{a}$, the resulting LU factors will be dense, leading to a significant increase in memory usage and computation time. pyHB addresses this issue by reformulating Eq.\eqref{eq:arc_length_NR_iteration_augmented} as the blocked form
\begin{equation}
      \label{eq:arc_length_NR_iteration_augmented_blocked}
      \begin{bmatrix}
            \mathbf{J} & \partial \text{vec}(\mathbf{R}) / \partial \omega \\
            (\mathbf{W}{\mathbf{v}^{(0)}})_{1:on} & (\mathbf{W}{\mathbf{v}^{(0)}})_{on+1}
      \end{bmatrix}
      \begin{bmatrix}
            \Delta \mathbf{q} \\ \Delta \omega
      \end{bmatrix}
      =
      \begin{bmatrix}
            \mathbf{J} & \mathbf{R}_{\omega} \\
            \mathbf{h}_{q} & h_{\omega}
      \end{bmatrix}
      \begin{bmatrix}
            \Delta \mathbf{q} \\ \Delta \omega
      \end{bmatrix}
      =
      \begin{bmatrix}
            -\text{vec}(\mathbf{R}) \\ -h(\mathbf{z})
      \end{bmatrix},
\end{equation}
which can be solved by
\begin{subequations}
      \label{eq:arc_length_NR_iteration_augmented_blocked_solution}
      \begin{align}
            \Delta \omega &= \frac{-h_{\omega} + \mathbf{h}_{q} \mathbf{J}^{-1} \text{vec}(\mathbf{R})}{\mathbf{h}_{q} \mathbf{J}^{-1} \mathbf{R}_{\omega} - h_{\omega}}, \\
      \Delta \mathbf{q} &= -\mathbf{J}^{-1} (\text{vec}(\mathbf{R}) + \Delta \omega \mathbf{R}_{\omega}).
      \end{align}
\end{subequations}
Note that Eq.\eqref{eq:arc_length_NR_iteration_augmented_blocked_solution} requires only one sparse LU decomposition of $\mathbf{J}$, which is same as that in Eq.\eqref{eq:NR_iteration}, and the remaining operations are sparse matrix-vector multiplications which can reuse the obtained LU factors. 

\subsection{Monodromy matrix computation using an implicit integrator}
The algorithm optimizations and programming implementations described above enable pyHB to solve for the periodic response of large-scale systems while maintaining acceptable memory usage and high computational efficiency. However, it is worth noting that applying the Floquet theory to assess solution stability is also a crucial part of the HB method's workflow. Most researchers use Hsu's method to compute the Floquet multipliers, which requires computing the matrix exponent $N_{2}$ times at each convergent point. For high-dimensional systems, this represents a significant computational overhead; in fact, the time spent on post-processing can far exceed that spent on preprocessing and solving \cite{peletanComparisonStabilityComputational2013}. A time integration scheme is more appropriate in this situation \cite{gaonkarComputingFloquetTransition1981}. To ensure stability over a wide range of nonlinear forces, pyHB provides a method for computing the monodromy matrix using an implicit integrator. 

Specifically, the obtained periodic solution is sampled at $N_{2}$ points over $T$ (interval is $\Delta \tau$), denoted as $\mathbf{x}_{i}$, $\mathbf{x}'_{i}$, and $\mathbf{x}''_{i}$, where $i = 1, 2, \cdots, N_{2}$. Then, $\bar{\mathbf{M}}(\tau)$, $\bar{\mathbf{C}}(\tau)$, and $\bar{\mathbf{K}}(\tau)$ are computed at each discrete point. Using an implicit trapezoidal integrator, we obtain
\begin{equation}
      \label{eq:monodromy_matrix_implicit_integrator}
      \mathbf{L}_{i} \begin{bmatrix}
            \delta \mathbf{x}_{i+1} \\ \delta \mathbf{x}'_{i+1}
      \end{bmatrix}
      =
      \mathbf{R}_{i} \begin{bmatrix}
            \delta \mathbf{x}_{i} \\ \delta \mathbf{x}'_{i}
      \end{bmatrix},
\end{equation}
where
\begin{subequations}
      \label{eq:monodromy_matrix_implicit_integrator_LR}
      \begin{align}
            \mathbf{L}_{i} &= \begin{bmatrix}
                  \mathbf{I}_{n} & -\frac{\Delta \tau}{2} \mathbf{I}_{n} \\
                  \frac{\Delta \tau}{2} \bar{\mathbf{K}}_{i} & \bar{\mathbf{M}}_{i} + \frac{\Delta \tau}{2} \bar{\mathbf{C}}_{i}
            \end{bmatrix}, \\
            \mathbf{R}_{i} &= \begin{bmatrix}
                  \mathbf{I}_{n} & \frac{\Delta \tau}{2} \mathbf{I}_{n} \\
                  -\frac{\Delta \tau}{2} \bar{\mathbf{K}}_{i} & \bar{\mathbf{M}}_{i} - \frac{\Delta \tau}{2} \bar{\mathbf{C}}_{i}
            \end{bmatrix}.
      \end{align}
\end{subequations}
Hence, the transition matrix from $\tau_{i}$ to $\tau_{i+1}$ is given by
\begin{equation}
      \label{eq:monodromy_matrix_implicit_integrator_transition}
      \mathbf{T}_{i} = \mathbf{L}_{i}^{-1} \mathbf{R}_{i}.
\end{equation}
The monodromy matrix is then obtained by multiplying the transition matrices over one period, formulated as
\begin{equation}
      \label{eq:monodromy_matrix_implicit_integrator_monodromy}
      \bm{\Lambda} = \prod_{i = 1}^{N_{2}} \mathbf{T}_{i}.
\end{equation}
pyHB also supports batch computation of $\bar{\mathbf{M}}_{i}$, $\bar{\mathbf{C}}_{i}$, and $\bar{\mathbf{K}}_{i}$ based on the AD and CUDA. At this point, computing the inverse of the matrix in Eq.\eqref{eq:monodromy_matrix_implicit_integrator_transition} is the primary computational bottleneck; however, for high-dimensional systems, it is generally still possible to employ the sparse matrix technique, which are more efficient than computing the matrix exponent. 

\subsection{Modular software architecture of pyHB}

\begin{figure}[htbp]
      \centering
      \includegraphics[width=1.0\textwidth]{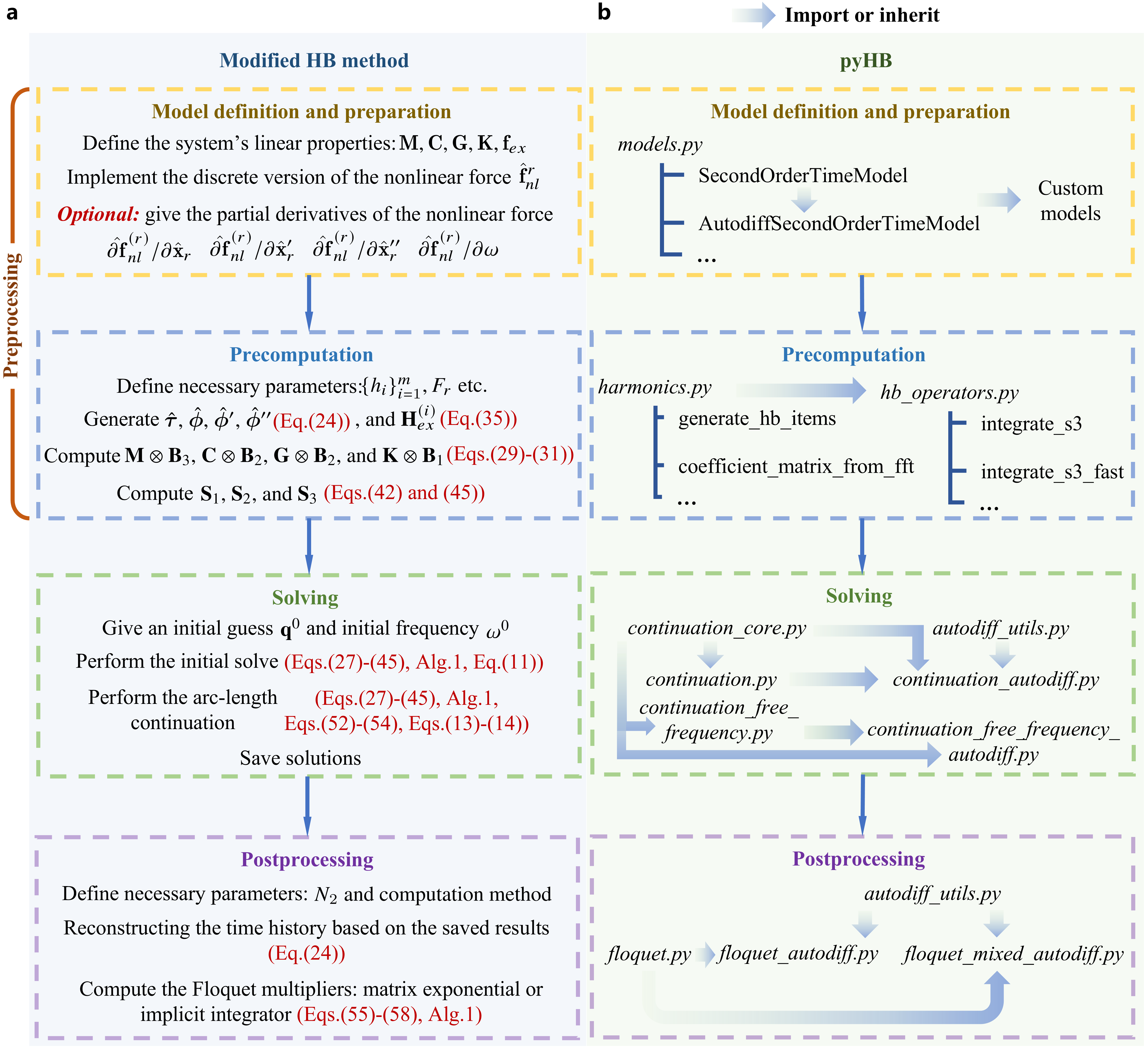}
      \caption{The structural design of pyHB and its correspondence with the HB method's workflow. \textbf{a} The HB method's entire workflow, including preprocessing, solving, and postprocessing. \textbf{b} The main modules of pyHB and the dependency relationships between the primary APIs.}
      \label{fig:pyhb_structure}
\end{figure}

The computational architecture of pyHB follows the workflow of the modified HB method described above. As shown in Fig.\ref{fig:pyhb_structure}\textbf{a}, the complete analysis is divided into model definition and preparation, precomputation, solving, and postprocessing; Fig.\ref{fig:pyhb_structure}\textbf{b} maps these four stages to the corresponding modules and primary APIs in pyHB. Rather than coupling the problem definition, harmonic operations, continuation algorithm, and stability analysis within a single program, pyHB organizes them into independent yet interoperable modules. Dependencies are established through explicit imports and inheritance, allowing mathematical operations shared across analysis procedures to be reused without duplicating their implementations.

For model definition and preparation, \texttt{models.py} provides \texttt{SecondOrderTimeModel} as the basic interface for the second-order system in Eq.\eqref{eq:MODF_nonlinear_system_dimensionless}. Through this interface, the user specifies the linear matrices $\mathbf{M}$, $\mathbf{C}$, $\mathbf{G}$, and $\mathbf{K}$, the external force $\mathbf{f}_{ex}$, and the discrete local nonlinear force $\hat{\mathbf{f}}_{nl}^{(r)}$ associated with the localized representation in Eq.\eqref{eq:localized_nonlinearity}. However, the partial derivatives of the nonlinear force are also need to be explicitly provided, which is inconvenient for the complex system. Alternatively, \texttt{AutodiffSecondOrderTimeModel} extends this interface by invoking the PyTorch-based AD procedure described in Alg.\ref{alg:tangent_stiffness}, thereby allowing the required tangent mass, damping, and stiffness matrices to be generated from the nonlinear force's implementation, which is one of the pyHB's highlights. Consequently, a custom system can be introduced by inheriting and rewriting the appropriate model class, while its physical definition remains separated from the subsequent HB operations.

The precomputation stage is mainly supported by \texttt{harmonics.py} and \texttt{hb\_operators.py}. \texttt{harmonics.py} constructs the discrete-time sequence, the Fourier basis matrices, and their first- and second-order derivatives, as given in Eq.\eqref{eq:fourier_series_discrete}. It is important to note that fractional harmonic components are necessary in the dual-rotor system at a specific speed ratio, or when calculating the system's subharmonic or period-doubling bifurcation responses \cite{FU2026112475, SU2026114471, CHEN2023104256}. pyHB addresses this issue by treating the frequency resolution $F_{r}$ as an additional parameter. Specifically, for example, if $F_{r} = 0.5$, the Fourier basis matrices are constructed with half-integer harmonics, thereby enabling the analysis of 1/2 subharmonic responses. In practice, this approach requires no additional computation; it simply involves setting the corresponding values of $T$ and $N_{1}$ based on $F_{r}$. Moreover, \texttt{hb\_operators.py} evaluates the three-basis integration operators used to construct $\mathbf{S}_{1}$, $\mathbf{S}_{2}$, and $\mathbf{S}_{3}$ in Eqs.\eqref{eq:jacobian_nonlinear_part_N1_permutation_final} and \eqref{eq:jacobian_nonlinear_part_final_S}. Two computation strategies are provided: \texttt{integrate\_s3} and its optimized counterpart \texttt{integrate\_s3\_fast}. The former uses batch numerical integration supported by \texttt{scipy.integrate.quad\_vec}, while the latter employs a more efficient approach based on the orthogonality of the Fourier basis and FFT, which is the default option.Together with the precomputed Kronecker-product terms in Eq.\eqref{eq:jacobian_linear_part_final}, these modules isolate the quantities that are independent of the current iteration and enable their reuse throughout the entire solving procedure. Besides, these precomputation results are saved as CSC format sparse matrices to conserve memory.

The solving stage is organized around \texttt{continuation\_core.py}, which contains the common numerical operations required for the initial Newton-Raphson solve and the predictor-corrector continuation process in Eqs.\eqref{eq:NR_iteration} and \eqref{eq:arc_length_NR_iteration}. The initial guess is set manually by the user and can be derived from a random number, an approximate linearized system response, or the FFT of a numerical solution. Based on this common core, four solvers are implemented to accommodate different scenarios. Specifically, \texttt{continuation.py} performs forced-response continuation when the nonlinear derivatives are provided by the model (use in conjunction with \texttt{SecondOrderTimeModel}), whereas \texttt{continuation\_autodiff.py} cooperates with \texttt{autodiff\_utils.py} to evaluate these derivatives automatically. The corresponding free-frequency procedures are exposed through \texttt{continuation\_free\_frequency.py} and \texttt{continuation\_free\_frequency\_autodiff.py}, which additionally incorporate the phase condition and the further-augmented continuation variable introduced in the free-frequency formulation. These solver modules also integrate the weighted arc-length constraint in Eq.\eqref{eq:arc_length_constraint_weighted} and the sparse block solution in Eqs.\eqref{eq:arc_length_NR_iteration_augmented_blocked_solution}, while storing the converged harmonic coefficients and continuation parameters for subsequent analysis.

For postprocessing, \texttt{floquet.py} and \texttt{floquet\_autodiff.py} provide alternative stability-analysis interfaces according to whether the tangent matrices are supplied by the model or obtained through AD. These modules reconstruct the sampled time histories from the saved harmonic coefficients, assemble the periodic tangent matrices $\bar{\mathbf{M}}(\tau)$, $\bar{\mathbf{C}}(\tau)$, and $\bar{\mathbf{K}}(\tau)$ in Eq.\eqref{eq:linearized_perturbation_equation}, and compute the Floquet multipliers from the resulting monodromy matrix. The AD-enabled variants reuse \texttt{autodiff\_utils.py}, while the propagation of perturbations can be carried out using either the matrix-exponential approach or the implicit integration procedure in Eqs.\eqref{eq:monodromy_matrix_implicit_integrator}-\eqref{eq:monodromy_matrix_implicit_integrator_monodromy}. Furthermore, some systems, such as piezoelectric energy harvesters and structure-thermal coupling, exhibit a combination of first- and second-order dynamics \cite{yuanHarmonicBalanceApproach2019, gaoNonlinearThermalBehaviors2020}. By appropriately filling the mass matrix with zeros, neither the preprocessing nor the solution is affected; however, since the filled mass matrix is negative definite, Eq.\eqref{eq:linearized_perturbation_equation_state_2} is not available. Hence, pyHB implements an additional API, \texttt{floquet\_mixed\_autodiff.py}, to handle stability analysis problems for such situations. 

This modular organization separates system-specific information from the reusable numerical machinery and provides consistent interfaces for both analytical-derivative and AD-enabled workflows. Although the AD-enhanced solver constitute a principal focus of pyHB, analytical-derivative counterparts are also provided for both forced-response and free-frequency continuation, allowing explicitly available derivataion information to be fully utilized. Moreover, \texttt{floquet\_mixed\_autodiff.py} extends the Floquet multiplier computation to systems combining first- and second-order dynamics. From this aspect, pyHB is not merely an AD-enhanced implementation of HB solver, but an open-source library that integrates the complete HB analysis workflow, from model definition and harmonic precomputation to the initial solution, arc-length continuation, solution storage, response reconstruction, and stability assessment, through a coherent and comprehensive set of APIs. After specifying the harmonic and continuation settings, users can concentrate on defining the system matrices, excitation, and nonlinear force without reimplementing the HB operators, Jacobian construction, continuation algorithm, or monodromy matrix. This architecture improves maintainability and extensibility while making pyHB applicable to a broad range of user-defined nonlinear systems. 

\section{Numerical examples}

This section gives several numerical examples with real-world contexts to demonstrate the effectiveness of pyHB, including: (1) A quasi-zero stiffness vibration isolator with multiple zero stiffness points (MZS-QZS); (2) A nonlinear piezoelectric energy harvester (PEH) with a magnetic oscillator; (3) An aeroengine dual-rotor-bearing-casing system considering the nonlinearity of the inter-shaft bearing; and (4) A forced Bernoulli beam with a nonlinear spring support. It should be emphasized that, due to space limitations, this section focuses only on the four examples mentioned above. In the pyHB open-source repository, we provide a broader range of examples, each accompanied by complete, executable code. Users can refer to these resources to easily apply pyHB to analyze systems of interest.

\subsection{Quasi-zero stiffness vibration isolator with multiple zero stiffness points}

\begin{figure}[htbp]
      \centering
      \includegraphics[width=1.0\textwidth]{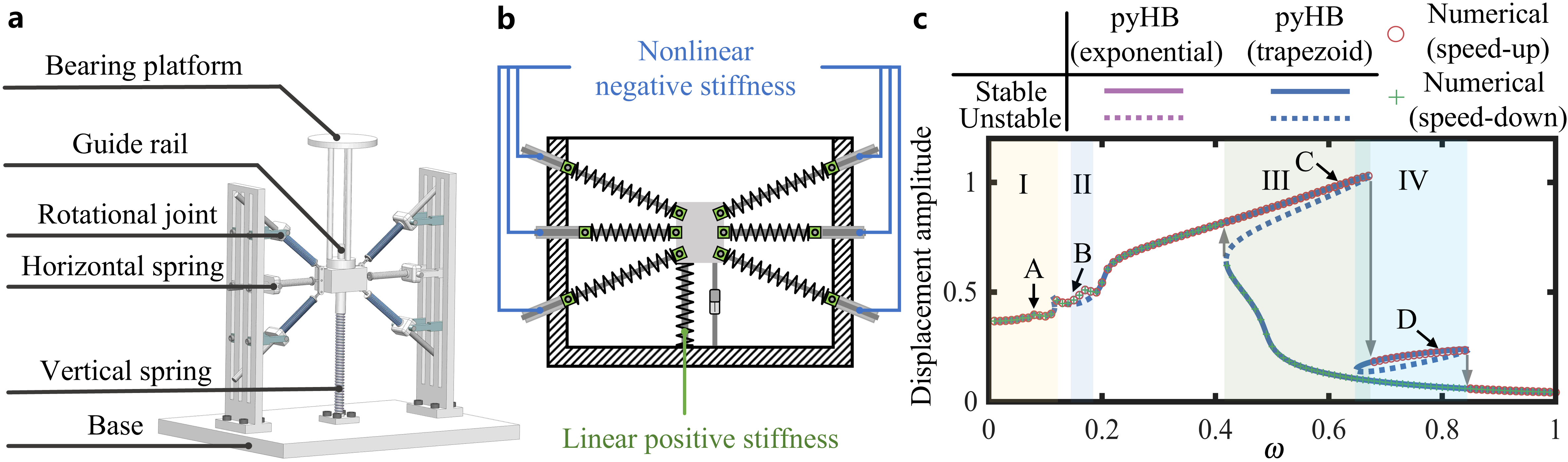}
      \caption{Adopting the pyHB to analyze the periodic response of the MZS-QZS isolator. \textbf{a} The 3D structure of the MZS-QZS isolator, consisting of a bearing platform, a guide rail, rotational joints, springs, and a base. \textbf{b} Schematic diagram of the MZS-QZS isolator's theoretical model, in which the linear positive stiffness component and multiple nonlinear negative stiffness components work together to achieve QZS characteristics, thereby widening the QZS range and enabling the isolator to accommodate potential payload deviations. \textbf{c} Amplitude-frequency response curves of the MZS-QZS isolator, where the solid and dashed lines represent stable and unstable solutions, respectively. The overall curve can be divided into four regions.}
      \label{fig:mzx_qzs_isolator}
\end{figure}

QZS isolators have important applications in the field of low- and ultra-low-frequency vibration isolation. The first example consider the MZS-QZS isolator proposed by Meng et al. \cite{mengQuasizeroStiffnessVibration2025}, whose 3D structure is shown in Fig.\ref{fig:mzx_qzs_isolator}\textbf{a}. The bearing platform is constrained to move vertically along the guide rails, while the rotational joints and symmetrically arranged springs form the nonlinear stiffness mechanisms. The theoretical model corresponding to Fig.\ref{fig:mzx_qzs_isolator}\textbf{a} is illustrated in Fig.\ref{fig:mzx_qzs_isolator}\textbf{b}, where the vertical spring provides linear positive stiffness, while the multiple horizontal springs form nonlinear negative stiffness. The dimensionless governing equation of the MZS-QZS isolator is given by
\begin{equation}
      \label{eq:mzx_qzs_isolator_governing_equation}
      \begin{aligned}
            \ddot{x} &+ 2 \zeta_{v} \dot{x} + \frac{8 \zeta_{2} \mu_{2}^{2}}{((\mu_{3} + x)^2 + \mu_{2}^{2})^2} \dot{x} + \frac{8 \zeta_{2} \mu_{5}^{2}}{(\mu_{5}^2 + x^{2})^2} \dot{x} + \frac{8 \zeta_{2} \mu_{2}^{2}}{((\mu_{3} - x)^2 + \mu_{2}^{2})^2} \dot{x} - \frac{(1 + \mu_{1})(\mu_{3} + x)}{\sqrt{\mu_{2}^{2} + (\mu_{3} + x)^2}} \\
            &- \frac{(1+\mu_{1})(x - \mu_{3})}{\sqrt{\mu_{2}^2 + (\mu_{3} - x)^{2}}} - \frac{\lambda_{1} \mu_{4} x}{\sqrt{\mu_{5}^{2} + x^{2}}} + (2 + \lambda_1 + \lambda_2) x = \bar{f} \cos(\omega t),
      \end{aligned}
\end{equation}
where seven structural parameters are contained, based on the geometric relationships and the zero stiffness conditions, only three of them are independent. More detailed derivations and explanations can be found in reference \cite{mengQuasizeroStiffnessVibration2025}. This isolator has multiple zero stiffness points, resulting in a wider quasi-zero stiffness range that allows it to better accommodate potential payload deviations.

Fig.\ref{fig:mzx_qzs_isolator}\textbf{c} presents the amplitude-frequency response of $omega$ over the range 0 to 1, including both stable and unstable periodic solutions. It can be observed that, owing to the complex nonlinear stiffness-damping coupling and the strongly non-monotonic restoring characteristic, the response curve contains successive bifurcation points and multiple coexisting solutions. The solid and dashed segments identify stable and unstable periodic solutions, respectively. The results obtained when the Floquet multipliers are evaluated using the matrix exponential and the implicit trapezoidal integrator are almost indistinguishable, demonstrating that the two postprocessing implementations provide consistent stability classifications. Hence, the implicit integration method is utlized by default in the following examples. The numerical results obtained by the 8th-order explicit Runge-Kutta method (supported by \texttt{scipy.integrate.DOP853}) are also provided in Fig.\ref{fig:mzx_qzs_isolator}\textbf{c}, including speed-up and speed-down procedures. It can be found that, since the numerical integration method cannot compute unstable periodic solutions, the results exhibit discontinuities near bifurcation points, thereby failing to capture the complete landscape of the system's solutions. In contrast, the arc-length continuation method passes through the bifurcation points and retains the unstable segments. In addition, the numerical results agree closely with the stable branches obtained by pyHB, confirming its accuracy. 

\begin{figure}[htbp]
      \centering
      \includegraphics[width=1.0\textwidth]{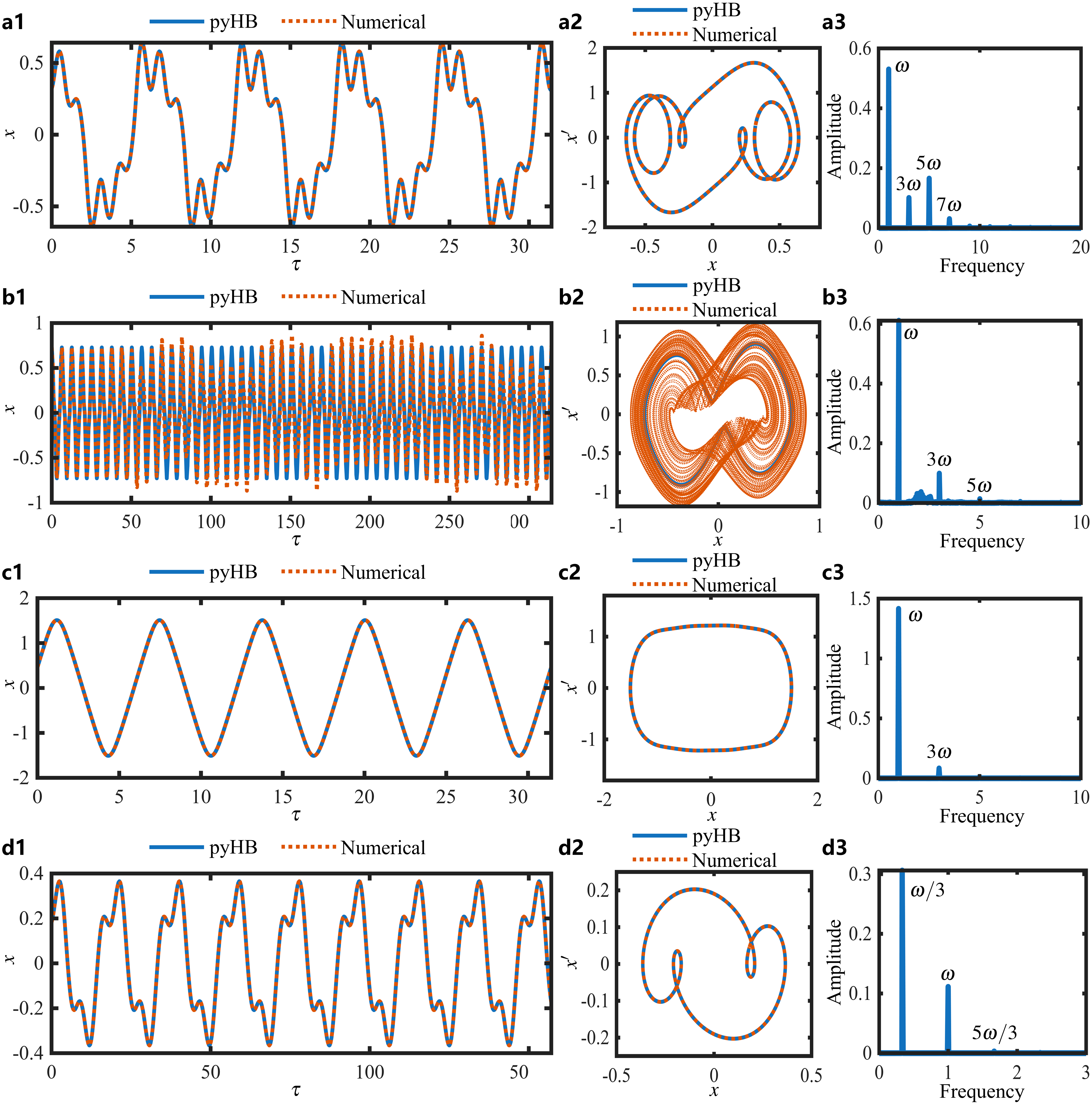}
      \caption{Characteristics of the solutions in each region of the MZS-QZS isolator's frequency-amplitude response curve. \textbf{a1}-\textbf{a3} Time history, phase portrait, and frequency spectrum of point A ($\omega = 0.0785$, region I). \textbf{b1}-\textbf{b3} Time history, phase portrait, and frequency spectrum of point B ($\omega = 0.1368$, region II). \textbf{c1}-\textbf{c3} Time history, phase portrait, and frequency spectrum of point C ($\omega = 0.6349$, region III). \textbf{d1}-\textbf{d3} Time history, phase portrait, and frequency spectrum of point D ($\omega = 0.7952$, region IV).}
      \label{fig:mzx_qzs_isolator_time_series}
\end{figure}

The frequency-amplitude response shown in Fig.\ref{fig:mzx_qzs_isolator}\textbf{c} can be divided into I-IV regions. To clarify the solutions' characteristics, Fig.\ref{fig:mzx_qzs_isolator_time_series} further compares the representative solutions A-D selected from the four regions. Note that the initial conditions of numerical results are set to the corresponding HB solutions. At point A ($\omega=0.0785$), the pyHB's time history and phase portrait coincide with the numerical results, while the frequency spectrum contains the excitation-frequency component $\omega$ and pronounced odd higher harmonics, indicating that region I has super-harmonic responses. At point B ($\omega=0.1368$), what stands out in Fig.\ref{fig:mzx_qzs_isolator_time_series}\textbf{b1} is that the results from HB and the numerical method only match in the early stages; thereafter, a discrepancy arises. This reflects the fact that the periodic solution at point B is unstable and cannot be sustained over the long term by the numerical method. Furthermore, Figs.\ref{fig:mzx_qzs_isolator_time_series}\textbf{b1} and \textbf{b2} show that the phase portrait is no longer closed, but instead oscillates repeatedly within a bounded region, while the spectrum exhibits a dense cluster of peaks near $2 \omega$. These results suggest the emergence of chaos in region II. At point C ($\omega=0.6349$), Fig.\ref{fig:mzx_qzs_isolator_time_series}\textbf{c} presents that the response is dominated by $\omega$, with a much smaller $3 \omega$. Here, the excitation-frequency is near the system's resonance, and region III is the main resonance region with hardening characteristics. Unlike the results described above, Fig.\ref{fig:mzx_qzs_isolator_time_series}\textbf{d} shows that although the phase portrait at point D still exhibits a multi-loop structure, the time history is now dominated by $\omega / 3$ and its odd multiples, which is a typical 1/3 subharmonic resonance phenomenon. Fortunately, by simply setting the frequency resolution to $1/3$ and using an appropriate initial guess, pyHB can produce subharmonic responses that agree with the numerical results and further fully characterize the unstable periodic solutions in region IV. Collectively, the four representative solutions demonstrate that pyHB can provide an accurate and comprehensive understanding of the MZS-QZS isolator's complex nonlinear dynamics.

\subsection{Nonlinear piezoelectric energy harvester with a magnetic oscillator}
\begin{figure}[htbp]
      \centering
      \includegraphics[width=1.0\textwidth]{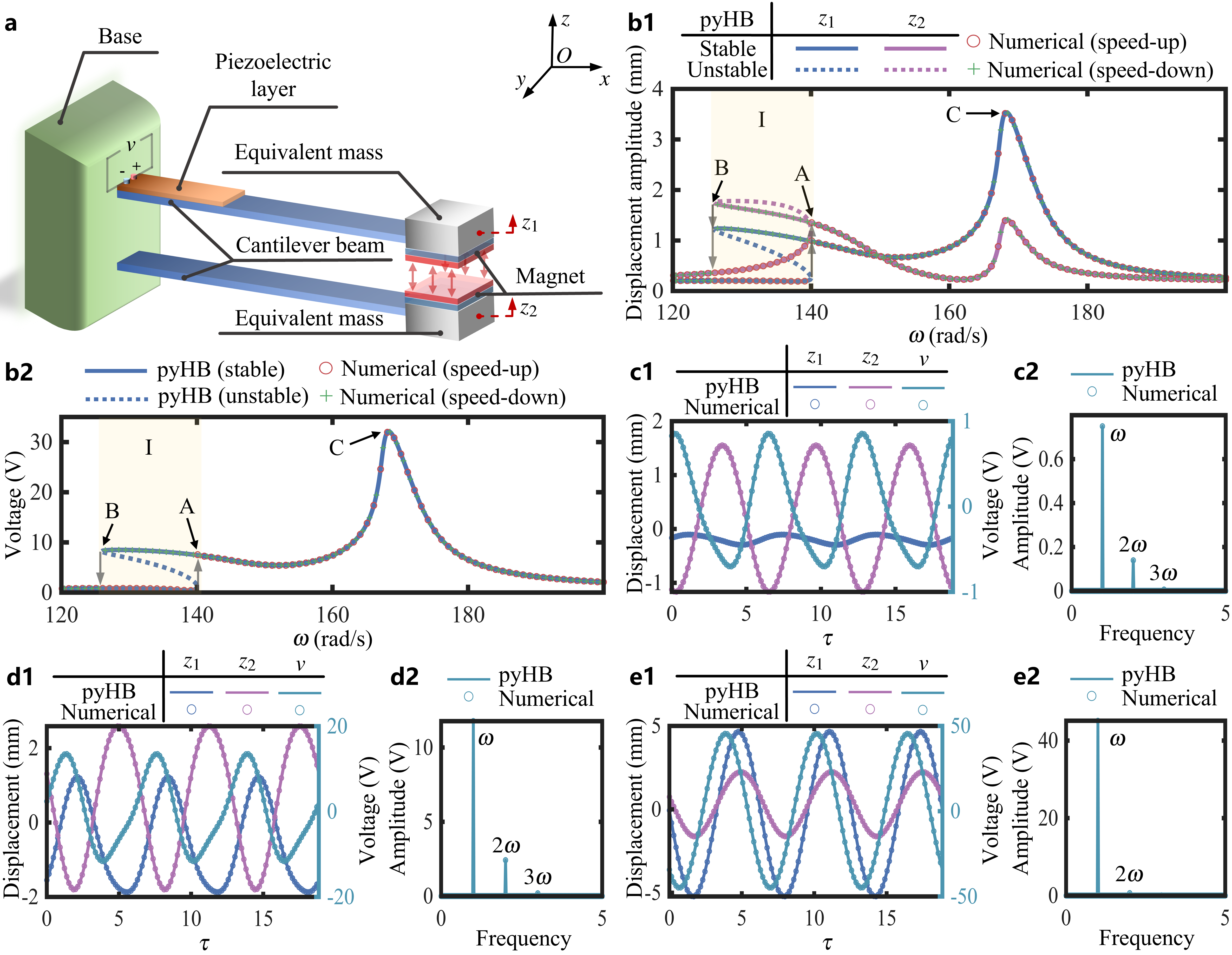}
      \caption{Adopting the pyHB to analyze the periodic response of the nonlinear PEH with a magnetic oscillator. \textbf{a} Schematic diagram of the nonlinear PEH with a magnetic oscillator. PEH and magnetic oscillator are coupled via magnetic interactions to enhance the bandwidth. \textbf{b1} and \textbf{b2} Amplitude-frequency response curves of the displacement and voltage, respectively. \textbf{c1} and \textbf{c2} Time history and the voltage's frequency spectrum of point A ($\omega = 139.5825$rad/s, region I). \textbf{d1} and \textbf{d2} Time history and the voltage's frequency spectrum of point B ($\omega = 129.2576$rad/s, region I). \textbf{e1} and \textbf{e2} Time history and the voltage's frequency spectrum of point C ($\omega = 168.5779$rad/s).}
      \label{fig:piezoelectric_magnetic_harvester}
\end{figure}

The second example considers the PEH with a magnetic oscillator proposed by Tang and Yang \cite{tangNonlinearPiezoelectricEnergy2012a}. As shown in Fig.\ref{fig:piezoelectric_magnetic_harvester}\textbf{a}, the system consists of two cantilever beams: the top one has a piezoelectric layer made of piezoelectric fiber composite at its fixed end, serving as a PEH; the bottom one is a magnetic oscillator. Both have mass blocks with embedded magnets at the free ends, resulting in a nonlinear magnetic interaction. 

The displacements of the PEH and the magnetic oscillator are denoted by $z_{1}$ and $z_{2}$, respectively, and $v$ is the output voltage. Under the dipole-dipole interaction assumption, the electromechanically coupled governing equations can be expressed as
\begin{subequations}
      \label{eq:piezoelectric_magnetic_harvester_governing_equation}
      \begin{align}
            M_{\mathrm{eq},1}\ddot{z}_{1}+C_{\mathrm{eq},1}\dot{z}_{1}+K_{\mathrm{eq},1}z_{1}+Hv-F_{\mathrm{mag}}(z_{1},z_{2})&=-\lambda_{1}M_{\mathrm{eq},1}\ddot{z}_{0}, \\
            M_{\mathrm{eq},2}\ddot{z}_{2}+C_{\mathrm{eq},2}\dot{z}_{2}+K_{\mathrm{eq},2}z_{2}+F_{\mathrm{mag}}(z_{1},z_{2})&=-\lambda_{2}M_{\mathrm{eq},2}\ddot{z}_{0}, \\
            \frac{v}{R}+C_{s}\dot{v}-H\dot{z}_{1}&=0,
      \end{align}
\end{subequations}
where
\begin{equation}
      \label{eq:piezoelectric_magnetic_harvester_governing_equation_magnetic_force}
      F_{\mathrm{mag}}(z_{1},z_{2})=\frac{3\mu_{0}m_{1}m_{2}}{2\pi(z_{1}-z_{2}+D_{0})^{4}}
\end{equation}
is the magnetic force. $M_{\mathrm{eq},i}$, $C_{\mathrm{eq},i}$, and $K_{\mathrm{eq},i}$ are the equivalent mass, damping, and stiffness of the two cantilevers, respectively. $H$ is the electromechanical coupling coefficient. $R$ and $C_{s}$ are the payload resistance and clamped capacitance. $m_{1}$ and $m_{2}$ are the magnetic dipole moments, and $D_{0}$ is the initial magnet spacing. $\mu_{0}$ is the vacuum permeability. $\ddot{z}_{0}$ is the base acceleration, and $\lambda_{i}$ is the forcing correction factor. More details on the model and parameters are available in reference \cite{tangNonlinearPiezoelectricEnergy2012a}. Note that the first two equations in Eq.\eqref{eq:piezoelectric_magnetic_harvester_governing_equation} are second-order mechanical equations, whereas the voltage equation is first-order. In pyHB, this mixed-order system is handled by setting the mass matrix entry associated with $v$ to zero, and stability is evaluated using \texttt{floquet\_mixed\_autodiff.py}.

Figs.\ref{fig:piezoelectric_magnetic_harvester}\textbf{b1} and \textbf{b2} present the displacement and voltage amplitude-frequency response curves over $\omega=120$-$200$ rad/s. The voltage amplitude-frequency profile is generally consistent with that of the PEH. Compared to the linear PEH and the PEH with a fixed magnetic dipole, the system in Eq.\eqref{eq:piezoelectric_magnetic_harvester_governing_equation} exhibits an additional peak near the magnetic oscillator's resonance (region I), thereby broadening its operating bandwidth. Region I contains multiple coexisting periodic solutions and exhibits typical softening behavior. The numerical speed-up and speed-down results follow different stable branches and exhibit jumps near the bifurcation points, whereas pyHB continuously traces both the stable and unstable solutions and overlaps the corresponding numerical results. In addition to region I, a second pronounced peak occurs near the PEH's resonance, where the PEH vibration and output voltage reach their largest amplitudes. In terms of stability, the \texttt{floquet\_mixed\_autodiff.py} provided by pyHB successfully computed the Floquet multipliers for this class of mixed-order systems, and the stability determination is consistent with the results in reference \cite{yuanHarmonicBalanceApproach2019}. 

To further illustrate the response characteristics, Fig.\ref{fig:piezoelectric_magnetic_harvester}\textbf{c}--\textbf{e} compare the pyHB and numerical solutions at points A-C. At point A ($\omega=139.5825$ rad/s), the magnetic oscillator has a larger displacement than the PEH, and the voltage spectrum is dominated by $\omega$, with smaller $2\omega$ and $3\omega$ components. Point B ($\omega=129.2576$ rad/s) lies on the high-amplitude branch in region I. Here, energy from the magnetic oscillator's resonance is partially transmitted to the PEH via magnetic interactions, thereby significantly enhancing the voltage output. At point C ($\omega=168.5779$ rad/s), the PEH response and output voltage reach their main resonance peaks, and the voltage spectrum is almost entirely dominated by $\omega$. The pyHB and numerical time histories coincide at all three points, further proving its effectiveness and generality. 

\subsection{Aeroengine dual-rotor-bearing-casing system}
\begin{figure}[htbp]
      \centering
      \includegraphics[width=1.0\textwidth]{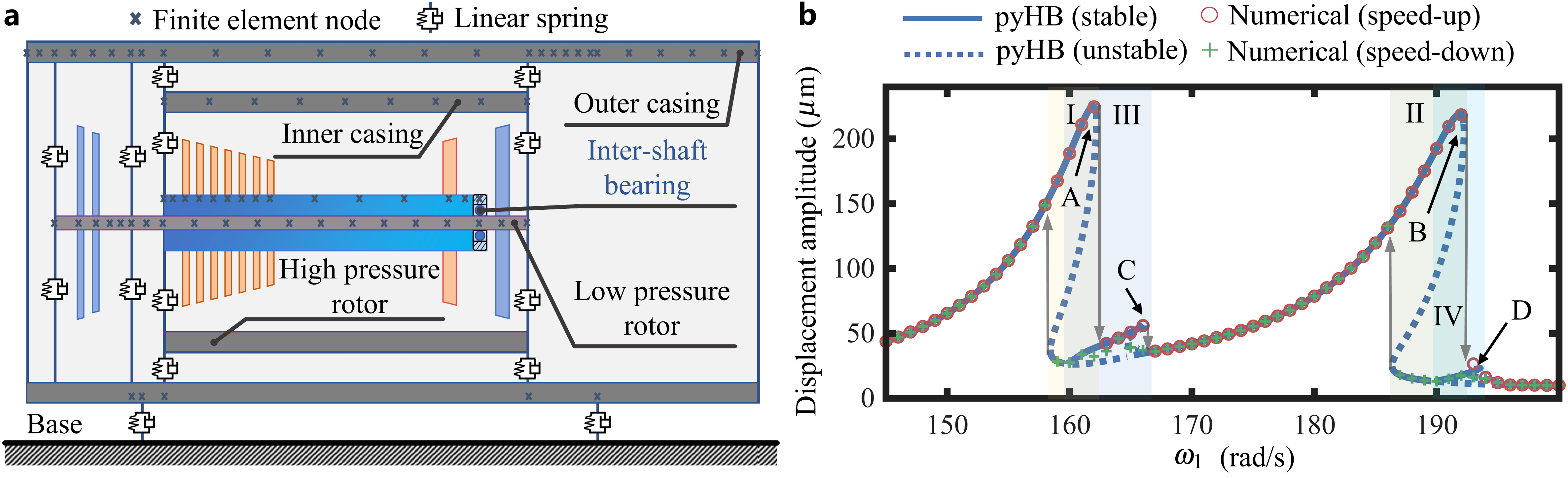}
      \caption{Adopting the pyHB to analyze the periodic response of the aeroengine dual-rotor-bearing-casing system. \textbf{a} The finite element model of the aeroengine dual-rotor-bearing-casing system consists of a low-pressure rotor, a high-pressure rotor, an inner casing, an outer casing, and an inter-shaft bearing. The rotors are connected to the casings, and the casings are connected to the base, via linear springs; the high- and low-pressure rotors are connected through the nonlinear inter-shaft bearing. The speed ratio, $\omega_{2}/\omega_{1}$, between the high- and low-pressure rotors is 1.2. \textbf{b} Amplitude-frequency response curves at the inter-shaft-bearing node, where the solid and dashed lines represent stable and unstable solutions, respectively. The overall curve can be divided into four multivalued regions.}
      \label{fig:aeroengine}
\end{figure}

The third example considers the aeroengine dual-rotor-bearing-casing system \cite{chenCombinationResonancesDualrotorbearingcasing2024}. As shown in Fig.\ref{fig:aeroengine}\textbf{a}, the theoretical model includes the inner and outer casings, the high and low pressure rotors modeled with Timoshenko beam elements, and the compressor and turbine disks, which are simplified as rigid disks. The casings and rotors are connected by linear springs, while an inter-shaft bearing connects the high and low pressure rotors. The nonlinear force of the latter is described by the Hertz contact theory as
\begin{subequations}
      \label{eq:inter_shaft_bearing_nonlinear_force}
      \begin{align}
            F_{x} = K_{b} \sum_{k = 1}^{N_{b}} \delta_{k}^{10/9} H(\delta_{k}) \cos(\theta_{k}), \\
            F_{y} = K_{b} \sum_{k = 1}^{N_{b}} \delta_{k}^{10/9} H(\delta_{k}) \sin(\theta_{k}),
      \end{align}
\end{subequations}
where $k_{b}$ is the contact stiffness, $N_{b}$ is the number of rollers, $\delta_{k}$ and $\theta_{k}$ is the radial deformation and the angular position of the $k$-th roller, respectively. $H(\cdot)$ is the Heaviside function. $\delta_{k}$ considers the radial clearance $\delta_{0}$ and the relative displacement between the two rotors, given by
\begin{equation}
      \label{eq:inter_shaft_bearing_nonlinear_force_radial_deformation}
      \delta_{k} = (x_{i} - x_{o}) \cos(\theta_{k}) + (y_{i} - y_{o}) \sin(\theta_{k}) - \delta_{0},
\end{equation}
where $(x_{i},y_{i})$ and $(x_{o},y_{o})$ are the displacements of the inner and outer rotors at the inter-shaft bearing node, respectively. The system is excited by the unbalance force. Notably, the two rotors rotate asynchronously with a fixed speed ratio of $\omega_{2}/\omega_{1}=1.2$; hence, the system is simultaneously excited at two base frequencies, $\omega_{1}$ and $\omega_{2}$.

The finite element model shown in Fig.\ref{fig:aeroengine}\textbf{a} has 284 DOFs and incorporates various nonlinearities, including gaps, fractional exponents, and the Heaviside function, making it a challenging problem for nonlinear analysis. Fig.\ref{fig:aeroengine}\textbf{b} presents the displacement amplitude-frequency response at the inter-shaft-bearing node as the low pressure rotor speed $\omega_{1}$ is varied. It can be observed that the system exhibits complex response patterns, including four multiple solution regions labeled I-IV. Among them, regions I and II have larger amplitudes and represent distinct hardening characteristics and resonance hysteresis phenomena, indicating that they are the primary resonance region. The numerical speed-up and speed-down results overlap the corresponding stable branches obtained by pyHB but jump between them near the bifurcation points. This behavior can easily lead to deterioration of the operating conditions for the inter-shaft bearing, thereby adversely affecting the system's health and stable operation. pyHB, on the other hand, can trace the complete response structure, including the unstable branches, thereby offering a more comprehensive understanding of the system's dynamics and potential guidance for avoiding or mitigating such vibration jumps. 

\begin{figure}[htbp]
      \centering
      \includegraphics[width=1.0\textwidth]{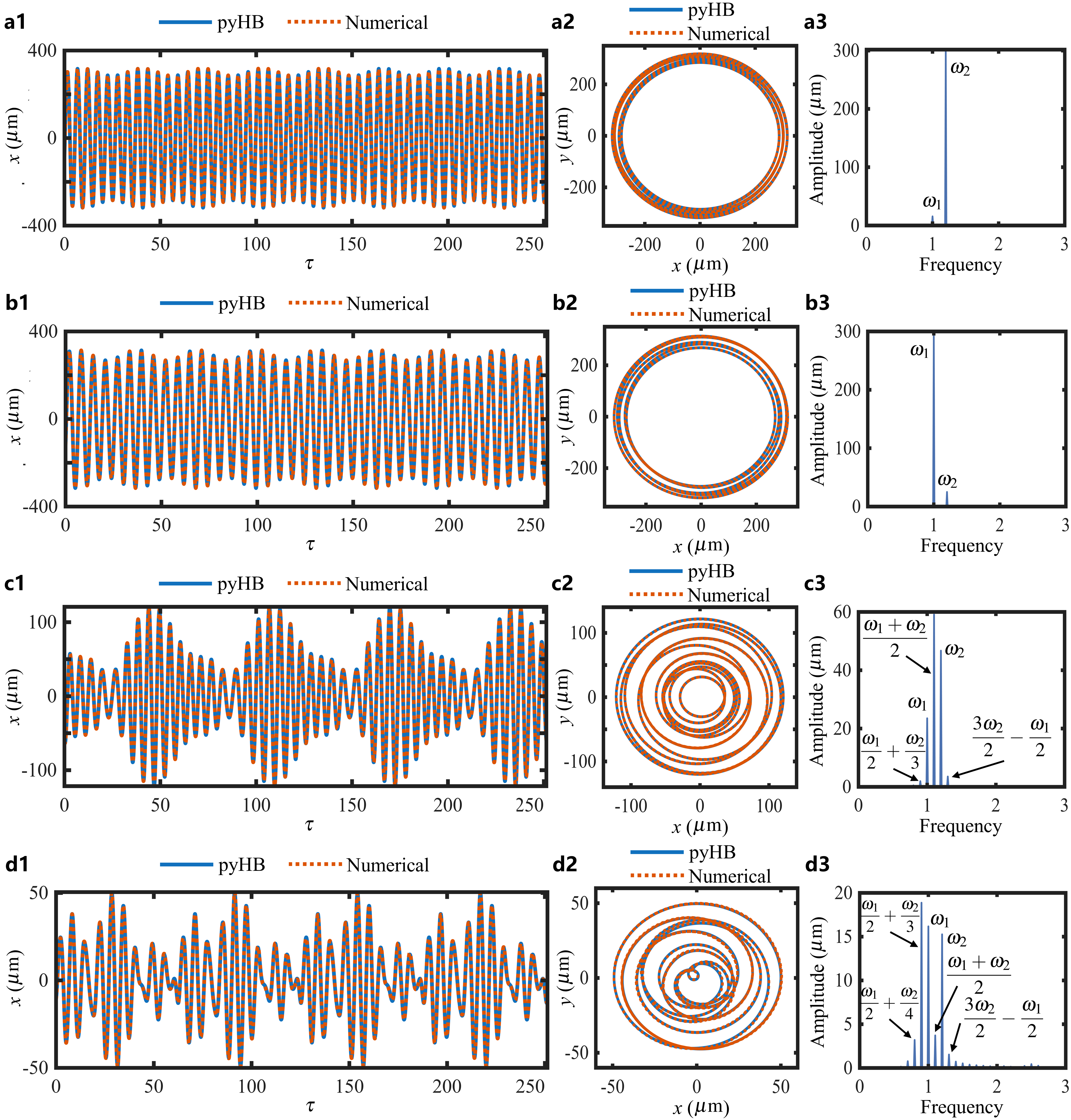}
      \caption{Characteristics of the solutions in each region of the aeroengine dual-rotor-bearing-casing system's amplitude-frequency response curve at the inter-shaft-bearing node. \textbf{a1}--\textbf{a3} Time history, rotor orbit, and frequency spectrum of point A ($\omega_{1}=162.1959$ rad/s, region I). \textbf{b1}--\textbf{b3} Time history, rotor orbit, and frequency spectrum of point B ($\omega_{1}=192.2347$ rad/s, region II). \textbf{c1}--\textbf{c3} Time history, rotor orbit, and frequency spectrum of point C ($\omega_{1}=165.8363$ rad/s, region III). \textbf{d1}--\textbf{d3} Time history, rotor orbit, and frequency spectrum of point D ($\omega_{1}=193.3422$ rad/s, region IV).}
      \label{fig:aeroengine_time_series}
\end{figure}

To further investigate the characteristics of the solutions in each region, Fig.\ref{fig:aeroengine_time_series} presents time histories, rotor orbits, and the frequency spectrum of the inter-shaft bearing node at points A-D. Figs.\ref{fig:aeroengine_time_series}\textbf{a} and \textbf{b} show similar patterns, with the frequency spectrum containing only two base components; however, point A is dominated by $\omega_{2}$, whereas point B is dominated by $\omega_{1}$. The dominant frequencies at the two points are numerically close. These results indicate that region I is excited when the high pressure rotor passes through the system's critical speed, while region II is excited when the low pressure rotor passes through the system's critical speed. This is a typical phenomenon in a dual-frequency excitation system. In contrast, the responses at points C and D are entirely different. The time histories exhibit a more complex multi-peak structure (shown in Figs.\ref{fig:aeroengine_time_series}\textbf{c1} and \textbf{d1}), and the rotor orbits have a greater number of loops (shown in Figs.\ref{fig:aeroengine_time_series}\textbf{c2} and \textbf{d2}). Of even greater interest is the frequency spectrum. As shown in Figs.\ref{fig:aeroengine_time_series}\textbf{c3} and \textbf{d3}, points C and D contain multiple frequency components in addition to the two base frequencies; these are not simply superharmonics or subharmonics, but fractional combinations of $\omega_{1}$ and $\omega_{2}$. Specifically, point C is dominated by $(\omega_{1} + \omega_{2}) / 2$, and point D is dominated by $\omega_{1} / 2 + \omega_{2} / 3$. This phenomenon, commonly referred to as combination resonance, arises from dual-frequency excitation and the system's intrinsic nonlinearity. Fortunately, pyHB can capture these complex frequency components and provide periodic solutions that closely match the numerical results. It should be emphasized that users need only provide the expression for the inter-shaft bearing's nonlinear force (shown in Eq.\eqref{eq:inter_shaft_bearing_nonlinear_force}). The calculation of its partial derivatives and the assembly of the Jacobian matrix are performed automatically within pyHB, which is very convenient. The above results fully demonstrate the effectiveness of pyHB in high-dimensional systems with complex nonlinearities. 

\subsection{forced Bernoulli beam with a nonlinear spring support}
\begin{figure}[htbp]
      \centering
      \includegraphics[width=1.0\textwidth]{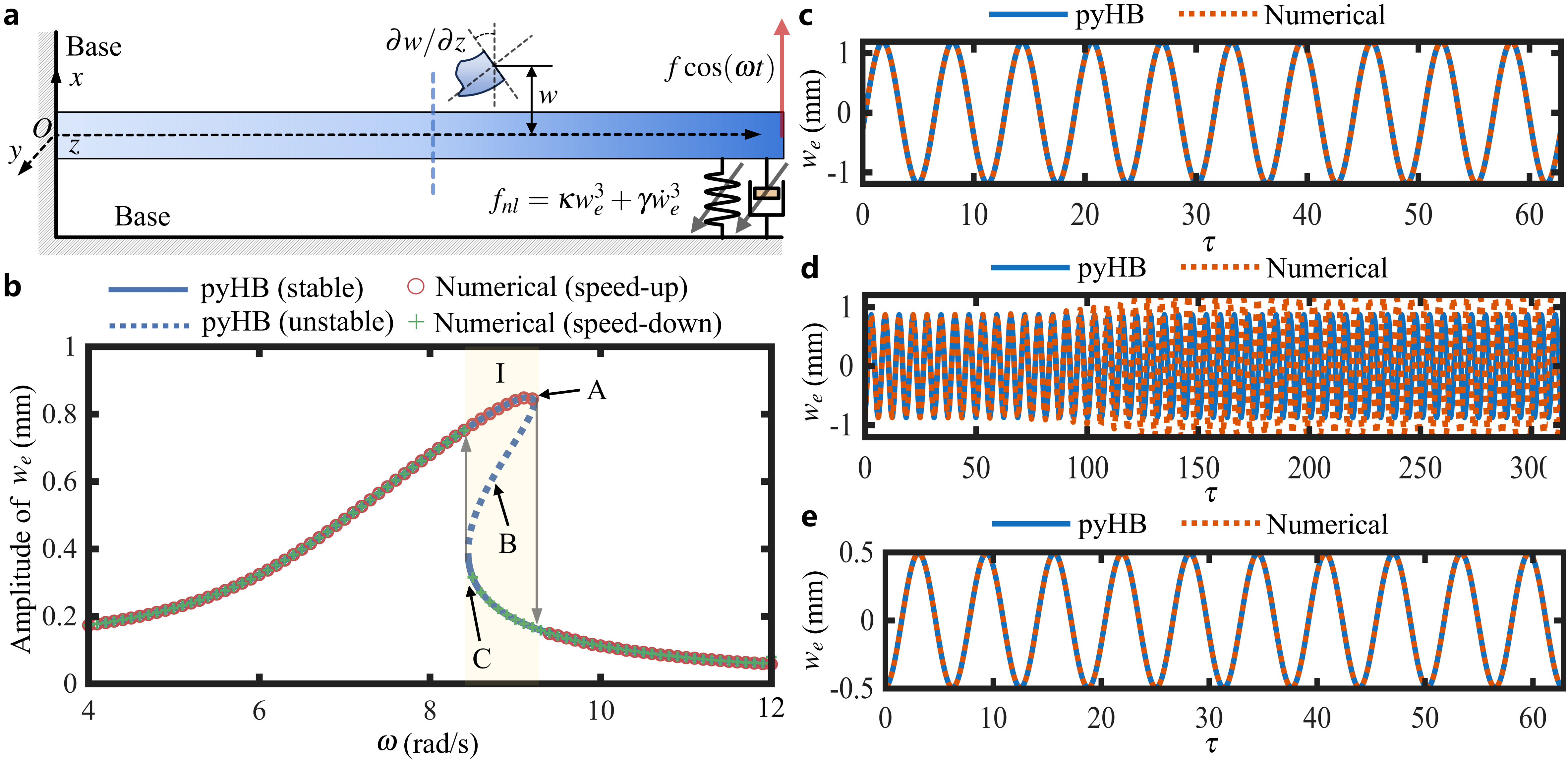}
      \caption{Adopting pyHB to analyze the periodic response of the forced Bernoulli beam with a nonlinear spring support. \textbf{a} Schematic diagram of the forced Bernoulli beam with a nonlinear spring support, including cubic stiffness and cubic damping. \textbf{b} Amplitude-frequency response curve of the free-end deflection. The system exhibits hardening characteristics, and multiple solutions coexist in region I. \textbf{c} Time history of the free-end deflection at point A ($\omega=9.1734$ rad/s, upper stable branch). \textbf{d} Time history of the free-end deflection at point B ($\omega=8.7141$ rad/s). The periodic solution obtained by pyHB is unstable; hence, the numerical solution agrees with it only during the initial stage and cannot remain on this orbit over the long term. It ultimately converges to a stable branch in region I, with the selected branch depending on the initial conditions. \textbf{e} Time history of the free-end deflection at point C ($\omega=8.4479$ rad/s, lower stable branch).}
      \label{fig:bernoulli_beam}
\end{figure}

The final example considers a cantilever beam with the free end subjected to harmonic excitation, as shown in Fig.\ref{fig:bernoulli_beam}\textbf{a}. The free end is supported by a nonlinear spring with cubic stiffness and damping. Based on the Bernoulli beam theory, the governing equation after finite element discretization can be expressed as \cite{ponsioenModelReductionSpectral2020}
\begin{subequations}
      \label{eq:bernoulli_beam_governing_equation}
      \begin{align}
            \mathbf{M}\ddot{\mathbf{x}}+\mathbf{C}\dot{\mathbf{x}}+\mathbf{K}\mathbf{x}+\mathbf{e}_{e}f_{\mathrm{nl}}(w_{e},\dot{w}_{e})&=\mathbf{e}_{e}f\cos(\omega t), \\
            f_{\mathrm{nl}}(w_{e},\dot{w}_{e})&=\kappa w_{e}^{3}+\gamma\dot{w}_{e}^{3},
      \end{align}
\end{subequations}
where $\mathbf{x}$ is the vector of generalized coordinates, consisting of the node deflection and deflection angle. $w_{e}$ denotes the deflection of the free end. $\kappa$ and $\gamma$ represent the nonlinear stiffness and damping coefficients, here we set $\kappa = 4.0$ and $\gamma = 10^{-3}$. More physical parameters and details are provided in reference \cite{ponsioenModelReductionSpectral2020}. 

Notably, a fine-mesh finite element discretization is employed here, with the system having 2000 DOFs, to fully evaluate the effectiveness and computational efficiency of pyHB in the high-dimensional system. Fig.\ref{fig:bernoulli_beam}\textbf{b} presents the amplitude-frequency response of $w_{e}$. The amplitude-frequency curve bends toward higher excitation frequencies due to nonlinearity and exhibits typical hardening behavior. Region I contains two stable solutions separated by an unstable solution, leading to hysteresis under frequency sweeps. The numerical speed-up and speed-down results follow the corresponding stable branches and jump near the bifurcation points. It should be emphasized that the explicit Runge-Kutta method adopted in the previous examples is unstable for this system, and the Newmark-$\beta$ method is used instead. For nonlinear systems, Newton-Raphson iterations must be performed at each time step, resulting in a significant increase in computational overhead compared to linear systems. Under identical hardware and software conditions, the Newmark-$\beta$ method takes more than 80s to compute the response over 10 periods at a single $\omega$ value, whereas pyHB, benefiting from its well-designed sparse matrix structure and full utilization of nonlinear locality, can generate the complete amplitude-frequency response curve in 15s, demonstrating exceptionally high computational efficiency. 

Figs.\ref{fig:bernoulli_beam}\textbf{c}-\textbf{e} further compare the pyHB and numerical time histories at representative points A-C. Points A and C lie on the upper and lower stable branches, respectively, and the two methods yield identical periodic responses. Remarkably, although the numerical trajectory is initialized with the pyHB solution, small perturbations grow over time and drive it away from the unstable periodic orbit toward one of the stable branches in region I. This behavior is consistent with the stability classification obtained by pyHB and illustrates an important advantage of the HB formulation: it provides a comprehensive landscape of the system's dynamics, including unstable ones, and a solid foundation for studying potential nonlinear resonance mechanisms.

\section{Comparisons with existing methods}
To further illustrate the effectiveness and superiority of pyHB, this section provides a detailed comparison of its performance with other methods in the numerical examples discussed above. The methods considered include: (1) the analytical solver provided by pyHB, which is essentially an efficient implementation of the traditional HB method; (2) the AD-enhanced solver provided by pyHB; (3) our previous work, denoted as HB-AD \cite{chenHarmonicBalanceautomaticDifferentiation2026}; and (4) numerical methods (Runge-Kutta or Newmark-$\beta$) for direct time integration. To ensure a fair comparison, all methods are run under the same hardware and software conditions (Python 3.13, PyTorch 2.12, Numpy 2.4, Scipy 1.17, Intel Core Ultra 7 270K Plus CPU, 32GB of RAM, and an NVIDIA GeForce RTX 5070 Ti GPU with 16GB of video memory) and the numerical methods are initialized based on the HB solutions to obtain steady-state responses directly within 10 periods, thereby avoiding an excessive exaggeration of the comparison results due to a long time integration interval.

Tab.\ref{tab:comparison} summarizes the computational time and memory usage of pyHB, HB-AD, and direct numerical integration for the four examples discussed above. First, the analytical solver of pyHB has the shortest average iteration and convergence times across three examples. This is expected because explicitly provided derivatives avoid the computational-graph construction and derivative propagation required by AD. However, this efficiency relies on manually deriving and implementing the Jacobian matrices of the nonlinear force. For complex or non-smooth nonlinearities, this procedure is often time-consuming and error-prone, and must be repeated whenever the nonlinear model is changed. By contrast, both AD-enhanced approaches eliminate this requirement and achieve better generality. Notably, in the first two cases, the difference in solution efficiency between the pyHB AD-enhanced solver and HB-AD is not significant. HB-AD is even the sub-optimal candidate for the MZS-QZS isolator. A reasonable explanation is that the pyHB AD-enhanced solver performs the main iteration and sparse linear solution on the CPU, while the sampled Jacobian matrices of the nonlinear force are generated on the GPU. Consequently, CPU-GPU data transfer and conversion between dense and sparse matrix formats introduce additional overhead. In low-dimensional problems, the cost of solving the linear system is small, so these overheads are relatively pronounced. However, as the problem dimension increases, solving the system of linear equations gradually becomes the principal computational cost, and the efficiency difference between the two pyHB solvers becomes negligible. 

\begin{table*}[htbp]
      \centering
      \begin{threeparttable}
      \caption{Comparison of computational performance and memory usage of different approaches in the discussed examples.}
      \label{tab:comparison}
      \begin{tabular*}{1.0\textwidth}{@{}LLLLLLLLLLL@{}}
            \toprule
            Example & $n$ & Method & Device & \makecell[l]{Explicit\\$\partial \mathbf{f}_{\mathrm{nl}}$\\required?} & $m$ & \makecell[l]{HB\\unknowns} & \makecell[l]{Iteration\\time\\(ms/step)} & \makecell[l]{Continuation\\time\\(ms/point)\tnote{a}} & \makecell[l]{RAM\\(MiB)\tnote{b}} & \makecell[l]{GPU\\(MiB)\tnote{b}} \\
            \midrule
            \multirow{6}{*}{\makecell[l]{MZS-QZS\\isolator}} & \multirow{6}{*}{1} & \makecell[l]{pyHB\\analytical} & CPU & \textcolor{red}{Yes} & 50 & 101 & \textbf{2.582}\tnote{c} & \textbf{10.844} & 405.9 & -- \\
            ~ & ~ & \makecell[l]{pyHB\\AD} & CPU+GPU & \textcolor{green!50!black}{No} & 50 & 101 & 8.234 & 34.583 & 472.8 & 243.5 \\
            ~ & ~ & HB-AD & Mainly GPU & \textcolor{green!50!black}{No} & 50 & 101 & \underline{6.558}\tnote{c} & \underline{26.230} & 1173.3 & 967.5 \\
            ~ & ~ & Runge-Kutta & CPU & \textcolor{green!50!black}{No} & -- & -- & -- & 185.730 & 1.2 & -- \\
            \midrule
            \multirow{6}{*}{\makecell[l]{Piezoelectric\\magnetic\\harvester}} & \multirow{6}{*}{3} & \makecell[l]{pyHB\\analytical} & CPU & \textcolor{red}{Yes} & 20 & 123 & \textbf{1.193} & \textbf{6.444} & 50.4 & -- \\
            ~ & ~ & \makecell[l]{pyHB\\AD} & CPU+GPU & \textcolor{green!50!black}{No} & 20 & 123 & \underline{2.893} & \underline{15.624} & 316.5 & 243.5 \\
            ~ & ~ & HB-AD & Mainly GPU & \textcolor{green!50!black}{No} & 20 & 123 & 6.752 & 36.461 & 1121.5 & 981.5 \\
            ~ & ~ & Runge-Kutta & CPU & \textcolor{green!50!black}{No} & -- & -- & -- & 27.697 & 3.1 & -- \\
            \midrule
            \multirow{6}{*}{Aeroengine} & \multirow{6}{*}{284} & \makecell[l]{pyHB\\analytical} & CPU & \textcolor{red}{Yes} & 27 & 15620 & \underline{32.115} & \underline{96.346} & 88.1 & -- \\
            ~ & ~ & \makecell[l]{pyHB\\AD} & CPU+GPU & \textcolor{green!50!black}{No} & 27 & 15620 & \textbf{28.946} & \textbf{86.839} & 414.6 & 265.5 \\
            ~ & ~ & HB-AD & Mainly GPU & \textcolor{green!50!black}{No} & 27 & 15620 & \multicolumn{4}{c}{\textcolor{red}{Out of memory}} \\
            ~ & ~ & Runge-Kutta & CPU & \textcolor{green!50!black}{No} & -- & -- & -- & 23168.791 & 133.0 & -- \\
            \midrule
            \multirow{10}{*}{\makecell[l]{Bernoulli\\beam}} & \multirow{10}{*}{2000} & \multirow{4}{*}{\makecell[l]{pyHB\\analytical}} & \multirow{4}{*}{CPU} & \multirow{4}{*}{\textcolor{red}{Yes}} & 5 & 22000 & \textbf{19.812} & \textbf{39.624} & 161.7 & -- \\
            ~ & ~ & ~ & ~ & ~ & 10 & 42000 & \textbf{38.486} & \textbf{76.973} & 151.5 & -- \\
            ~ & ~ & ~ & ~ & ~ & 20 & 82000 & \textbf{79.621} & \textbf{159.242} & 200.9 & -- \\
            ~ & ~ & ~ & ~ & ~ & 50 & 202000 & \textbf{217.730} & \textbf{435.461} & 530.6 & -- \\
            ~ & ~ & \multirow{4}{*}{\makecell[l]{pyHB\\AD}} & \multirow{4}{*}{CPU+GPU} & \multirow{4}{*}{\textcolor{green!50!black}{No}} & 5 & 22000 & \underline{21.905} & \underline{43.811} & 350.8 & 243.5 \\
            ~ & ~ & ~ & ~ & ~ & 10 & 42000 & \underline{40.662} & \underline{81.324} & 375.1 & 243.5 \\
            ~ & ~ & ~ & ~ & ~ & 20 & 82000 & \underline{83.402} & \underline{166.803} & 430.1 & 243.5 \\
            ~ & ~ & ~ & ~ & ~ & 50 & 202000 & \underline{219.452} & \underline{438.904} & 637.8 & 243.5 \\
            ~ & ~ & HB-AD & Mainly GPU & \textcolor{green!50!black}{No} & 5 & 22000 & \multicolumn{4}{c}{\textcolor{red}{Out of memory}} \\
            ~ & ~ & Newmark-$\beta$ & CPU & \textcolor{red}{Yes} & -- & -- & -- & 81550.061 & 975.6 & -- \\
            \bottomrule
      \end{tabular*}
      \begin{tablenotes}
            \item[a] For direct numerical-integration rows, the ninth column reports the measured wall time for a 10-period trajectory initialized from the HB response.
            \item[b] Memory is the maximum solver- or integration-stage increment above the runtime baseline across three runs; video memory is the Windows process dedicated GPU peak.
            \item[c] The fastest iteration and continuation times are highlighted in bold, while the second-fastest are underlined.
      \end{tablenotes}
      \end{threeparttable}
\end{table*}

HB-AD is executed primarily on the GPU, thereby avoiding cross-device communication. However, the price of this implementation is a rapid, hard-to-control increase in video memory consumption. Its AD strategy is fundamentally different from that of pyHB: HB-AD applies AD directly to the complete HB residual equation; hence, the computational graph includes the response reconstruction, residual evaluation, and FFT. Its size grows with the number of DOFs and harmonics \cite{chenHarmonicBalanceautomaticDifferentiation2026}. Accordingly, HB-AD already requires approximately 1GB of GPU memory for the MZS-QZS isolator and the PEH examples, and it runs out of memory for both the aeroengine model and the Bernoulli beam. In contrast, pyHB exploits the localized representation in Eq.\eqref{eq:localized_nonlinearity} and applies AD only to the reduced nonlinear force $\mathbf{f}_{nl}^{(r)}$. The resulting computational graph depends mainly on the local dimensions $q$ and $r$, rather than directly on the global number of DOFs. Besides, the subsequent FFT, tensor contraction, Jacobian assembly, and sparse linear solution remain outside the computational graph. The dedicated GPU memory peak of the pyHB AD-enhanced solver remains between 243.5MB and 265.5MB for all examples in Tab.\ref{tab:comparison}, demonstrating substantially more manageable memory usage in high-dimensional systems.

The Bernoulli beam example further examines pyHB's scalability at higher harmonic orders. For the discrete model with 2000 DOFs, increasing $m$ from 5 to 10, 20, and 50 raises the number of HB unknowns from 22000 to 42000, 82000, and 202000, respectively. Nevertheless, RAM usage remains within a manageable range: even at $m=50$, the analytical and AD-enhanced solvers require only 530.6 MB and 637.8 MB of additional RAM. Remarkably, it is noteworthy that since the dimension of the local nonlinear force is identical, video memory usage is exactly the same across these harmonic settings. These results benefit from the sparse treatment of the necessary matrices and the blocked solving strategy for the augmented system in Eqs.\eqref{eq:arc_length_NR_iteration_augmented_blocked} and \eqref{eq:arc_length_NR_iteration_augmented_blocked_solution}. Moreover, iteration and continuation times increase approximately linearly with the number of HB unknowns. At $m=50$, the analytical and AD-enhanced solvers require 217.730ms and 219.452ms per iteration, respectively. This means that even for a large-scale problem with more than 200000 unknowns, pyHB can generate a complete amplitude-frequency response curve within 3 minutes. By contrast, the Newmark-$\beta$ method requires more than 1 minute to compute a 10-period trajectory at a single excitation frequency. Besides, Ponsioen et al. \cite{ponsioenModelReductionSpectral2020} reported the computation results performed on the same problem using the COCO \cite{dankowiczRecipesContinuation2013} and NLvib \cite{nlvib} (another HB-based solver developed by Malte and Johann) tools. It can be seen that when the number of DOFs is 500, the computational cost of both methods exceeds 24 hours, in stark contrast to pyHB. The above results clearly demonstrate the superiority of pyHB, particularly its advantages in solution efficiency and memory usage in high-dimensional systems.

\section{Discussions}
This work presents pyHB, an open-source computational framework that integrates the key stages of HB analysis, including model definition, harmonic precomputation, periodic-solution calculation, arc-length continuation, and Floquet stability assessment. Its central contribution is an efficient, general strategy for constructing the HB Jacobian without requiring users to explicitly derive the tangent matrices of the nonlinear force. In particular, the linear part of the Jacobian is precomputed wherever possible, while the nonlinear part is evaluated via FFT-based harmonic projection and tensor contraction. By applying AD only to the reduced nonlinear force in local physical coordinates, pyHB decouples the computational graph from the global number of DOFs and harmonic components. Together with sparse matrix assembly, blocked solution of the augmented continuation equations, weighted arc-length continuation, and dedicated stability-analysis interfaces, this treatment provides a complete and modular workflow for nonlinear systems ranging from low-dimensional strongly nonlinear oscillators to a finite element model with more than $2\times10^{5}$ HB unknowns. The numerical examples demonstrate that the framework can trace both stable and unstable branches, identify complex resonance responses, and maintain controllable computational time and memory usage as the problem dimension and harmonic order increase.

The present implementation employs Newton-Raphson iterations with an explicitly assembled sparse Jacobian, whereas existing approaches that reduce the dependence on an exact Jacobian can be broadly divided into two categories. The first category comprises quasi-Newton schemes, such as Broyden's method, which update an approximate Jacobian or its inverse using secant information from successive iterations \cite{wangModifiedIncrementalHarmonic2015a, sunComputationallyEfficientMethod2024, liEnhancedIncrementalHarmonic2025, wangDynamicAnalysisAutomotive2017}. This scheme can reduce the repeated construction and factorization of the exact Jacobian, but its robustness and efficiency depend strongly on the quality of the initial approximation. In addition, if the iteration fails to converge, an appropriate backtracking and restarting strategy is also required. The structured Jacobian generated by pyHB is a natural candidate for such initialization and restart. The second category employs the Krylov-subspace method to iteratively solve the Newton correction equation via the matrix-vector product, which can be evaluated without explicitly assembling the complete Jacobian matrix. Although this matrix-free treatment substantially reduces storage requirements, an effective preconditioner is generally essential for high-dimensional and ill-conditioned HB systems, and its construction commonly relies on the tangent-like operator \cite{1257664, lindbladConvergenceAccelerationHarmonic2020a, 11132845, kuetherLargescaleHarmonicBalance2024}. The localized derivatives based on AD technology proposed in this work could therefore be transferred to construct the preconditioner required by the above methods. Fundamentally, these two categories modify the linear system solving procedure, whereas this work addresses how accurate derivative information is generated and organized; they are therefore complementary rather than competing strategies. Incorporating quasi-Newton updates could reduce the need for repeated Jacobian assembly and refactorization, while a preconditioned matrix-free Newton-Krylov implementation would avoid storing the full Jacobian matrix and provide a practical route toward systems with millions of HB unknowns. One of our important future research directions is to further integrate these two solution paradigms into the framework of this work.

Moreover, several additional developments can further extend the efficiency and scope of pyHB. First, the sparse linear solver remains the dominant computational and memory cost for large-scale problems, and future work may investigate the iterative sparse solver and adaptive switching between the solving strategies. In weakly nonlinear and non-resonant regions, a Picard iteration can exploit a reusable linear operator and avoid repeatedly updating the nonlinear tangent matrices; once residual reduction stagnates, the solver could automatically switch to the more robust Newton-Raphson iteration \cite{jainFastComputationSteadystate2019a}. Second, the current single-time-scale formulation represents responses using harmonics of one fundamental frequency. Fractional frequency resolution enables commensurate components, such as the dual-frequency and subharmonic responses considered in the above examples, to be reduced to a common period. However, it cannot represent genuinely incommensurate frequencies within a finite period. Introducing a two-time-scale HB formulation would extend pyHB to quasi-periodic responses on a multidimensional frequency basis \cite{huangQuasiperiodicMotionsHighdimensional2019, liTwotimescaleConvergenceenhancedEfficient2026a, huangIncrementalHarmonicBalance2021}. Finally, although the current framework supports systems combining first- and second-order evolution equations, it is formulated primarily for ODEs and does not explicitly address general algebraic constraints. Extending the pyHB to differential-algebraic equations would broaden its applicability to constrained multibody systems \cite{juSteadyStateRotaryPeriodic2024, zhouConstraintEliminationbasedHarmonic2026a, juEfficientGalerkinAveragingincremental2021a}. 
 
\section{Conclusions}
This work presents pyHB, an open-source computational framework for analyzing the nonlinear system's periodic responses. It introduces automatic differentiation with the HB method, providing a completely new paradigm for computing the Jacobian matrix to solve the HB residual equation and addressing the shortcomings of symbolic and numerical differentiation. Integrating the sparse matrix technique, the weighted arc-length continuation method, and the Floquet theory, pyHB can efficiently compute the system's complete periodic solutions of (including unstable orbits).

Various examples, including the quasi-zero stiffness vibration isolator with multiple zero stiffness points, the piezoelectric-magnetic harvester, the aeroengine dual-rotor-bearing-casing system with 284 DOFs, and the forced 2000 DOFs Bernoulli beam with a nonlinear spring support, demonstrate that pyHB can efficiently and accurately capture the system's complex nonlinear behaviors (such as subharmonic resonance, combination resonance, and mixed-order electromechanical responses), and offer excellent generalization to high-dimensional systems with complex nonlinearities. The detailed comparison with existing methods further highlights the advantages of pyHB in terms of computational efficiency and memory usage, particularly for large-scale problems: for the Bernoulli beam with 2000 DOFs and 50 harmonics (over $2 \times 10^{5}$ HB unknowns), pyHB can generate the complete amplitude-frequency response curve in less than 3 minutes, while the Newmark-$\beta$ method requires more than 1 minute to compute a 10-period trajectory at a single excitation frequency. HB-AD runs out of memory for this problem, while pyHB's dedicated GPU memory peak remains below 250MB. 

This work provides a comprehensive methodology and code implementation to solve the periodic response of complex nonlinear systems with arbitrary user-defined support, and offers a versatile and easy-to-use benchmark platform for the HB method.

\section*{Acknowledgements}
It is very grateful for the financial supports from National Natural Science Foundation of China (Grant No. 12502004), the Postdoctoral Fellowship Program (Grade C) of China Postdoctoral Science Foundation (Grant No. GZC20252749), the Heilongjiang Postdoctoral Fund (Grant No. LBH-Z25129), the New Era Longjiang Outstanding Master's and Doctoral Dissertation Project (Grant No. LJYXLZR2025-031), the Natural Science Foundation of Henan Province (Grant No. 262300421989, 262300421374), and the Key Technologies Research and Development Program of Henan Province (Grant No. 252102240044, 262102321064).




\clearpage 





\printcredits

\bibliographystyle{model1-num-names}

\bibliography{cas-refs}

\begin{thebibliography}{137}
\expandafter\ifx\csname natexlab\endcsname\relax\def\natexlab#1{#1}\fi
\providecommand{\url}[1]{\texttt{#1}}
\providecommand{\href}[2]{#2}
\providecommand{\path}[1]{#1}
\providecommand{\DOIprefix}{doi:}
\providecommand{\ArXivprefix}{arXiv:}
\providecommand{\URLprefix}{URL: }
\providecommand{\Pubmedprefix}{pmid:}
\providecommand{\doi}[1]{\href{http://dx.doi.org/#1}{\path{#1}}}
\providecommand{\Pubmed}[1]{\href{pmid:#1}{\path{#1}}}
\providecommand{\bibinfo}[2]{#2}
\ifx\xfnm\relax \def\xfnm[#1]{\unskip,\space#1}\fi
\bibitem[{Mei et~al.(2026)Mei, Caprani, and Cantero}]{MEI2026111521}
\bibinfo{author}{S.~Mei}, \bibinfo{author}{C.~Caprani}, \bibinfo{author}{D.~Cantero},
\newblock \bibinfo{title}{Dynamic response of a combined beam and spring-mass system under moving load: A closed-form solution using higher-order ordinary differential equations},
\newblock \bibinfo{journal}{Structures} \bibinfo{volume}{87} (\bibinfo{year}{2026}) \bibinfo{pages}{111521}.
\bibitem[{Zhang et~al.(2026)Zhang, Teng, Feng, and Zhang}]{ZHANG2026115115}
\bibinfo{author}{Y.~Zhang}, \bibinfo{author}{Y.~Teng}, \bibinfo{author}{H.~Feng}, \bibinfo{author}{W.~Zhang},
\newblock \bibinfo{title}{Nonlinear vibration analysis and experimental research on rotating pre-twisted tc4 blades under combined transverse aerodynamic loads and tip clearance airflow},
\newblock \bibinfo{journal}{Thin-Walled Structures} \bibinfo{volume}{228} (\bibinfo{year}{2026}) \bibinfo{pages}{115115}.
\bibitem[{Theodosiou et~al.(2009)Theodosiou, Sikelis, and Natsiavas}]{THEODOSIOU20093565}
\bibinfo{author}{C.~Theodosiou}, \bibinfo{author}{K.~Sikelis}, \bibinfo{author}{S.~Natsiavas},
\newblock \bibinfo{title}{Periodic steady state response of large scale mechanical models with local nonlinearities},
\newblock \bibinfo{journal}{International Journal of Solids and Structures} \bibinfo{volume}{46} (\bibinfo{year}{2009}) \bibinfo{pages}{3565--3576}.
\bibitem[{Su et~al.(2026)Su, Guo, and Zhao}]{SU20261}
\bibinfo{author}{R.-X. Su}, \bibinfo{author}{Y.-F. Guo}, \bibinfo{author}{Y.-P. Zhao},
\newblock \bibinfo{title}{Steady-state dynamical behavior analysis of an improved fitzhugh-nagumo neuron system based on the generalized cell mapping method},
\newblock \bibinfo{journal}{Chinese Journal of Physics} \bibinfo{volume}{103} (\bibinfo{year}{2026}) \bibinfo{pages}{1--17}.
\bibitem[{Dong et~al.(2022)Dong, Hu, and Wang}]{dongComprehensiveStudyCoupled2022}
\bibinfo{author}{Y.~Dong}, \bibinfo{author}{H.~Hu}, \bibinfo{author}{L.~Wang},
\newblock \bibinfo{title}{A comprehensive study on the coupled multi-mode vibrations of cylindrical shells},
\newblock \bibinfo{journal}{Mechanical Systems and Signal Processing} \bibinfo{volume}{169} (\bibinfo{year}{2022}) \bibinfo{pages}{108730}.
\bibitem[{Karam et~al.(2021)Karam, Sutherland, and Saad}]{karamLowcostRungeKuttaIntegrators2021}
\bibinfo{author}{M.~Karam}, \bibinfo{author}{J.~C. Sutherland}, \bibinfo{author}{T.~Saad},
\newblock \bibinfo{title}{Low-cost {{Runge-Kutta}} integrators for incompressible flow simulations},
\newblock \bibinfo{journal}{Journal of Computational Physics} \bibinfo{volume}{443} (\bibinfo{year}{2021}) \bibinfo{pages}{110518}.
\bibitem[{Wang et~al.(2026)Wang, Tang, Zhao, Qian, and Jiang}]{Wang2026115162}
\bibinfo{author}{K.~Wang}, \bibinfo{author}{C.~Tang}, \bibinfo{author}{D.~Zhao}, \bibinfo{author}{X.~Qian}, \bibinfo{author}{S.~Jiang},
\newblock \bibinfo{title}{Simulation of transient heat conduction in granular materials using the numerical manifold method with a high-accuracy explicit time integration scheme},
\newblock \bibinfo{journal}{Journal of Computational Physics} \bibinfo{volume}{564} (\bibinfo{year}{2026}) \bibinfo{pages}{115162}.
\bibitem[{Sokolov et~al.(2026)Sokolov, Barkhayev, and Turek}]{SOKOLOV2026117017}
\bibinfo{author}{A.~Sokolov}, \bibinfo{author}{P.~Barkhayev}, \bibinfo{author}{S.~Turek},
\newblock \bibinfo{title}{Numerical study of the rbf-fd parallel-in-time contour integration method for convection–diffusion equations},
\newblock \bibinfo{journal}{Journal of Computational and Applied Mathematics} \bibinfo{volume}{475} (\bibinfo{year}{2026}) \bibinfo{pages}{117017}.
\bibitem[{Zhang et~al.(2022)Zhang, Zhang, and zhong Sun}]{ZHANG2022108048}
\bibinfo{author}{Q.~Zhang}, \bibinfo{author}{J.~Zhang}, \bibinfo{author}{Z.~zhong Sun},
\newblock \bibinfo{title}{Optimal convergence rate of the explicit euler method for convection-diffusion equations},
\newblock \bibinfo{journal}{Applied Mathematics Letters} \bibinfo{volume}{131} (\bibinfo{year}{2022}) \bibinfo{pages}{108048}.
\bibitem[{Hanna(1988)}]{HANNA19881083}
\bibinfo{author}{O.~Hanna},
\newblock \bibinfo{title}{New explicit and implicit "improved euler" methods for the integration of ordinary differential equations},
\newblock \bibinfo{journal}{Computers \& Chemical Engineering} \bibinfo{volume}{12} (\bibinfo{year}{1988}) \bibinfo{pages}{1083--1086}.
\bibitem[{Faleichik and Moisa(2026)}]{FALEICHIK2026117061}
\bibinfo{author}{B.~Faleichik}, \bibinfo{author}{A.~Moisa},
\newblock \bibinfo{title}{Explicit runge-kutta-chebyshev methods of second order with monotonic stability polynomial},
\newblock \bibinfo{journal}{Journal of Computational and Applied Mathematics} \bibinfo{volume}{476} (\bibinfo{year}{2026}) \bibinfo{pages}{117061}.
\bibitem[{Mei and Chen(2012)}]{MEI20129547}
\bibinfo{author}{L.~Mei}, \bibinfo{author}{Y.~Chen},
\newblock \bibinfo{title}{Explicit multistep method for the numerical solution of rlw equation},
\newblock \bibinfo{journal}{Applied Mathematics and Computation} \bibinfo{volume}{218} (\bibinfo{year}{2012}) \bibinfo{pages}{9547--9554}.
\bibitem[{Chang et~al.(2025)Chang, Hou, Lin, Chen, and Masarati}]{changMonolithicApproachesTransient2025}
\bibinfo{author}{Z.~Chang}, \bibinfo{author}{L.~Hou}, \bibinfo{author}{R.~Lin}, \bibinfo{author}{Y.~Chen}, \bibinfo{author}{P.~Masarati},
\newblock \bibinfo{title}{Monolithic approaches to transient thermo-mechanical interaction in nonlinear rotor systems},
\newblock \bibinfo{journal}{International Journal of Mechanical Sciences} \bibinfo{volume}{290} (\bibinfo{year}{2025}) \bibinfo{pages}{110066}.
\bibitem[{Qin et~al.(2024)Qin, Yu, Jiang, Huang, and Yan}]{QIN2024106089}
\bibinfo{author}{X.~Qin}, \bibinfo{author}{J.~Yu}, \bibinfo{author}{Z.~Jiang}, \bibinfo{author}{L.~Huang}, \bibinfo{author}{C.~Yan},
\newblock \bibinfo{title}{Explicit strong stability preserving second derivative multistep methods for the euler and navier-stokes equations},
\newblock \bibinfo{journal}{Computers \& Fluids} \bibinfo{volume}{268} (\bibinfo{year}{2024}) \bibinfo{pages}{106089}.
\bibitem[{El~Abbassi et~al.(2025)El~Abbassi, Bichri, Achenani, Souinida, and Tilioua}]{elabbassiControlNonlinearDynamic2025}
\bibinfo{author}{A.~El~Abbassi}, \bibinfo{author}{A.~Bichri}, \bibinfo{author}{Y.~Achenani}, \bibinfo{author}{L.~Souinida}, \bibinfo{author}{A.~Tilioua},
\newblock \bibinfo{title}{Control of a nonlinear dynamic system by backstepping: Numerical simulation and modelling using electronic circuits},
\newblock \bibinfo{journal}{Procedia Computer Science} \bibinfo{volume}{265} (\bibinfo{year}{2025}) \bibinfo{pages}{728--733}.
\bibitem[{Li et~al.(2021)Li, Yu, and Li}]{liIdenticalSecondorderSingle2021}
\bibinfo{author}{J.~Li}, \bibinfo{author}{K.~Yu}, \bibinfo{author}{X.~Li},
\newblock \bibinfo{title}{An identical second-order single step explicit integration algorithm with dissipation control for structural dynamics},
\newblock \bibinfo{journal}{International Journal for Numerical Methods in Engineering} \bibinfo{volume}{122} (\bibinfo{year}{2021}) \bibinfo{pages}{1089--1132}.
\bibitem[{Liu et~al.(2012)Liu, Zhao, Li, and Zhang}]{LIU201220}
\bibinfo{author}{T.~Liu}, \bibinfo{author}{C.~Zhao}, \bibinfo{author}{Q.~Li}, \bibinfo{author}{L.~Zhang},
\newblock \bibinfo{title}{An efficient backward euler time-integration method for nonlinear dynamic analysis of structures},
\newblock \bibinfo{journal}{Computers \& Structures} \bibinfo{volume}{106-107} (\bibinfo{year}{2012}) \bibinfo{pages}{20--28}.
\bibitem[{Gui and Du(2026)}]{GUI2026110088}
\bibinfo{author}{Y.~Gui}, \bibinfo{author}{R.~Du},
\newblock \bibinfo{title}{Second-order semi-implicit imex runge-kutta methods for micromagnetic simulations in antiferromagnetics},
\newblock \bibinfo{journal}{Communications in Nonlinear Science and Numerical Simulation} \bibinfo{volume}{161} (\bibinfo{year}{2026}) \bibinfo{pages}{110088}.
\bibitem[{Li et~al.(2025)Li, Du, Si, Wang, and Xiong}]{liRigidflexibleCouplingDynamic2025}
\bibinfo{author}{Z.~Li}, \bibinfo{author}{H.~Du}, \bibinfo{author}{J.~Si}, \bibinfo{author}{Z.~Wang}, \bibinfo{author}{W.~Xiong},
\newblock \bibinfo{title}{Rigid-flexible coupling dynamic modeling and validation of a helicopter rescue simulator based on an inverted {{Stewart}} platform},
\newblock \bibinfo{journal}{Ocean Engineering} \bibinfo{volume}{319} (\bibinfo{year}{2025}) \bibinfo{pages}{120164}.
\bibitem[{Liu and Ji(2024)}]{liuNonlinearDynamicsThreedimensional2024}
\bibinfo{author}{B.~Liu}, \bibinfo{author}{Y.~Ji},
\newblock \bibinfo{title}{Nonlinear dynamics of three-dimensional curved geometrically exact beams by a quadrature element formulation},
\newblock \bibinfo{journal}{Nonlinear Dynamics} \bibinfo{volume}{112} (\bibinfo{year}{2024}) \bibinfo{pages}{14925--14958}.
\bibitem[{Chowdhury and Adhikari(2025)}]{chowdhuryNonlinearStiffenedInertial2025}
\bibinfo{author}{S.~Chowdhury}, \bibinfo{author}{S.~Adhikari},
\newblock \bibinfo{title}{Nonlinear stiffened inertial amplifier tuned mass friction dampers},
\newblock \bibinfo{journal}{Soil Dynamics and Earthquake Engineering} \bibinfo{volume}{191} (\bibinfo{year}{2025}) \bibinfo{pages}{109264}.
\bibitem[{Chen et~al.(2026)Chen, Lü, Hu, Zhang, Liu, Xie, and Li}]{CHEN2026113858}
\bibinfo{author}{R.~Chen}, \bibinfo{author}{Y.~Lü}, \bibinfo{author}{R.~Hu}, \bibinfo{author}{Y.~Zhang}, \bibinfo{author}{G.~Liu}, \bibinfo{author}{Z.~Xie}, \bibinfo{author}{C.~Li},
\newblock \bibinfo{title}{Nonlinear dynamic characteristics of flexible bolted rotor system supported by journal bearings considering time varying bending stiffness of flange interface},
\newblock \bibinfo{journal}{Mechanical Systems and Signal Processing} \bibinfo{volume}{245} (\bibinfo{year}{2026}) \bibinfo{pages}{113858}.
\bibitem[{Wang et~al.(2026)Wang, Gao, liang Xu, fang Zheng, and ping Chen}]{WANG2026110403}
\bibinfo{author}{F.~Wang}, \bibinfo{author}{J.~Gao}, \bibinfo{author}{L.~liang Xu}, \bibinfo{author}{Y.~fang Zheng}, \bibinfo{author}{C.~ping Chen},
\newblock \bibinfo{title}{Analysis of nonlinear dynamic response of magneto-electro-elastic laminated beams with weak interface bonding},
\newblock \bibinfo{journal}{Communications in Nonlinear Science and Numerical Simulation} \bibinfo{volume}{162} (\bibinfo{year}{2026}) \bibinfo{pages}{110403}.
\bibitem[{Yin et~al.(2025)Yin, Quan, Wu, Kaiersaer, and Huang}]{YIN2025104022}
\bibinfo{author}{X.~Yin}, \bibinfo{author}{Y.~Quan}, \bibinfo{author}{L.~Wu}, \bibinfo{author}{T.~Kaiersaer}, \bibinfo{author}{Z.~Huang},
\newblock \bibinfo{title}{A 3d vehicle-bridge interaction framework integrating energy-conserving hamilton's principle and stabilized newmark-β method},
\newblock \bibinfo{journal}{Advances in Engineering Software} \bibinfo{volume}{211} (\bibinfo{year}{2025}) \bibinfo{pages}{104022}.
\bibitem[{Piterskaya and Mortensen(2026)}]{PITERSKAYA2026100683}
\bibinfo{author}{A.~Piterskaya}, \bibinfo{author}{M.~Mortensen},
\newblock \bibinfo{title}{A comparison of implicit-explicit runge-kutta time integration schemes in numerical solvers based on the galerkin and petrov-galerkin spectral methods for two-dimensional magneto-hydrodynamic problems},
\newblock \bibinfo{journal}{Results in Applied Mathematics} \bibinfo{volume}{29} (\bibinfo{year}{2026}) \bibinfo{pages}{100683}.
\bibitem[{Yin et~al.(2026)Yin, Quan, Zhou, Huang, Liu, Tang, and Wei}]{YIN2026126751}
\bibinfo{author}{X.~Yin}, \bibinfo{author}{Y.~Quan}, \bibinfo{author}{Y.~Zhou}, \bibinfo{author}{Z.~Huang}, \bibinfo{author}{J.~Liu}, \bibinfo{author}{H.~Tang}, \bibinfo{author}{P.~Wei},
\newblock \bibinfo{title}{A substructure method with coupled explicit-implicit strategy for long-span wave-bridge fsi analysis},
\newblock \bibinfo{journal}{Ocean Engineering} \bibinfo{volume}{363} (\bibinfo{year}{2026}) \bibinfo{pages}{126751}.
\bibitem[{Zhang et~al.(2026)Zhang, Zhang, Chen, and Hong}]{ZHANG2026106977}
\bibinfo{author}{J.~Zhang}, \bibinfo{author}{Y.~Zhang}, \bibinfo{author}{S.~Chen}, \bibinfo{author}{G.~Hong},
\newblock \bibinfo{title}{A hybrid implicit-explicit time integration for stiff chemically reacting flows based on adaptive component-splitting method},
\newblock \bibinfo{journal}{Computers \& Fluids} \bibinfo{volume}{307} (\bibinfo{year}{2026}) \bibinfo{pages}{106977}.
\bibitem[{Cardona et~al.(1994)Cardona, Coune, Lerusse, and Geradin}]{cardonaMultiharmonicMethodNonlinear1994}
\bibinfo{author}{A.~Cardona}, \bibinfo{author}{T.~Coune}, \bibinfo{author}{A.~Lerusse}, \bibinfo{author}{M.~Geradin},
\newblock \bibinfo{title}{A multiharmonic method for non-linear vibration analysis},
\newblock \bibinfo{journal}{International Journal for Numerical Methods in Engineering} \bibinfo{volume}{37} (\bibinfo{year}{1994}) \bibinfo{pages}{1593--1608}.
\bibitem[{Ke et~al.(2024)Ke, Wang, Chen, Li, and Lin}]{keQuantitativeAnalysisLimit2024}
\bibinfo{author}{Q.~Ke}, \bibinfo{author}{H.~Wang}, \bibinfo{author}{Z.~Chen}, \bibinfo{author}{J.~Li}, \bibinfo{author}{Y.~Lin},
\newblock \bibinfo{title}{Quantitative analysis of limit cycles in two-stroke oscillators with exponential functions based on {{Perturbation Incremental Method}}},
\newblock \bibinfo{journal}{Chaos, Solitons \& Fractals} \bibinfo{volume}{188} (\bibinfo{year}{2024}) \bibinfo{pages}{115549}.
\bibitem[{Su et~al.(2027)Su, Kang, Zhang, Hu, Cong, and Liu}]{SU2027117138}
\bibinfo{author}{X.~Su}, \bibinfo{author}{H.~Kang}, \bibinfo{author}{W.~Zhang}, \bibinfo{author}{T.~Hu}, \bibinfo{author}{Y.~Cong}, \bibinfo{author}{T.~Liu},
\newblock \bibinfo{title}{Nonlinear coupled vibrations of a resonant two-cable system with springs under the support motion},
\newblock \bibinfo{journal}{Applied Mathematical Modelling} \bibinfo{volume}{161} (\bibinfo{year}{2027}) \bibinfo{pages}{117138}.
\bibitem[{Kang et~al.(2026)Kang, Zhang, Zhang, Zhang, and Liu}]{kangInvestigationTunedClutch2026a}
\bibinfo{author}{Y.~Kang}, \bibinfo{author}{Z.~Zhang}, \bibinfo{author}{X.~Zhang}, \bibinfo{author}{Z.~Zhang}, \bibinfo{author}{Q.~Liu},
\newblock \bibinfo{title}{Investigation of a tuned clutch inerter mass damper for vibration reduction: {{An ANN-based}} harmonic balance method and parameter optimization},
\newblock \bibinfo{journal}{Applied Mathematical Modelling} \bibinfo{volume}{149} (\bibinfo{year}{2026}) \bibinfo{pages}{116321}.
\bibitem[{Chang et~al.(2022)Chang, Hou, and Chen}]{changInvestigation122022}
\bibinfo{author}{Z.~Chang}, \bibinfo{author}{L.~Hou}, \bibinfo{author}{Y.~Chen},
\newblock \bibinfo{title}{Investigation on the 1:2 internal resonance of an {{FGM}} blade},
\newblock \bibinfo{journal}{Nonlinear Dynamics} \bibinfo{volume}{107} (\bibinfo{year}{2022}) \bibinfo{pages}{1937--1964}.
\bibitem[{Li et~al.(2026)Li, Li, Lan, Wang, and Hu}]{liStudySuperharmonicparametricCombined2026}
\bibinfo{author}{Z.~Li}, \bibinfo{author}{G.~Li}, \bibinfo{author}{S.~Lan}, \bibinfo{author}{H.~Wang}, \bibinfo{author}{Y.~Hu},
\newblock \bibinfo{title}{Study on superharmonic-parametric combined resonance and multistability of translating piezoelectric sandwich plate},
\newblock \bibinfo{journal}{Thin-Walled Structures} \bibinfo{volume}{231} (\bibinfo{year}{2026}) \bibinfo{pages}{115370}.
\bibitem[{Wu et~al.(2020)Wu, Zhang, and Yao}]{WU2020112056}
\bibinfo{author}{Z.~Wu}, \bibinfo{author}{Y.~Zhang}, \bibinfo{author}{G.~Yao},
\newblock \bibinfo{title}{3/2 superharmonic resonance and 1/2 subharmonic resonance of functionally graded carbon nanotube reinforced composite beams},
\newblock \bibinfo{journal}{Composite Structures} \bibinfo{volume}{241} (\bibinfo{year}{2020}) \bibinfo{pages}{112056}.
\bibitem[{Lin et~al.(2024)Lin, Hou, Zhong, and Chen}]{linNonlinearVibrationStability2024a}
\bibinfo{author}{R.~Lin}, \bibinfo{author}{L.~Hou}, \bibinfo{author}{S.~Zhong}, \bibinfo{author}{Y.~Chen},
\newblock \bibinfo{title}{Nonlinear vibration and stability analysis of a dual-disk rotor-bearing system under multiple frequency excitations},
\newblock \bibinfo{journal}{Nonlinear Dynamics} \bibinfo{volume}{112} (\bibinfo{year}{2024}) \bibinfo{pages}{12815--12846}.
\bibitem[{Li et~al.(2025)Li, Huang, Liao, and Zhu}]{liCoupledmodeSubharmonicResonance2025}
\bibinfo{author}{Y.~Li}, \bibinfo{author}{J.~Huang}, \bibinfo{author}{F.~Liao}, \bibinfo{author}{W.~Zhu},
\newblock \bibinfo{title}{Coupled-mode subharmonic resonance of a piecewise-linear gear transmission system with 1:2 internal resonance},
\newblock \bibinfo{journal}{Mechanical Systems and Signal Processing} \bibinfo{volume}{240} (\bibinfo{year}{2025}) \bibinfo{pages}{113383}.
\bibitem[{Xu et~al.(2020)Xu, Cai, Li, Zhang, Liu, He, and Zhang}]{xuDynamicCharacteristicsReliability2020}
\bibinfo{author}{M.~Xu}, \bibinfo{author}{B.~Cai}, \bibinfo{author}{C.~Li}, \bibinfo{author}{H.~Zhang}, \bibinfo{author}{Z.~Liu}, \bibinfo{author}{D.~He}, \bibinfo{author}{Y.~Zhang},
\newblock \bibinfo{title}{Dynamic characteristics and reliability analysis of ball screw feed system on a lathe},
\newblock \bibinfo{journal}{Mechanism and Machine Theory} \bibinfo{volume}{150} (\bibinfo{year}{2020}) \bibinfo{pages}{103890}.
\bibitem[{Yang and Zhang(2014)}]{yangNonlinearDynamicsAxially2014}
\bibinfo{author}{X.-D. Yang}, \bibinfo{author}{W.~Zhang},
\newblock \bibinfo{title}{Nonlinear dynamics of axially moving beam with coupled longitudinal--transversal vibrations},
\newblock \bibinfo{journal}{Nonlinear Dynamics} \bibinfo{volume}{78} (\bibinfo{year}{2014}) \bibinfo{pages}{2547--2556}.
\bibitem[{Wang et~al.(2024)Wang, Hou, Meng, Cui, and Wang}]{wangThreemagnetringQuasizeroStiffness2024b}
\bibinfo{author}{S.~Wang}, \bibinfo{author}{L.~Hou}, \bibinfo{author}{Q.~Meng}, \bibinfo{author}{G.~Cui}, \bibinfo{author}{X.~Wang},
\newblock \bibinfo{title}{Three-magnet-ring quasi-zero stiffness isolator for low-frequency vibration isolation},
\newblock \bibinfo{journal}{International Journal of Mechanical System Dynamics} \bibinfo{volume}{4} (\bibinfo{year}{2024}) \bibinfo{pages}{153--170}.
\bibitem[{{AL-Shudeifat} et~al.(2010){AL-Shudeifat}, Butcher, and Stern}]{al-shudeifatGeneralHarmonicBalance2010}
\bibinfo{author}{M.~A. {AL-Shudeifat}}, \bibinfo{author}{E.~A. Butcher}, \bibinfo{author}{C.~R. Stern},
\newblock \bibinfo{title}{General harmonic balance solution of a cracked rotor-bearing-disk system for harmonic and sub-harmonic analysis: {{Analytical}} and experimental approach},
\newblock \bibinfo{journal}{International Journal of Engineering Science} \bibinfo{volume}{48} (\bibinfo{year}{2010}) \bibinfo{pages}{921--935}.
\bibitem[{Mohamed et~al.(2025)Mohamed, Br{\"o}dling, and Duddeck}]{mohamedAutoencoderbasedReductionHarmonic2025}
\bibinfo{author}{H.~Mohamed}, \bibinfo{author}{N.~Br{\"o}dling}, \bibinfo{author}{F.~Duddeck},
\newblock \bibinfo{title}{Autoencoder-based reduction of harmonic balance {{Fourier}} coefficients for uncertainty analysis of multi-{{DOF}} gear systems},
\newblock \bibinfo{journal}{Results in Engineering} \bibinfo{volume}{28} (\bibinfo{year}{2025}) \bibinfo{pages}{107902}.
\bibitem[{Lau and Cheung(1981)}]{lauAmplitudeIncrementalVariational1981}
\bibinfo{author}{S.~L. Lau}, \bibinfo{author}{Y.~K. Cheung},
\newblock \bibinfo{title}{Amplitude {{Incremental Variational Principle}} for {{Nonlinear Vibration}} of {{Elastic Systems}}},
\newblock \bibinfo{journal}{Journal of Applied Mechanics} \bibinfo{volume}{48} (\bibinfo{year}{1981}) \bibinfo{pages}{959--964}.
\bibitem[{Kim and Noah(1991)}]{kimStabilityBifurcationAnalysis1991}
\bibinfo{author}{Y.~B. Kim}, \bibinfo{author}{S.~T. Noah},
\newblock \bibinfo{title}{Stability and {{Bifurcation Analysis}} of {{Oscillators With Piecewise-Linear Characteristics}}: {{A General Approach}}},
\newblock \bibinfo{journal}{Journal of Applied Mechanics} \bibinfo{volume}{58} (\bibinfo{year}{1991}) \bibinfo{pages}{545--553}.
\bibitem[{Chopra and Dugundji(1979)}]{chopraNonlinearDynamicResponse1979}
\bibinfo{author}{I.~Chopra}, \bibinfo{author}{J.~Dugundji},
\newblock \bibinfo{title}{Non-linear dynamic response of a wind turbine blade},
\newblock \bibinfo{journal}{Journal of Sound and Vibration} \bibinfo{volume}{63} (\bibinfo{year}{1979}) \bibinfo{pages}{265--286}.
\bibitem[{Yan et~al.(2023)Yan, Dai, Wang, and N.~Atluri}]{yanHarmonicBalanceMethods2023}
\bibinfo{author}{Z.~Yan}, \bibinfo{author}{H.~Dai}, \bibinfo{author}{Q.~Wang}, \bibinfo{author}{S.~N.~Atluri},
\newblock \bibinfo{title}{Harmonic {{Balance Methods}}: {{A Review}} and {{Recent Developments}}},
\newblock \bibinfo{journal}{Computer Modeling in Engineering \& Sciences} \bibinfo{volume}{137} (\bibinfo{year}{2023}) \bibinfo{pages}{1419--1459}.
\bibitem[{Zhang et~al.(2024)Zhang, Lu, Chai, Cheng, Fu, and Guo}]{zhangDynamicModelingParameter2024}
\bibinfo{author}{K.~Zhang}, \bibinfo{author}{K.~Lu}, \bibinfo{author}{S.~Chai}, \bibinfo{author}{H.~Cheng}, \bibinfo{author}{C.~Fu}, \bibinfo{author}{D.~Guo},
\newblock \bibinfo{title}{Dynamic modeling and parameter sensitivity analysis of {{AUV}} by using the {{POD}} method and the {{HB-AFT}} method},
\newblock \bibinfo{journal}{Ocean Engineering} \bibinfo{volume}{293} (\bibinfo{year}{2024}) \bibinfo{pages}{116693}.
\bibitem[{Litewka and Lewandowski(2026)}]{litewkaNonlinearHarmonicVibrations2026}
\bibinfo{author}{P.~Litewka}, \bibinfo{author}{R.~Lewandowski},
\newblock \bibinfo{title}{Non-linear harmonic vibrations of laminate plates with viscoelastic layers described by extended fractional models},
\newblock \bibinfo{journal}{Thin-Walled Structures} \bibinfo{volume}{228} (\bibinfo{year}{2026}) \bibinfo{pages}{115091}.
\bibitem[{Chang et~al.(2024)Chang, Hou, Masarati, Lin, Li, and Chen}]{changModelingNonlinearAnalysis2024}
\bibinfo{author}{Z.~Chang}, \bibinfo{author}{L.~Hou}, \bibinfo{author}{P.~Masarati}, \bibinfo{author}{R.~Lin}, \bibinfo{author}{Z.~Li}, \bibinfo{author}{Y.~Chen},
\newblock \bibinfo{title}{Modeling and nonlinear analysis of a coupled thermo-mechanical dual-rotor system},
\newblock \bibinfo{journal}{Nonlinear Dynamics} \bibinfo{volume}{112} (\bibinfo{year}{2024}) \bibinfo{pages}{17811--17842}.
\bibitem[{Lin et~al.(2023)Lin, Hou, Dun, Cai, Sun, and Chen}]{linSynchronousImpactPhenomenon2023}
\bibinfo{author}{R.~Lin}, \bibinfo{author}{L.~Hou}, \bibinfo{author}{S.~Dun}, \bibinfo{author}{Y.~Cai}, \bibinfo{author}{C.~Sun}, \bibinfo{author}{Y.~Chen},
\newblock \bibinfo{title}{Synchronous impact phenomenon of a high-dimension complex nonlinear dual-rotor system subjected to multi-frequency excitations},
\newblock \bibinfo{journal}{Science China Technological Sciences} \bibinfo{volume}{66} (\bibinfo{year}{2023}) \bibinfo{pages}{1757--1768}.
\bibitem[{Meng et~al.(2025)Meng, Hou, Lin, Chen, Saeed, Fouly, and Awwad}]{mengQuasizeroStiffnessVibration2025}
\bibinfo{author}{Q.~Meng}, \bibinfo{author}{L.~Hou}, \bibinfo{author}{R.~Lin}, \bibinfo{author}{Y.~Chen}, \bibinfo{author}{N.~A. Saeed}, \bibinfo{author}{A.~Fouly}, \bibinfo{author}{E.~Awwad},
\newblock \bibinfo{title}{On a quasi-zero stiffness vibration isolator with multiple zero stiffness points for mass load deviation},
\newblock \bibinfo{journal}{Applied Mathematical Modelling} \bibinfo{volume}{145} (\bibinfo{year}{2025}) \bibinfo{pages}{116112}.
\bibitem[{Lei et~al.(2025)Lei, Ren, Zhou, Guo, and Wang}]{leiEmbeddingIntelligentExcitation2025}
\bibinfo{author}{H.~Lei}, \bibinfo{author}{H.~Ren}, \bibinfo{author}{P.~Zhou}, \bibinfo{author}{F.~Guo}, \bibinfo{author}{Y.~Wang},
\newblock \bibinfo{title}{Embedding intelligent excitation adaptability for resonance-suppressed {{QZS}} vibration isolation},
\newblock \bibinfo{journal}{International Journal of Mechanical Sciences} \bibinfo{volume}{307} (\bibinfo{year}{2025}) \bibinfo{pages}{110931}.
\bibitem[{Wu et~al.(2026)Wu, Wang, Xiao, Wu, Feng, Cao, and Han}]{wuRotationalinertiaenhancedQuasizerostiffnessChiral2026}
\bibinfo{author}{W.~Wu}, \bibinfo{author}{Y.~Wang}, \bibinfo{author}{S.~Xiao}, \bibinfo{author}{Q.~Wu}, \bibinfo{author}{Y.~Feng}, \bibinfo{author}{D.~Cao}, \bibinfo{author}{H.~Han},
\newblock \bibinfo{title}{Rotational-inertia-enhanced quasi-zero-stiffness chiral metamaterial for high-performance low-frequency vibration isolation},
\newblock \bibinfo{journal}{International Journal of Mechanical Sciences} \bibinfo{volume}{324} (\bibinfo{year}{2026}) \bibinfo{pages}{111787}.
\bibitem[{Wei et~al.(2020)Wei, Wang, and Wang}]{WEI2020105433}
\bibinfo{author}{L.-S. Wei}, \bibinfo{author}{Y.-Z. Wang}, \bibinfo{author}{Y.-S. Wang},
\newblock \bibinfo{title}{Nonreciprocal transmission of nonlinear elastic wave metamaterials by incremental harmonic balance method},
\newblock \bibinfo{journal}{International Journal of Mechanical Sciences} \bibinfo{volume}{173} (\bibinfo{year}{2020}) \bibinfo{pages}{105433}.
\bibitem[{Chen et~al.(2023)Chen, Gong, and Zheng}]{CHEN2023104256}
\bibinfo{author}{Y.~Chen}, \bibinfo{author}{B.~Gong}, \bibinfo{author}{Z.~Zheng},
\newblock \bibinfo{title}{On the subcritical period doubling of a non-smooth network system by incremental harmonic balance method},
\newblock \bibinfo{journal}{International Journal of Non-Linear Mechanics} \bibinfo{volume}{148} (\bibinfo{year}{2023}) \bibinfo{pages}{104256}.
\bibitem[{Jin et~al.(2026)Jin, Liu, Ma, Shen, Wang, Qian, and Jiang}]{jinPhysicsinformedHarmonicBalance2026}
\bibinfo{author}{R.~Jin}, \bibinfo{author}{Q.~Liu}, \bibinfo{author}{Y.~Ma}, \bibinfo{author}{X.~Shen}, \bibinfo{author}{Y.~Wang}, \bibinfo{author}{H.~Qian}, \bibinfo{author}{D.~Jiang},
\newblock \bibinfo{title}{Physics-informed harmonic balance identification of high-dimensional structures with nonlinear stiffness and damping},
\newblock \bibinfo{journal}{International Journal of Non-Linear Mechanics} \bibinfo{volume}{181} (\bibinfo{year}{2026}) \bibinfo{pages}{105282}.
\bibitem[{Dai et~al.(2014)Dai, Yue, Yuan, and Atluri}]{daiTimeDomainCollocation2014}
\bibinfo{author}{H.~Dai}, \bibinfo{author}{X.~Yue}, \bibinfo{author}{J.~Yuan}, \bibinfo{author}{S.~N. Atluri},
\newblock \bibinfo{title}{A time domain collocation method for studying the aeroelasticity of a two dimensional airfoil with a structural nonlinearity},
\newblock \bibinfo{journal}{Journal of Computational Physics} \bibinfo{volume}{270} (\bibinfo{year}{2014}) \bibinfo{pages}{214--237}.
\bibitem[{Dai et~al.(2023)Dai, Yan, Wang, Yue, and Atluri}]{daiCollocationbasedHarmonicBalance2023a}
\bibinfo{author}{H.~Dai}, \bibinfo{author}{Z.~Yan}, \bibinfo{author}{X.~Wang}, \bibinfo{author}{X.~Yue}, \bibinfo{author}{S.~N. Atluri},
\newblock \bibinfo{title}{Collocation-based harmonic balance framework for highly accurate periodic solution of nonlinear dynamical system},
\newblock \bibinfo{journal}{International Journal for Numerical Methods in Engineering} \bibinfo{volume}{124} (\bibinfo{year}{2023}) \bibinfo{pages}{458--481}.
\bibitem[{Liang et~al.(2024)Liang, Zang, Wei, and Wang}]{liangIntegratedApproachResponse2024}
\bibinfo{author}{H.~Liang}, \bibinfo{author}{C.~Zang}, \bibinfo{author}{X.~Wei}, \bibinfo{author}{H.~Wang},
\newblock \bibinfo{title}{An integrated approach for response prediction in large-scale rotor-bearing system with local nonlinear joints based on {{FRF-based}} harmonic balance method},
\newblock \bibinfo{journal}{Journal of Sound and Vibration} \bibinfo{volume}{583} (\bibinfo{year}{2024}) \bibinfo{pages}{118450}.
\bibitem[{Chen et~al.(2025)Chen, Cao, and Hou}]{chenIndirectHarmonicBalance2025a}
\bibinfo{author}{N.~Chen}, \bibinfo{author}{S.~Cao}, \bibinfo{author}{Y.~Hou},
\newblock \bibinfo{title}{An indirect harmonic balance method based on frequency response functions simplification for periodical response analysis of local nonlinearity systems},
\newblock \bibinfo{journal}{Computers \& Structures} \bibinfo{volume}{310} (\bibinfo{year}{2025}) \bibinfo{pages}{107663}.
\bibitem[{Cola{\"i}tis and Batailly(2021)}]{colaitisHarmonicBalanceMethod2021}
\bibinfo{author}{Y.~Cola{\"i}tis}, \bibinfo{author}{A.~Batailly},
\newblock \bibinfo{title}{The harmonic balance method with arc-length continuation in blade-tip/casing contact problems},
\newblock \bibinfo{journal}{Journal of Sound and Vibration} \bibinfo{volume}{502} (\bibinfo{year}{2021}) \bibinfo{pages}{116070}.
\bibitem[{Li et~al.(2025{\natexlab{a}})Li, Huang, and Zhu}]{liGeneralizedIncrementalHarmonic2025c}
\bibinfo{author}{Y.~Li}, \bibinfo{author}{J.~Huang}, \bibinfo{author}{W.~Zhu},
\newblock \bibinfo{title}{A generalized incremental harmonic balance method by combining a data-driven framework for initial value selection of strongly nonlinear dynamic systems},
\newblock \bibinfo{journal}{International Journal of Non-Linear Mechanics} \bibinfo{volume}{169} (\bibinfo{year}{2025}{\natexlab{a}}) \bibinfo{pages}{104951}.
\bibitem[{Li et~al.(2025{\natexlab{b}})Li, Huang, and Zhu}]{liEnhancedIncrementalHarmonic2025}
\bibinfo{author}{Y.~L. Li}, \bibinfo{author}{J.~L. Huang}, \bibinfo{author}{W.~D. Zhu},
\newblock \bibinfo{title}{An enhanced incremental harmonic balance method to improve the computational efficiency and convergence for systems with non-polynomial nonlinearities},
\newblock \bibinfo{journal}{Nonlinear Dynamics} \bibinfo{volume}{113} (\bibinfo{year}{2025}{\natexlab{b}}) \bibinfo{pages}{8265--8294}.
\bibitem[{Wu et~al.(2025)Wu, Wang, and Cao}]{wuImprovingSolutionProcedure2025}
\bibinfo{author}{Q.~Wu}, \bibinfo{author}{Y.~Wang}, \bibinfo{author}{D.~Cao},
\newblock \bibinfo{title}{Improving the {{Solution Procedure}} of {{Incremental Harmonic Balance Method}} for {{Multi-Degree-of-Freedom Self-Excited Vibration Systems}}},
\newblock \bibinfo{journal}{International Journal of Applied Mechanics} \bibinfo{volume}{17} (\bibinfo{year}{2025}) \bibinfo{pages}{2450125}.
\bibitem[{Ju et~al.(2021)Ju, Fan, and Zhu}]{juEfficientGalerkinAveragingincremental2021a}
\bibinfo{author}{R.~Ju}, \bibinfo{author}{W.~Fan}, \bibinfo{author}{W.~D. Zhu},
\newblock \bibinfo{title}{An efficient {{Galerkin}} averaging-incremental harmonic balance method for nonlinear dynamic analysis of rigid multibody systems governed by differential--algebraic equations},
\newblock \bibinfo{journal}{Nonlinear Dynamics} \bibinfo{volume}{105} (\bibinfo{year}{2021}) \bibinfo{pages}{475--498}.
\bibitem[{Zhou et~al.(2026)Zhou, Heya, and Inoue}]{zhouConstraintEliminationbasedHarmonic2026b}
\bibinfo{author}{X.~Zhou}, \bibinfo{author}{A.~Heya}, \bibinfo{author}{T.~Inoue},
\newblock \bibinfo{title}{A constraint elimination-based harmonic balance method for multibody systems},
\newblock \bibinfo{journal}{International Journal of Mechanical Sciences} \bibinfo{volume}{317} (\bibinfo{year}{2026}) \bibinfo{pages}{111490}.
\bibitem[{Huang et~al.(2021)Huang, Wang, and Zhu}]{HUANG2021103767}
\bibinfo{author}{J.~Huang}, \bibinfo{author}{T.~Wang}, \bibinfo{author}{W.~Zhu},
\newblock \bibinfo{title}{An incremental harmonic balance method with two time-scales for quasi-periodic responses of a van der pol-mathieu equation},
\newblock \bibinfo{journal}{International Journal of Non-Linear Mechanics} \bibinfo{volume}{135} (\bibinfo{year}{2021}) \bibinfo{pages}{103767}.
\bibitem[{Li et~al.(2026)Li, Huang, and Zhu}]{liTwotimescaleConvergenceenhancedEfficient2026a}
\bibinfo{author}{Y.~Li}, \bibinfo{author}{J.~Huang}, \bibinfo{author}{W.~Zhu},
\newblock \bibinfo{title}{A two-time-scale convergence-enhanced and efficient incremental harmonic balance method for solving quasi-periodic responses of nonlinear systems},
\newblock \bibinfo{journal}{Communications in Nonlinear Science and Numerical Simulation} \bibinfo{volume}{152} (\bibinfo{year}{2026}) \bibinfo{pages}{109042}.
\bibitem[{{Ansys Inc.}(2025)}]{ansys}
\bibinfo{author}{{Ansys Inc.}}, \bibinfo{title}{Ansys help}, \bibinfo{howpublished}{\url{https://ansyshelp.ansys.com/public/account/secured?returnurl=/Views/Secured/main_page.html}}, \bibinfo{year}{2025}. \bibinfo{note}{Ansys documentation.}
\bibitem[{Malte and Johann(2019)}]{nlvib}
\bibinfo{author}{K.~Malte}, \bibinfo{author}{G.~Johann}, \bibinfo{title}{Nlvib-a matlab tool for nonlinear vibration problems.}, \bibinfo{howpublished}{\url{https://www.ila.uni-stuttgart.de/nlvib}}, \bibinfo{year}{2019}.
\bibitem[{Chen et~al.(2026)Chen, Jin, Lin, Jiang, Mei, Hou, Wang, Yong, and Guo}]{chenHarmonicBalanceautomaticDifferentiation2026}
\bibinfo{author}{Y.~Chen}, \bibinfo{author}{Y.~Jin}, \bibinfo{author}{R.~Lin}, \bibinfo{author}{Y.~Jiang}, \bibinfo{author}{X.~Mei}, \bibinfo{author}{L.~Hou}, \bibinfo{author}{Y.~Wang}, \bibinfo{author}{N.~T. Yong}, \bibinfo{author}{A.~Guo},
\newblock \bibinfo{title}{Harmonic balance-automatic differentiation method: {{A}} practical nonlinear dynamics solver},
\newblock \bibinfo{journal}{International Journal of Mechanical Sciences} \bibinfo{volume}{312} (\bibinfo{year}{2026}) \bibinfo{pages}{111192}.
\bibitem[{Chen et~al.(2024)Chen, Hou, Lin, Wang, Saeed, and Chen}]{chenCombinationResonancesDualrotorbearingcasing2024}
\bibinfo{author}{Y.~Chen}, \bibinfo{author}{L.~Hou}, \bibinfo{author}{R.~Lin}, \bibinfo{author}{Y.~Wang}, \bibinfo{author}{N.~A. Saeed}, \bibinfo{author}{Y.~Chen},
\newblock \bibinfo{title}{Combination resonances of a dual-rotor-bearing-casing system},
\newblock \bibinfo{journal}{Nonlinear Dynamics} \bibinfo{volume}{112} (\bibinfo{year}{2024}) \bibinfo{pages}{4063--4083}.
\bibitem[{Wang et~al.(2026)Wang, Zhou, Liang, Ding, Zhou, Dong, and Lim}]{wangDynamicCharacteristicsHorizontal2026}
\bibinfo{author}{H.~Wang}, \bibinfo{author}{S.~Zhou}, \bibinfo{author}{X.~Liang}, \bibinfo{author}{T.~Ding}, \bibinfo{author}{T.~Zhou}, \bibinfo{author}{C.~Dong}, \bibinfo{author}{K.~M. Lim},
\newblock \bibinfo{title}{Dynamic {{Characteristics}} of {{Horizontal Rotor}}--{{Stator Nonlinear Interaction Under Additional Acceleration}}},
\newblock \bibinfo{journal}{AIAA Journal}  (\bibinfo{year}{2026}) \bibinfo{pages}{1--20}.
\bibitem[{Arango~Montoya et~al.(2026)Arango~Montoya, Chiello, Sinou, and Tufano}]{arangomontoyaHarmonicBalanceMethod2026a}
\bibinfo{author}{J.~Arango~Montoya}, \bibinfo{author}{O.~Chiello}, \bibinfo{author}{J.-J. Sinou}, \bibinfo{author}{R.~Tufano},
\newblock \bibinfo{title}{A {{Harmonic Balance Method}} with contact condensation for the frequency-domain computation of self-sustained nonlinear vibration related to railway curve squeal},
\newblock \bibinfo{journal}{Journal of Sound and Vibration} \bibinfo{volume}{626} (\bibinfo{year}{2026}) \bibinfo{pages}{119628}.
\bibitem[{Lee and Ahn(2026)}]{leeComputationalFrameworkBased2026a}
\bibinfo{author}{G.-Y. Lee}, \bibinfo{author}{K.~Ahn},
\newblock \bibinfo{title}{A computational framework based on the harmonic balance method for nonlinear bulk-flow analysis},
\newblock \bibinfo{journal}{Tribology International} \bibinfo{volume}{214} (\bibinfo{year}{2026}) \bibinfo{pages}{111336}.
\bibitem[{Zhang et~al.(2025)Zhang, Xu, Wang, and Zhou}]{zhangTheoreticalAnalysisCoupled2025a}
\bibinfo{author}{H.~Zhang}, \bibinfo{author}{H.~Xu}, \bibinfo{author}{S.~Wang}, \bibinfo{author}{S.~Zhou},
\newblock \bibinfo{title}{Theoretical analysis of coupled thermo-electric-elastic rotational piezoelectric energy harvesters based on {{Green}}'s function and harmonic balance method},
\newblock \bibinfo{journal}{Applied Mathematical Modelling} \bibinfo{volume}{139} (\bibinfo{year}{2025}) \bibinfo{pages}{115815}.
\bibitem[{Saadatmand and Kook(2025)}]{saadatmandNonlinearVibrationAnalysis2025a}
\bibinfo{author}{M.~Saadatmand}, \bibinfo{author}{J.~Kook},
\newblock \bibinfo{title}{Nonlinear vibration analysis of circular multilayer graphene-based {{NEMS}} sensors using harmonic balance and pseudo-arclength continuation methods},
\newblock \bibinfo{journal}{International Journal of Non-Linear Mechanics} \bibinfo{volume}{178} (\bibinfo{year}{2025}) \bibinfo{pages}{105187}.
\bibitem[{Zhou et~al.(2024)Zhou, Ji, Wu, Shen, Guo, and Lu}]{zhouMultiPassageHarmonicBalance2024a}
\bibinfo{author}{D.~Zhou}, \bibinfo{author}{Z.~Ji}, \bibinfo{author}{J.~Wu}, \bibinfo{author}{E.~Shen}, \bibinfo{author}{T.~Guo}, \bibinfo{author}{Z.~Lu},
\newblock \bibinfo{title}{Multi-{{Passage}} harmonic balance method for flutter prediction in multi-row configurations},
\newblock \bibinfo{journal}{Aerospace Science and Technology} \bibinfo{volume}{150} (\bibinfo{year}{2024}) \bibinfo{pages}{109212}.
\bibitem[{Mohamed et~al.(2025)Mohamed, Br{\"o}dling, and Duddeck}]{mohamedMultifidelityHarmonicBalance2025a}
\bibinfo{author}{H.~Mohamed}, \bibinfo{author}{N.~Br{\"o}dling}, \bibinfo{author}{F.~Duddeck},
\newblock \bibinfo{title}{Multi-fidelity harmonic balance method: {{A}} non-intrusive approach},
\newblock \bibinfo{journal}{Mechanism and Machine Theory} \bibinfo{volume}{217} (\bibinfo{year}{2025}) \bibinfo{pages}{106251}.
\bibitem[{Wang and Zhu(2015)}]{wangModifiedIncrementalHarmonic2015a}
\bibinfo{author}{X.~F. Wang}, \bibinfo{author}{W.~D. Zhu},
\newblock \bibinfo{title}{A modified incremental harmonic balance method based on the fast {{Fourier}} transform and {{Broyden}}'s method},
\newblock \bibinfo{journal}{Nonlinear Dynamics} \bibinfo{volume}{81} (\bibinfo{year}{2015}) \bibinfo{pages}{981--989}.
\bibitem[{Huang et~al.(2019)Huang, Zhou, and Zhu}]{huangQuasiperiodicMotionsHighdimensional2019}
\bibinfo{author}{J.~Huang}, \bibinfo{author}{W.~Zhou}, \bibinfo{author}{W.~Zhu},
\newblock \bibinfo{title}{Quasi-periodic motions of high-dimensional nonlinear models of a translating beam with a stationary load subsystem under harmonic boundary excitation},
\newblock \bibinfo{journal}{Journal of Sound and Vibration} \bibinfo{volume}{462} (\bibinfo{year}{2019}) \bibinfo{pages}{114870}.
\bibitem[{Yuan et~al.(2019)Yuan, Yang, and Chen}]{yuanHarmonicBalanceApproach2019}
\bibinfo{author}{T.-C. Yuan}, \bibinfo{author}{J.~Yang}, \bibinfo{author}{L.-Q. Chen},
\newblock \bibinfo{title}{A harmonic balance approach with alternating frequency/time domain progress for piezoelectric mechanical systems},
\newblock \bibinfo{journal}{Mechanical Systems and Signal Processing} \bibinfo{volume}{120} (\bibinfo{year}{2019}) \bibinfo{pages}{274--289}.
\bibitem[{Sun et~al.(2024)Sun, Yin, Shangguan, and Rakheja}]{sunComputationallyEfficientMethod2024}
\bibinfo{author}{Y.~Sun}, \bibinfo{author}{Z.~Yin}, \bibinfo{author}{W.-B. Shangguan}, \bibinfo{author}{S.~Rakheja},
\newblock \bibinfo{title}{A computationally efficient method for dynamic response analyses of accessory drive systems using harmonic balance method together with alternating frequency/time domain technique},
\newblock \bibinfo{journal}{Mechanical Systems and Signal Processing} \bibinfo{volume}{213} (\bibinfo{year}{2024}) \bibinfo{pages}{111307}.
\bibitem[{Chen et~al.(2023)Chen, Hou, Chen, Song, Lin, Jin, and Chen}]{chenNonlinearDynamicsAnalysis2023a}
\bibinfo{author}{Y.~Chen}, \bibinfo{author}{L.~Hou}, \bibinfo{author}{G.~Chen}, \bibinfo{author}{H.~Song}, \bibinfo{author}{R.~Lin}, \bibinfo{author}{Y.~Jin}, \bibinfo{author}{Y.~Chen},
\newblock \bibinfo{title}{Nonlinear dynamics analysis of a dual-rotor-bearing-casing system based on a modified {{HB-AFT}} method},
\newblock \bibinfo{journal}{Mechanical Systems and Signal Processing} \bibinfo{volume}{185} (\bibinfo{year}{2023}) \bibinfo{pages}{109805}.
\bibitem[{Ju et~al.(2021)Ju, Fan, and Zhu}]{juComparisonIncrementalHarmonic2021}
\bibinfo{author}{R.~Ju}, \bibinfo{author}{W.~Fan}, \bibinfo{author}{W.~D. Zhu},
\newblock \bibinfo{title}{Comparison {{Between}} the {{Incremental Harmonic Balance Method}} and {{Alternating Frequency}}/{{Time-Domain Method}}},
\newblock \bibinfo{journal}{Journal of Vibration and Acoustics} \bibinfo{volume}{143} (\bibinfo{year}{2021}) \bibinfo{pages}{024501}.
\bibitem[{Lin et~al.(2023)Lin, Hou, Chen, Jin, Saeed, and Chen}]{linNovelAdaptiveHarmonic2023}
\bibinfo{author}{R.~Lin}, \bibinfo{author}{L.~Hou}, \bibinfo{author}{Y.~Chen}, \bibinfo{author}{Y.~Jin}, \bibinfo{author}{N.~A. Saeed}, \bibinfo{author}{Y.~Chen},
\newblock \bibinfo{title}{A novel adaptive harmonic balance method with an asymptotic harmonic selection},
\newblock \bibinfo{journal}{Applied Mathematics and Mechanics} \bibinfo{volume}{44} (\bibinfo{year}{2023}) \bibinfo{pages}{1887--1910}.
\bibitem[{Speksnijder et~al.(2025)Speksnijder, Karacadagli, Seyffert, and Grammatikopoulos}]{speksnijderApplicationHarmonicBalance2025a}
\bibinfo{author}{A.~Speksnijder}, \bibinfo{author}{U.~Karacadagli}, \bibinfo{author}{H.~Seyffert}, \bibinfo{author}{A.~Grammatikopoulos},
\newblock \bibinfo{title}{Application of the harmonic balance method for ship-cargo interaction with intermittent contact nonlinearities},
\newblock \bibinfo{journal}{Journal of Sound and Vibration} \bibinfo{volume}{601} (\bibinfo{year}{2025}) \bibinfo{pages}{118925}.
\bibitem[{Li and Cai(2026)}]{liModelingDynamicAnalysis2026a}
\bibinfo{author}{T.~Li}, \bibinfo{author}{Y.~Cai},
\newblock \bibinfo{title}{Modeling and dynamic analysis of belt drive systems using a harmonic balance and alternating frequency/time domain method},
\newblock \bibinfo{journal}{Engineering Computations} \bibinfo{volume}{43} (\bibinfo{year}{2026}) \bibinfo{pages}{452--482}.
\bibitem[{Zhang et~al.(2024)Zhang, Zhang, and Zhang}]{zhangNonlinearVibrationsPoroushyperelastic2024}
\bibinfo{author}{J.~Zhang}, \bibinfo{author}{W.~Zhang}, \bibinfo{author}{Y.~Zhang},
\newblock \bibinfo{title}{Nonlinear vibrations of porous-hyperelastic cylindrical shell under harmonic force using harmonic balance and pseudo-arc length continuation methods},
\newblock \bibinfo{journal}{Thin-Walled Structures} \bibinfo{volume}{198} (\bibinfo{year}{2024}) \bibinfo{pages}{111767}.
\bibitem[{Woiwode et~al.(2020)Woiwode, Balaji, Kappauf, Tubita, Guillot, Vergez, Cochelin, Grolet, and Krack}]{woiwodeComparisonTwoAlgorithms2020}
\bibinfo{author}{L.~Woiwode}, \bibinfo{author}{N.~N. Balaji}, \bibinfo{author}{J.~Kappauf}, \bibinfo{author}{F.~Tubita}, \bibinfo{author}{L.~Guillot}, \bibinfo{author}{C.~Vergez}, \bibinfo{author}{B.~Cochelin}, \bibinfo{author}{A.~Grolet}, \bibinfo{author}{M.~Krack},
\newblock \bibinfo{title}{Comparison of two algorithms for {{Harmonic Balance}} and path continuation},
\newblock \bibinfo{journal}{Mechanical Systems and Signal Processing} \bibinfo{volume}{136} (\bibinfo{year}{2020}) \bibinfo{pages}{106503}.
\bibitem[{Lee and Park(2023)}]{leeProperGeneralizedDecompositionbased2023}
\bibinfo{author}{G.-Y. Lee}, \bibinfo{author}{Y.-H. Park},
\newblock \bibinfo{title}{A proper generalized decomposition-based harmonic balance method with arc-length continuation for nonlinear frequency response analysis},
\newblock \bibinfo{journal}{Computers \& Structures} \bibinfo{volume}{275} (\bibinfo{year}{2023}) \bibinfo{pages}{106913}.
\bibitem[{Wu et~al.(2025)Wu, Wang, and Cao}]{wuNonlinearDynamicAnalysis2025}
\bibinfo{author}{Q.~Wu}, \bibinfo{author}{Y.~Wang}, \bibinfo{author}{D.~Cao},
\newblock \bibinfo{title}{Nonlinear dynamic analysis of high aspect ratio wings via {{IHB}} method},
\newblock \bibinfo{journal}{Nonlinear Dynamics} \bibinfo{volume}{113} (\bibinfo{year}{2025}) \bibinfo{pages}{16225--16244}.
\bibitem[{Guo et~al.(2026)Guo, Ren, Yao, Fan, Ju, and Zhou}]{guoSemianalyticalHopfBifurcation2026}
\bibinfo{author}{F.~Guo}, \bibinfo{author}{H.~Ren}, \bibinfo{author}{Y.~Yao}, \bibinfo{author}{W.~Fan}, \bibinfo{author}{R.~Ju}, \bibinfo{author}{P.~Zhou},
\newblock \bibinfo{title}{Semi-analytical {{Hopf}} bifurcation analysis of self-excited hunting motion in railway vehicles},
\newblock \bibinfo{journal}{Applied Mathematical Modelling}  (\bibinfo{year}{2026}) \bibinfo{pages}{117144}.
\bibitem[{Zhou et~al.(2015)Zhou, Thouverez, and Lenoir}]{zhouVariablecoefficientHarmonicBalance2015}
\bibinfo{author}{B.~Zhou}, \bibinfo{author}{F.~Thouverez}, \bibinfo{author}{D.~Lenoir},
\newblock \bibinfo{title}{A variable-coefficient harmonic balance method for the prediction of quasi-periodic response in nonlinear systems},
\newblock \bibinfo{journal}{Mechanical Systems and Signal Processing} \bibinfo{volume}{64--65} (\bibinfo{year}{2015}) \bibinfo{pages}{233--244}.
\bibitem[{Huang et~al.(2021)Huang, Wang, and Zhu}]{huangIncrementalHarmonicBalance2021}
\bibinfo{author}{J.~Huang}, \bibinfo{author}{T.~Wang}, \bibinfo{author}{W.~Zhu},
\newblock \bibinfo{title}{An incremental harmonic balance method with two time-scales for quasi-periodic responses of a {{Van}} der {{Pol}}--{{Mathieu}} equation},
\newblock \bibinfo{journal}{International Journal of Non-Linear Mechanics} \bibinfo{volume}{135} (\bibinfo{year}{2021}) \bibinfo{pages}{103767}.
\bibitem[{Klinger et~al.(1995)Klinger, Greiner, Rohde, Piel, and Koepke}]{klingerVanPolBehavior1995}
\bibinfo{author}{T.~Klinger}, \bibinfo{author}{F.~Greiner}, \bibinfo{author}{A.~Rohde}, \bibinfo{author}{A.~Piel}, \bibinfo{author}{M.~E. Koepke},
\newblock \bibinfo{title}{Van der {{Pol}} behavior of relaxation oscillations in a periodically driven thermionic discharge},
\newblock \bibinfo{journal}{Physical Review E} \bibinfo{volume}{52} (\bibinfo{year}{1995}) \bibinfo{pages}{4316--4327}.
\bibitem[{Seydel(2010)}]{seydelPracticalBifurcationStability2010}
\bibinfo{author}{R.~Seydel}, \bibinfo{title}{Practical {{Bifurcation}} and {{Stability Analysis}}}, volume~\bibinfo{volume}{5} of \textit{\bibinfo{series}{Interdisciplinary {{Applied Mathematics}}}}, \bibinfo{publisher}{Springer New York}, \bibinfo{address}{New York, NY}, \bibinfo{year}{2010}. \DOIprefix\doi{10.1007/978-1-4419-1740-9}.
\bibitem[{Doedel et~al.(1997)Doedel, Paenroth, Champneys, and Fairgrieve}]{Doedel1997AUTO2}
\bibinfo{author}{E.~J. Doedel}, \bibinfo{author}{R.~C. Paenroth}, \bibinfo{author}{A.~R. Champneys}, \bibinfo{author}{T.~F. Fairgrieve},
\newblock \bibinfo{title}{Auto 2000 : Continuation and bifurcation software for ordinary differential equations (with homcont)},
\newblock \bibinfo{year}{1997}. \URLprefix \url{https://api.semanticscholar.org/CorpusID:117201224}.
\bibitem[{Taege et~al.(2025)Taege, Schiefhauer, Ali, and M\"{u}ller}]{TAEGE2025680}
\bibinfo{author}{S.~Taege}, \bibinfo{author}{H.~Schiefhauer}, \bibinfo{author}{A.~Ali}, \bibinfo{author}{B.~M\"{u}ller},
\newblock \bibinfo{title}{Floquet theory based stability analysis of low-speed sensorless control},
\newblock \bibinfo{journal}{IFAC-PapersOnLine} \bibinfo{volume}{59} (\bibinfo{year}{2025}) \bibinfo{pages}{680--685}.
\bibitem[{Wang et~al.(2025)Wang, Cai, Liao, Hu, Zhang, Sun, Zhong, and Li}]{WANG2025136279}
\bibinfo{author}{Y.~Wang}, \bibinfo{author}{C.~Cai}, \bibinfo{author}{C.~Liao}, \bibinfo{author}{Z.~Hu}, \bibinfo{author}{L.~Zhang}, \bibinfo{author}{X.~Sun}, \bibinfo{author}{X.~Zhong}, \bibinfo{author}{Q.~Li},
\newblock \bibinfo{title}{Aeroelastic stability analysis of large-scale wind turbine blades under different operating conditions based on system identification and floquet theory},
\newblock \bibinfo{journal}{Energy} \bibinfo{volume}{326} (\bibinfo{year}{2025}) \bibinfo{pages}{136279}.
\bibitem[{Riahi et~al.(2025)Riahi, {Hayani Choujaa}, and Wang}]{RIAHI2025109507}
\bibinfo{author}{M.~Riahi}, \bibinfo{author}{M.~{Hayani Choujaa}}, \bibinfo{author}{S.~Wang},
\newblock \bibinfo{title}{Flow reversal and neimark–sacker bifurcations in zero-mean time-modulated temperatures in a hele-shaw cell: Floquet theory and wentzel–kramers–brillouin analysis},
\newblock \bibinfo{journal}{International Communications in Heat and Mass Transfer} \bibinfo{volume}{169} (\bibinfo{year}{2025}) \bibinfo{pages}{109507}.
\bibitem[{Ricci and Pennacchi(2012)}]{ricciDiscussionDynamicStability2012}
\bibinfo{author}{R.~Ricci}, \bibinfo{author}{P.~Pennacchi},
\newblock \bibinfo{title}{Discussion of the dynamic stability of a multi-degree-of-freedom rotor system affected by a transverse crack},
\newblock \bibinfo{journal}{Mechanism and Machine Theory} \bibinfo{volume}{58} (\bibinfo{year}{2012}) \bibinfo{pages}{82--100}.
\bibitem[{Hsu and Cheng(1973)}]{Hsu1973ApplicationsOT}
\bibinfo{author}{C.~S. Hsu}, \bibinfo{author}{W.~Cheng},
\newblock \bibinfo{title}{Applications of the theory of impulsive parametric excitation and new treatments of general parametric excitation problems},
\newblock \bibinfo{journal}{Journal of Applied Mechanics} \bibinfo{volume}{40} (\bibinfo{year}{1973}) \bibinfo{pages}{78--86}.
\bibitem[{Cheung et~al.(1990)Cheung, Chen, and Lau}]{cheungApplicationIncrementalHarmonic1990}
\bibinfo{author}{Y.~Cheung}, \bibinfo{author}{S.~Chen}, \bibinfo{author}{S.~Lau},
\newblock \bibinfo{title}{Application of the incremental harmonic balance method to cubic non-linearity systems},
\newblock \bibinfo{journal}{Journal of Sound and Vibration} \bibinfo{volume}{140} (\bibinfo{year}{1990}) \bibinfo{pages}{273--286}.
\bibitem[{Yang et~al.(2025)Yang, Huang, and Zhu}]{YANG2025116240}
\bibinfo{author}{D.~Yang}, \bibinfo{author}{J.~Huang}, \bibinfo{author}{W.~Zhu},
\newblock \bibinfo{title}{Nonlinear dynamics of a misaligned and eccentric rotor system with three base motions},
\newblock \bibinfo{journal}{Applied Mathematical Modelling} \bibinfo{volume}{148} (\bibinfo{year}{2025}) \bibinfo{pages}{116240}.
\bibitem[{Jiang et~al.(2025)Jiang, Liu, Yuan, Cao, Shi, and Qin}]{JIANG2025110482}
\bibinfo{author}{W.~Jiang}, \bibinfo{author}{K.~Liu}, \bibinfo{author}{X.~Yuan}, \bibinfo{author}{H.~Cao}, \bibinfo{author}{J.~Shi}, \bibinfo{author}{Q.~Qin},
\newblock \bibinfo{title}{Nonlinear dynamics of rotor-support-casing system with support looseness fault},
\newblock \bibinfo{journal}{International Journal of Mechanical Sciences} \bibinfo{volume}{300} (\bibinfo{year}{2025}) \bibinfo{pages}{110482}.
\bibitem[{Yao et~al.(2026)Yao, Huang, Yang, and J\'{e}z\'{e}quel}]{YAO2026104180}
\bibinfo{author}{Y.~Yao}, \bibinfo{author}{X.~Huang}, \bibinfo{author}{X.~Yang}, \bibinfo{author}{L.~J\'{e}z\'{e}quel},
\newblock \bibinfo{title}{Nonlinear dynamics analysis of rotor systems based on extended nonlinear modal theory with gyroscopic coupling},
\newblock \bibinfo{journal}{Chinese Journal of Aeronautics}  (\bibinfo{year}{2026}) \bibinfo{pages}{104180}.
\bibitem[{Baglioni et~al.(2016)Baglioni, Cianetti, Braccesi, and De~Micheli}]{baglioniMultibodyModellingDOF2016}
\bibinfo{author}{S.~Baglioni}, \bibinfo{author}{F.~Cianetti}, \bibinfo{author}{C.~Braccesi}, \bibinfo{author}{D.~M. De~Micheli},
\newblock \bibinfo{title}{Multibody modelling of {{N DOF}} robot arm assigned to milling manufacturing: {{Dynamic}} analysis and position errors evaluation},
\newblock \bibinfo{journal}{Journal of Mechanical Science and Technology} \bibinfo{volume}{30} (\bibinfo{year}{2016}) \bibinfo{pages}{405--420}.
\bibitem[{Cao et~al.(2025)Cao, Chen, Wang, Salles, and Zheng}]{caoPerturbationFunctionIteration2025}
\bibinfo{author}{L.~Cao}, \bibinfo{author}{Y.~Chen}, \bibinfo{author}{L.~Wang}, \bibinfo{author}{L.~Salles}, \bibinfo{author}{Z.~Zheng}, \bibinfo{title}{Perturbation {{Function Iteration Method}}: {{A New Framework}} for {{Solving Periodic Solutions}} of {{Non-linear}} and {{Non-smooth Systems}}}, \bibinfo{year}{2025}. \href{http://arxiv.org/abs/2510.23071}{\tt arXiv:2510.23071}.
\bibitem[{Deshpande et~al.(2006)Deshpande, Mehta, and Jazar}]{deshpandeOptimizationSecondarySuspension2006}
\bibinfo{author}{S.~Deshpande}, \bibinfo{author}{S.~Mehta}, \bibinfo{author}{G.~N. Jazar},
\newblock \bibinfo{title}{Optimization of secondary suspension of piecewise linear vibration isolation systems},
\newblock \bibinfo{journal}{International Journal of Mechanical Sciences} \bibinfo{volume}{48} (\bibinfo{year}{2006}) \bibinfo{pages}{341--377}.
\bibitem[{Teloli and Da~Silva(2019)}]{teloliNewWayHarmonic2019}
\bibinfo{author}{R.~D.~O. Teloli}, \bibinfo{author}{S.~Da~Silva},
\newblock \bibinfo{title}{A new way for harmonic probing of hysteretic systems through nonlinear smooth operators},
\newblock \bibinfo{journal}{Mechanical Systems and Signal Processing} \bibinfo{volume}{121} (\bibinfo{year}{2019}) \bibinfo{pages}{856--875}.
\bibitem[{Wang et~al.(2023)Wang, Zhang, Guo, Pi, and Li}]{wangVibrationAnalysisNonlinear2023}
\bibinfo{author}{S.~Wang}, \bibinfo{author}{Y.~Zhang}, \bibinfo{author}{W.~Guo}, \bibinfo{author}{T.~Pi}, \bibinfo{author}{X.~Li},
\newblock \bibinfo{title}{Vibration analysis of nonlinear damping systems by the discrete incremental harmonic balance method},
\newblock \bibinfo{journal}{Nonlinear Dynamics} \bibinfo{volume}{111} (\bibinfo{year}{2023}) \bibinfo{pages}{2009--2028}.
\bibitem[{Wang et~al.(2019)Wang, Hua, Yang, Han, and Su}]{wangApplicationsIncrementalHarmonic2019}
\bibinfo{author}{S.~Wang}, \bibinfo{author}{L.~Hua}, \bibinfo{author}{C.~Yang}, \bibinfo{author}{X.~Han}, \bibinfo{author}{Z.~Su},
\newblock \bibinfo{title}{Applications of incremental harmonic balance method combined with equivalent piecewise linearization on vibrations of nonlinear stiffness systems},
\newblock \bibinfo{journal}{Journal of Sound and Vibration} \bibinfo{volume}{441} (\bibinfo{year}{2019}) \bibinfo{pages}{111--125}.
\bibitem[{Wengert(1964)}]{wengertSimpleAutomaticDerivative1964}
\bibinfo{author}{R.~E. Wengert},
\newblock \bibinfo{title}{A simple automatic derivative evaluation program},
\newblock \bibinfo{journal}{Communications of the ACM} \bibinfo{volume}{7} (\bibinfo{year}{1964}) \bibinfo{pages}{463--464}.
\bibitem[{Hanson et~al.(1962)Hanson, Caviness, and Joseph}]{hansonAnalyticDifferentiationComputer1962}
\bibinfo{author}{J.~W. Hanson}, \bibinfo{author}{J.~S. Caviness}, \bibinfo{author}{C.~Joseph},
\newblock \bibinfo{title}{Analytic differentiation by computer},
\newblock \bibinfo{journal}{Communications of the ACM} \bibinfo{volume}{5} (\bibinfo{year}{1962}) \bibinfo{pages}{349--355}.
\bibitem[{Baydin et~al.(2017)Baydin, Pearlmutter, Radul, and Siskind}]{3122009_3242010}
\bibinfo{author}{A.~G. Baydin}, \bibinfo{author}{B.~A. Pearlmutter}, \bibinfo{author}{A.~A. Radul}, \bibinfo{author}{J.~M. Siskind},
\newblock \bibinfo{title}{Automatic differentiation in machine learning: a survey},
\newblock \bibinfo{journal}{Journal of Machine Learning Research} \bibinfo{volume}{18} (\bibinfo{year}{2017}) \bibinfo{pages}{5595--5637}.
\bibitem[{Paszke et~al.(2019)Paszke, Gross, Massa, Lerer, Bradbury, Chanan, Killeen, Lin, Gimelshein, Antiga, Desmaison, Köpf, Yang, DeVito, Raison, Tejani, Chilamkurthy, Steiner, Fang, Bai, and Chintala}]{paszke2019pytorchimperativestylehighperformance}
\bibinfo{author}{A.~Paszke}, \bibinfo{author}{S.~Gross}, \bibinfo{author}{F.~Massa}, \bibinfo{author}{A.~Lerer}, \bibinfo{author}{J.~Bradbury}, \bibinfo{author}{G.~Chanan}, \bibinfo{author}{T.~Killeen}, \bibinfo{author}{Z.~Lin}, \bibinfo{author}{N.~Gimelshein}, \bibinfo{author}{L.~Antiga}, \bibinfo{author}{A.~Desmaison}, \bibinfo{author}{A.~Köpf}, \bibinfo{author}{E.~Yang}, \bibinfo{author}{Z.~DeVito}, \bibinfo{author}{M.~Raison}, \bibinfo{author}{A.~Tejani}, \bibinfo{author}{S.~Chilamkurthy}, \bibinfo{author}{B.~Steiner}, \bibinfo{author}{L.~Fang}, \bibinfo{author}{J.~Bai}, \bibinfo{author}{S.~Chintala}, \bibinfo{title}{Pytorch: An imperative style, high-performance deep learning library}, \bibinfo{year}{2019}. \href{http://arxiv.org/abs/1912.01703}{\tt arXiv:1912.01703}.
\bibitem[{{PyTorch Contributors}(2026)}]{pytorch_autograd_mechanics_2026}
\bibinfo{author}{{PyTorch Contributors}}, \bibinfo{title}{Autograd mechanics}, \bibinfo{howpublished}{\url{https://docs.pytorch.org/docs/2.12/notes/autograd.html}}, \bibinfo{year}{2026}. \bibinfo{note}{PyTorch 2.12 documentation, last updated Jan. 6, 2026; accessed Jul. 4, 2026}.
\bibitem[{Chen et~al.(2024)Chen, Hou, Lin, Toh, Ng, and Chen}]{chenGeneralEfficientHarmonic2024}
\bibinfo{author}{Y.~Chen}, \bibinfo{author}{L.~Hou}, \bibinfo{author}{R.~Lin}, \bibinfo{author}{W.~Toh}, \bibinfo{author}{T.~Ng}, \bibinfo{author}{Y.~Chen},
\newblock \bibinfo{title}{A general and efficient harmonic balance method for nonlinear dynamic simulation},
\newblock \bibinfo{journal}{International Journal of Mechanical Sciences} \bibinfo{volume}{276} (\bibinfo{year}{2024}) \bibinfo{pages}{109388}.
\bibitem[{Riks(1979)}]{riksIncrementalApproachSolution1979}
\bibinfo{author}{E.~Riks},
\newblock \bibinfo{title}{An incremental approach to the solution of snapping and buckling problems},
\newblock \bibinfo{journal}{International Journal of Solids and Structures} \bibinfo{volume}{15} (\bibinfo{year}{1979}) \bibinfo{pages}{529--551}.
\bibitem[{Crisfield(1981)}]{crisfieldFastIncrementalIterative1981}
\bibinfo{author}{M.~Crisfield},
\newblock \bibinfo{title}{A fast incremental/iterative solution procedure that handles ``snap-through''},
\newblock \bibinfo{journal}{Computers \& Structures} \bibinfo{volume}{13} (\bibinfo{year}{1981}) \bibinfo{pages}{55--62}.
\bibitem[{Doedel et~al.(2007)Doedel, Champneys, Dercole, Fairgrieve, Kuznetsov, Oldeman, Paffenroth, Sandstede, Wang, and Zhang}]{doedel2007auto07p}
\bibinfo{author}{E.~J. Doedel}, \bibinfo{author}{A.~R. Champneys}, \bibinfo{author}{F.~Dercole}, \bibinfo{author}{T.~F. Fairgrieve}, \bibinfo{author}{Y.~A. Kuznetsov}, \bibinfo{author}{B.~E. Oldeman}, \bibinfo{author}{R.~C. Paffenroth}, \bibinfo{author}{B.~Sandstede}, \bibinfo{author}{X.~Wang}, \bibinfo{author}{C.~Zhang}, \bibinfo{title}{{AUTO-07P}: Continuation and Bifurcation Software for Ordinary Differential Equations}, \bibinfo{type}{Software manual}, Concordia University, \bibinfo{address}{Montreal, Canada}, \bibinfo{year}{2007}. \URLprefix \url{https://cmvl.cs.concordia.ca/auto/}.
\bibitem[{Peletan et~al.(2013)Peletan, Baguet, Torkhani, and {Jacquet-Richardet}}]{peletanComparisonStabilityComputational2013}
\bibinfo{author}{L.~Peletan}, \bibinfo{author}{S.~Baguet}, \bibinfo{author}{M.~Torkhani}, \bibinfo{author}{G.~{Jacquet-Richardet}},
\newblock \bibinfo{title}{A comparison of stability computational methods for periodic solution of nonlinear problems with application to rotordynamics},
\newblock \bibinfo{journal}{Nonlinear Dynamics} \bibinfo{volume}{72} (\bibinfo{year}{2013}) \bibinfo{pages}{671--682}.
\bibitem[{Gaonkar et~al.(1981)Gaonkar, Simha~Prasad, and Sastry}]{gaonkarComputingFloquetTransition1981}
\bibinfo{author}{G.~H. Gaonkar}, \bibinfo{author}{D.~S. Simha~Prasad}, \bibinfo{author}{D.~Sastry},
\newblock \bibinfo{title}{On {{Computing Floquet Transition Matrices}} of {{Rotorcraft}}},
\newblock \bibinfo{journal}{Journal of the American Helicopter Society} \bibinfo{volume}{26} (\bibinfo{year}{1981}) \bibinfo{pages}{56--61}.
\bibitem[{Fu et~al.(2026)Fu, Miao, Hong, Li, Qu, Ma, and Wang}]{FU2026112475}
\bibinfo{author}{J.~Fu}, \bibinfo{author}{H.~Miao}, \bibinfo{author}{J.~Hong}, \bibinfo{author}{C.~Li}, \bibinfo{author}{X.~Qu}, \bibinfo{author}{Y.~Ma}, \bibinfo{author}{Y.~Wang},
\newblock \bibinfo{title}{Adaptive optimization design framework for vibration transfer suppression in aero-engine dual-rotor systems with shared bearing frame},
\newblock \bibinfo{journal}{Aerospace Science and Technology} \bibinfo{volume}{176} (\bibinfo{year}{2026}) \bibinfo{pages}{112475}.
\bibitem[{Su et~al.(2026)Su, Zhou, Yan, Qiu, and Gao}]{SU2026114471}
\bibinfo{author}{H.~Su}, \bibinfo{author}{W.~Zhou}, \bibinfo{author}{L.~Yan}, \bibinfo{author}{N.~Qiu}, \bibinfo{author}{B.~Gao},
\newblock \bibinfo{title}{Theoretical modeling and experimental validation of a dual-disk rotor system with coupled spatial crack and bearing outer race defects},
\newblock \bibinfo{journal}{Mechanical Systems and Signal Processing} \bibinfo{volume}{255} (\bibinfo{year}{2026}) \bibinfo{pages}{114471}.
\bibitem[{Gao et~al.(2020)Gao, Chen, and Hou}]{gaoNonlinearThermalBehaviors2020}
\bibinfo{author}{P.~Gao}, \bibinfo{author}{Y.~Chen}, \bibinfo{author}{L.~Hou},
\newblock \bibinfo{title}{Nonlinear thermal behaviors of the inter-shaft bearing in a dual-rotor system subjected to the dynamic load},
\newblock \bibinfo{journal}{Nonlinear Dynamics} \bibinfo{volume}{101} (\bibinfo{year}{2020}) \bibinfo{pages}{191--209}.
\bibitem[{Tang and Yang(2012)}]{tangNonlinearPiezoelectricEnergy2012a}
\bibinfo{author}{L.~Tang}, \bibinfo{author}{Y.~Yang},
\newblock \bibinfo{title}{A nonlinear piezoelectric energy harvester with magnetic oscillator},
\newblock \bibinfo{journal}{Applied Physics Letters} \bibinfo{volume}{101} (\bibinfo{year}{2012}) \bibinfo{pages}{094102}.
\bibitem[{Ponsioen et~al.(2020)Ponsioen, Jain, and Haller}]{ponsioenModelReductionSpectral2020}
\bibinfo{author}{S.~Ponsioen}, \bibinfo{author}{S.~Jain}, \bibinfo{author}{G.~Haller},
\newblock \bibinfo{title}{Model reduction to spectral submanifolds and forced-response calculation in high-dimensional mechanical systems},
\newblock \bibinfo{journal}{Journal of Sound and Vibration} \bibinfo{volume}{488} (\bibinfo{year}{2020}) \bibinfo{pages}{115640}.
\bibitem[{Dankowicz and Schilder(2013)}]{dankowiczRecipesContinuation2013}
\bibinfo{author}{H.~Dankowicz}, \bibinfo{author}{F.~Schilder}, \bibinfo{title}{Recipes for {{Continuation}}}, \bibinfo{publisher}{{Society for Industrial and Applied Mathematics}}, \bibinfo{address}{Philadelphia, PA}, \bibinfo{year}{2013}. \DOIprefix\doi{10.1137/1.9781611972573}.
\bibitem[{Wang and Zhu(2017)}]{wangDynamicAnalysisAutomotive2017}
\bibinfo{author}{X.~F. Wang}, \bibinfo{author}{W.~D. Zhu},
\newblock \bibinfo{title}{Dynamic {{Analysis}} of an {{Automotive Belt-Drive System With}} a {{Noncircular Sprocket}} by a {{Modified Incremental Harmonic Balance Method}}},
\newblock \bibinfo{journal}{Journal of Vibration and Acoustics} \bibinfo{volume}{139} (\bibinfo{year}{2017}) \bibinfo{pages}{011009}.
\bibitem[{Veerse(2003)}]{1257664}
\bibinfo{author}{F.~Veerse},
\newblock \bibinfo{title}{Efficient iterative time preconditioners for harmonic balance rf circuit simulation},
\newblock in: \bibinfo{booktitle}{ICCAD-2003. International Conference on Computer Aided Design (IEEE Cat. No.03CH37486)}, \bibinfo{year}{2003}, pp. \bibinfo{pages}{251--254}. \DOIprefix\doi{10.1109/ICCAD.2003.159698}.
\bibitem[{Lindblad and Andersson(2020)}]{lindbladConvergenceAccelerationHarmonic2020a}
\bibinfo{author}{D.~Lindblad}, \bibinfo{author}{N.~Andersson},
\newblock \bibinfo{title}{Convergence {{Acceleration}} of the {{Harmonic Balance Method Using}} a {{Time-Level Preconditioner}}},
\newblock \bibinfo{journal}{AIAA Journal} \bibinfo{volume}{58} (\bibinfo{year}{2020}) \bibinfo{pages}{4908--4922}.
\bibitem[{Tan et~al.(2025)Tan, Su, Yang, and Zeng}]{11132845}
\bibinfo{author}{C.~Tan}, \bibinfo{author}{Y.~Su}, \bibinfo{author}{F.~Yang}, \bibinfo{author}{X.~Zeng},
\newblock \bibinfo{title}{New time-domain preconditioners for hb jacobian of rf circuits},
\newblock in: \bibinfo{booktitle}{2025 62nd ACM/IEEE Design Automation Conference (DAC)}, \bibinfo{year}{2025}, pp. \bibinfo{pages}{1--7}. \DOIprefix\doi{10.1109/DAC63849.2025.11132845}.
\bibitem[{Kuether and Steyer(2024)}]{kuetherLargescaleHarmonicBalance2024}
\bibinfo{author}{R.~J. Kuether}, \bibinfo{author}{A.~Steyer},
\newblock \bibinfo{title}{Large-scale harmonic balance simulations with {{Krylov}} subspace and preconditioner recycling},
\newblock \bibinfo{journal}{Nonlinear Dynamics} \bibinfo{volume}{112} (\bibinfo{year}{2024}) \bibinfo{pages}{3377--3398}.
\bibitem[{Jain et~al.(2019)Jain, Breunung, and Haller}]{jainFastComputationSteadystate2019a}
\bibinfo{author}{S.~Jain}, \bibinfo{author}{T.~Breunung}, \bibinfo{author}{G.~Haller},
\newblock \bibinfo{title}{Fast computation of steady-state response for high-degree-of-freedom nonlinear systems},
\newblock \bibinfo{journal}{Nonlinear Dynamics} \bibinfo{volume}{97} (\bibinfo{year}{2019}) \bibinfo{pages}{313--341}.
\bibitem[{Ju et~al.(2024)Ju, Yang, Ren, Fan, Ni, and Gu}]{juSteadyStateRotaryPeriodic2024}
\bibinfo{author}{R.~Ju}, \bibinfo{author}{S.~M. Yang}, \bibinfo{author}{H.~Ren}, \bibinfo{author}{W.~Fan}, \bibinfo{author}{R.~C. Ni}, \bibinfo{author}{P.~Gu},
\newblock \bibinfo{title}{Steady-{{State Rotary Periodic Solutions}} of {{Rigid}} and {{Flexible Mechanisms With Large Spatial Rotations Using}} the {{Incremental Harmonic Balance Method}} for {{Differential-Algebraic Equations}}},
\newblock \bibinfo{journal}{Journal of Computational and Nonlinear Dynamics} \bibinfo{volume}{19} (\bibinfo{year}{2024}) \bibinfo{pages}{121001}.
\bibitem[{Zhou et~al.(2026)Zhou, Heya, and Inoue}]{zhouConstraintEliminationbasedHarmonic2026a}
\bibinfo{author}{X.~Zhou}, \bibinfo{author}{A.~Heya}, \bibinfo{author}{T.~Inoue},
\newblock \bibinfo{title}{A constraint elimination-based harmonic balance method for multibody systems},
\newblock \bibinfo{journal}{International Journal of Mechanical Sciences} \bibinfo{volume}{317} (\bibinfo{year}{2026}) \bibinfo{pages}{111490}.

\end{thebibliography}



\end{document}